\documentclass[useAMS,usenatbib]{mn2e}

\usepackage{times}

\usepackage[T1]{fontenc} 
\usepackage{aecompl} 

\usepackage{graphicx}
\usepackage{times}
\newcommand{\cthead}[1]{\multicolumn{1}{c}{#1}}
\newcommand{\kss}{km~s$^{-1}$ }
\newcommand{\ks}{km~s$^{-1}$}
\title{Southern class~I methanol masers at 36 and 44 GHz}
\author[
M. A. Voronkov et al.
]{M. A. Voronkov$^{1,2,3}$\thanks{E-mail:
Maxim.Voronkov@csiro.au}, J. L. Caswell$^1$, S. P. Ellingsen$^{3}$, J. A. Green$^{4,1}$ and S. L. Breen$^{1,3}$\\
$^{1}$Australia Telescope National Facility, CSIRO Astronomy and Space Science, PO Box 76, Epping,
NSW 1710, Australia\\
$^{2}$Astro Space Centre, Profsouznaya st. 84/32, 117997 Moscow, Russia\\
$^{3}$School of Mathematics and Physics, University of Tasmania, GPO Box
252-37, Hobart, Tasmania 7000, Australia\\
$^{4}$SKA Organisation, Jodrell Bank Observatory, Lower Withington, Macclesfield, Cheshire SK11 9DL, UK\\}
\begin{document}

\date{}

\pagerange{\pageref{firstpage}--\pageref{lastpage}} \pubyear{2010}

\maketitle

\label{firstpage}

\begin{abstract}
The Australia Telescope Compact Array (ATCA) has been used for high angular resolution imaging of
71 southern class~I methanol maser sources quasi-simultaneously at 36 and 44~GHz. The data 
reveal a high level of
morphological and kinematical complexity, and allow us to demonstrate 
associations, at arcsecond precision, of the class~I maser emission with outflows, expanding H{\sc ii} regions, 
dark clouds, shocks traced by the 4.5-$\mu$m emission and 8.0-$\mu$m filaments.
More than 700 maser component features were found at each of the two methanol transitions, but
with only 23~per cent recognisable at both transitions; the morphology of class~I emission is much better 
revealed by our survey of both transitions, compared with either one alone.
We found that the number of masers falls exponentially with the projected linear distance from the
associated class~II 6.7-GHz methanol maser. This distribution has a scale of 263$\pm$15~mpc, irrespective 
of the transition. The class~I masers associated with OH masers were found to have a tendency to be
more spread out, both spatially and in the velocity domain. This is consistent with the expectation that such
sources are more evolved. 
Apart from a small number of high-velocity components (which are largely blue-shifted and predominantly
seen at 36~GHz), the velocity distribution was found to be Gaussian, peaking near the systemic
velocity of the region, which had been estimated as the middle of the velocity interval of the associated class~II methanol 
maser at 6.7~GHz. The mean indicated a small, but significant blue shift asymmetry of $-$0.57~\kss (uncertainties
are 0.06 and 0.07~\kss for the 36- and 44-GHz masers, respectively) with respect to the 6.7-GHz masers. The standard deviation of the 
velocity distribution was found to be 3.65$\pm$0.05 and 3.32$\pm$0.07~\kss for the 
36- and 44-GHz masers, respectively. We also suggest a refined rest frequency 
value of 36169.238$\pm$0.011~MHz for the 4$_{-1}-$3$_0$~E methanol transition.

\end{abstract}


\section{Introduction}
Methanol masers are common phenomena in the regions of active star formation,
with more than twenty different centimetre and millimetre wavelength
maser transitions discovered to date \citep[e.g.,][]{mul04}. Early studies 
have empirically divided methanol maser transitions into two classes based
on the observed spatial distribution \citep[e.g.,][]{bat87}.
Class~I methanol masers (e.g. the most widespread masers at 36 and 44~GHz) usually occur in
multiple locations across the star forming region and are often scattered around an area
up to 1~parsec in linear extent \citep[e.g.,][]{kur04,vor06,cyg09}. In contrast, the class~II
methanol masers (e.g. well-known masers at 6.7-GHz) reside in close proximity to  individual
young stellar objects (YSOs) and are rarely found in multiple locations across the
star forming region at arcsec resolution \citep[e.g.,][]{cas97,phi98,wal98}. Exclusively associated 
with high-mass star formation (see, e.g., \citet{bre13} and references therein), class~II methanol 
masers are, therefore, excellent signposts of the exact location of high-mass YSOs. In this capacity,
they have been the subject of numerous observational studies in the last two decades, which culminated in
the Methanol Multibeam survey \citep[MMB; e.g.,][]{gre09,cas10,cas11a,gre12a} providing the most
accurate systematic census to date of the 6.7-GHz masers visible from the southern hemisphere,
including measurement of their absolute positions with arcsecond accuracy. 
It is worth noting that the empirical classification of methanol maser transitions and the drastically different observational
properties are now well understood through modelling of maser pumping \citep[e.g.,][]{cra92,vor05a}.
These models strongly suggest that the class~I masers are pumped via collisions with molecular 
hydrogen, in contrast to the class~II masers which are pumped by infrared radiation \citep[see also][]{sob94,vor06}.

To date, a few hundred class~I methanol maser sources are known, but the majority still have only single dish data 
\citep[for more details, see the review by][]{vor12}. 
The major published interferometric surveys are those by 
\citet{kur04} and \citet{cyg09}, which together provided maps for 56 mostly northern class~I methanol masers at 44~GHz.
Seven additional 44-GHz masers were imaged by \citet{kog98}, but phase referencing was not used in their experiment
making the absolute positions uncertain.
The reported interferometric observations of the 36-GHz methanol transition are scarce and limited to single source
papers only \citep{sjo10,vor10a,fis11,yus13}. At the frequencies of these molecular 
transitions the beam size of a 20-m class single dish is comparable to the typical angular spread of the class~I maser
emission \citep[e.g.,][]{kur04,vor06}. Therefore, interferometric observations, which allow us to measure accurate positions (and
also to correct for the beam attenuation to the first order), are required even for meaningful detection statistics. A search
for associations with other phenomena that are commonly observed in the crowded environment of high-mass star forming regions is 
another important application of interferometric observations. 

The majority of the class~I masers are believed to trace interface regions 
of outflows, although direct observational evidence of this has only been obtained at high angular resolution for a limited number of sources 
\citep[e.g.,][]{kur04,vor06,cyg12}. Also based on interferometric data, alternative scenarios have been 
demonstrated to occur in some sources, in particular the class~I masers have been shown to arise due to 
cloud-cloud collisions \citep{sjo10} and due to interaction of expanding H{\sc ii} regions with the ambient molecular gas \citep{vor10b,cyg12}.
The common feature of all these scenarios is the presence of shocked gas. Weak shocks are known to greatly enhance the methanol 
abundance by facilitating methanol release from dust grain mantles into the gas phase through heating \citep[e.g.,][]{har95,gib98,gar02a}.
At the same time, the shocks compress and heat up the gas, which is usually considered to be a favourable condition for the pumping of the class~I methanol masers
due to an increased rate of collisions between methanol and hydrogen molecules. Therefore, as discussed by \citet{vor10b}, the class~I masers
should be treated as merely shock tracers, where shocks could be caused by a range of phenomena. It is worth noting that \citet{yus13} suggested an
alternative methanol production mechanism due to cosmic rays and molecular gas interactions. This mechanism is most likely specific to the inner few 
hundred pc of the Galaxy (known as the central molecular zone).

In addition to the maser interferometry, in depth study of the morphology requires comparison datasets of similar angular resolution. The outflow identifications mentioned
above \citep{kur04,vor06} relied upon the images of the 2.12-$\mu$m molecular hydrogen emission, which is a well-known shock tracer \citep{hab05}. However, such data are 
significantly affected by extinction and have a quality sufficient for meaningful interpretation \citep[e.g. as in][]{bro03} for only a handful of sources. In contrast, the images 
obtained with the {\em Spitzer Space Telescope}'s Infrared Array Camera (IRAC) are now available for the majority of known class~I masers. It was shown by \citet{cyg09} that the 
class~I maser emission is often associated with the extended features showing a prominent excess in the 4.5-$\mu$m emission band in the {\em Spitzer} images. 
These features are also known as extended green objects \citep[EGOs;][]{cyg09} or `green fuzzies' \citep{cha09} because of the colour usually used to represent the 
4.5-$\mu$m emission in three-colour IRAC images. The excess of emission in this band is generally attributed to increased contribution from the emission of molecular hydrogen and carbon 
monoxide excited in shocks \citep[e.g.,][]{cha09,cyg09,deb10}. Other noteworthy structures often present in IRAC images include the 8.0-$\mu$m filaments and bubbles \citep[e.g.,][]{chu06}.
This IRAC band contains emission features of polycyclic aromatic hydrocarbons (PAHs), which often trace photodissociation regions (PDRs) around H{\sc ii} regions, although
near hot stars the emission in this band may be dominated by thermal dust emission with little or no contribution from PAHs \citep[e.g.,][]{wat08}.

In this paper, we present an interferometric survey at both 36 and 44~GHz of 71 southern class~I masers, and attempt to interpret the observed morphologies.
These are the strongest and the most common class~I maser transitions, which belong to two different transition series \citep[see][]{vor12}.
We also present a basic statistical analysis
of observed maser properties. Note that this is the largest interferometric survey of common class~I methanol masers available to date, and 
the first of this kind in the southern hemisphere. In addition, this is the first observation at 36~GHz for 
the majority of our targets and the first high angular resolution study of a large number of 36-GHz masers.


\section{Observations}

 
 \begin{table*}
 \caption{Observation details for all targets. Sources marked by an asterisk were observed using the same frequency configuration as the primary calibrator observations and
 potentially have better than average flux density accuracy.}
 \label{obsdetails}
 \begin{tabular}{l@{\hskip 7mm}rcc@{ }c@{}c@{}c@{}c@{}c}
 \cthead{Source} & \multicolumn{2}{c}{Pointing centre} &\multicolumn{3}{c}{36 GHz} & \multicolumn{3}{c}{44 GHz} \\
 \cthead{name}  & & &
 \cthead{Observation} & \cthead{$1\sigma$} & \cthead{Velocity} & \cthead{Observation} & \cthead{$1\sigma$} & 
 \cthead{Velocity}\\ 
 &\cthead{$\alpha_{2000}$}&\cthead{$\delta_{2000}$} & \cthead{date}&\cthead{noise} &\cthead{range} & \cthead{date}&
 \cthead{noise} &\cthead{range}\\
&\cthead{(h:m:s)}&\cthead{(\degr:arcmin:arcsec)} & \cthead{(year 2007)} & \cthead{(mJy)}&\cthead{(\ks)} &\cthead{(year 2007)}& 
\cthead{(mJy)}&\cthead{(\ks)}\\
  \hline
 G269.15$-$1.13 & 09:03:34.2 & $-$48:28:04 & May 21 & \hphantom{2}77 & $-$24, $+$35 & May 20 & 110 & $-$17, $+$32 \\
 G270.26$+$0.84 & 09:16:37.6 & $-$47:56:10 & May 21 & \hphantom{2}70 & $-$24, $+$35 & May 20 & 100 & $-$17, $+$31 \\
 G294.98$-$1.73 & 11:39:12.6 & $-$63:28:47 & May 21 & \hphantom{2}92 & $-$35, $+$24 & May 20 & \hphantom{2}86 & $-$30, $+$18 \\
 G300.97$+$1.15 & 12:34:52.5 & $-$61:40:02 & May 21 & \hphantom{2}57 & $-$74, $-$14 & May 20 & \hphantom{2}97 & $-$62, $-$13 \\
 G301.14$-$0.23 & 12:35:34.3 & $-$63:02:38 & May 21 & \hphantom{2}59 & $-$73, $-$14 & May 20 & \hphantom{2}92 & $-$61, $-$13 \\
 G305.21$+$0.21 & 13:11:09.6 & $-$62:34:41 & May 21 & \hphantom{2}57 & $-$71, $-$12 & May 20 & \hphantom{2}81 & $-$59, $-$10 \\
 G305.25$+$0.25 & 13:11:32.7 & $-$62:32:03 & May 21 & \hphantom{2}61 & $-$71, $-$12 & May 20 & \hphantom{2}84 & $-$59, $-$10 \\
 G305.37$+$0.21 & 13:12:33.8 & $-$62:33:25 & May 21 & \hphantom{2}65 & $-$71, $-$12 & May 20 & \hphantom{2}97 & $-$58, $-$10 \\
 G309.38$-$0.13 & 13:47:26.0 & $-$62:18:12 & May 25 & \hphantom{2}65 & $-$94, $-$35 & May 24 & 100 & $-$91, $-$43 \\
 G316.76$-$0.01 & 14:44:56.2 & $-$59:48:12 & May 21 & \hphantom{2}61 & $-$63, $-$4\hphantom{0} & May 20 & \hphantom{2}87 & $-$51, $-$2\hphantom{0} \\
 G318.05$+$0.09 & 14:53:42.6 & $-$59:09:01 & May 23 & \hphantom{2}69 & $-$54, $-$46 & May 22 & 136 & $-$54, $-$46 \\
 G318.95$-$0.20 & 15:00:53.2 & $-$58:59:06 & May 21 & \hphantom{2}54 & $-$62, $-$2\hphantom{0} & May 20 & \hphantom{2}84 & $-$49, $-$1\hphantom{0} \\
 G320.29$-$0.31 & 15:10:18.8 & $-$58:25:25 & May 23 & \hphantom{2}80 & $-$74, $-$56 & May 22 & 140 & $-$74, $-$56 \\
 G322.16$+$0.64 & 15:18:40.5 & $-$56:38:46 & May 25 & \hphantom{2}66 & $-$86, $-$27 & May 24 & 110 & $-$83, $-$35 \\
 G323.74$-$0.26 & 15:31:46.0 & $-$56:30:59 & May 25 & \hphantom{2}70 & $-$85, $-$26 & May 24 & 120 & $-$82, $-$33 \\
 G324.72$+$0.34 & 15:34:58.0 & $-$55:27:28 & May 25 & \hphantom{2}77 & $-$84, $-$25 & May 24 & 120 & $-$81, $-$33 \\
 G326.48$+$0.70\makebox[0mm]{~~$^*$} & 15:43:16.3 & $-$54:07:17 & May 23 & 110 & $-$68, $-$21 & May 22 & 190 & $-$54, $-$36 \\
 G326.64$+$0.61\makebox[0mm]{~~$^*$} & 15:44:32.7 & $-$54:05:31 & May 23 & \hphantom{2}82 & $-$74, $-$15 & May 22 & 150 & $-$66, $-$17 \\
 G326.66$+$0.57\makebox[0mm]{~~$^*$} & 15:44:48.9 & $-$54:06:49 & May 23 & \hphantom{2}70 & $-$74, $-$15 & May 22 & 140 & $-$66, $-$17 \\
 G326.86$-$0.68 & 15:51:14.2 & $-$54:58:05 & May 23 & \hphantom{2}88 & $-$90, $-$31 & May 22 & 160 & $-$92, $-$44 \\
 G327.29$-$0.58\makebox[0mm]{~~$^*$} & 15:53:06.2 & $-$54:37:06 & May 23 & \hphantom{2}79 & $-$73, $-$14 & May 22 & 150 & $-$65, $-$16 \\
 G327.39$+$0.20 & 15:50:18.5 & $-$53:57:06 & May 23 & \hphantom{2}88 & $-$123, $-$64 & May 22 & 170 & $-$113, $-$64 \\
 G327.62$-$0.11 & 15:52:50.2 & $-$54:03:01 & May 23 & \hphantom{2}68 & $-$123, $-$64 & May 22 & 180 & $-$112, $-$64 \\
 G328.21$-$0.59\makebox[0mm]{~~$^*$} & 15:57:54.8 & $-$54:02:13 & May 23 & \hphantom{2}80 & $-$67, $-$20 & May 22 & 120 & $-$59, $-$21 \\
 G328.24$-$0.55\makebox[0mm]{~~$^*$} & 15:57:58.3 & $-$53:59:17 & May 23 & \hphantom{2}77 & $-$73, $-$14 & May 22 & 150 & $-$64, $-$16 \\
 G328.25$-$0.53\makebox[0mm]{~~$^*$} & 15:57:59.8 & $-$53:57:59 & May 23 & \hphantom{2}81 & $-$73, $-$14 & May 22 & 140 & $-$64, $-$16 \\
 G328.81$+$0.63\makebox[0mm]{~~$^*$} & 15:55:48.5 & $-$52:43:19 & May 23 & \hphantom{2}82 & $-$73, $-$14 & May 22 & 150 & $-$64, $-$16 \\
 G329.03$-$0.20\makebox[0mm]{~~$^*$} & 16:00:31.1 & $-$53:12:37 & May 23 & \hphantom{2}77 & $-$72, $-$13 & May 22 & 160 & $-$64, $-$15 \\
 G329.07$-$0.31\makebox[0mm]{~~$^*$} & 16:01:08.9 & $-$53:16:08 & May 23 & \hphantom{2}83 & $-$72, $-$13 & May 22 & 120 & $-$64, $-$16 \\
 G329.18$-$0.31 & 16:01:47.0 & $-$53:11:44 & May 25 & \hphantom{2}82 & $-$81, $-$22 & May 24 & 120 & $-$78, $-$30 \\
 G329.47$+$0.50 & 15:59:40.7 & $-$52:23:28 & May 23 & \hphantom{2}79 & $-$89, $-$30 & May 22 & 140 & $-$91, $-$43 \\
 G331.13$-$0.24 & 16:10:59.9 & $-$51:50:39 & May 23 & \hphantom{2}70 & $-$120, $-$62 & May 22 & 160 & $-$110, $-$62 \\
 G331.34$-$0.35 & 16:12:26.3 & $-$51:46:13 & May 23 & \hphantom{2}97 & $-$87, $-$28 & May 22 & 150 & $-$90, $-$41 \\
 G331.44$-$0.19 & 16:12:12.5 & $-$51:35:10 & May 23 & \hphantom{2}83 & $-$121, $-$61 & May 22 & 190 & $-$110, $-$62 \\
 G332.30$-$0.09 & 16:15:45.4 & $-$50:55:54 & May 25 & \hphantom{2}83 & $-$79, $-$20 & May 24 & 170 & $-$76, $-$28 \\
 G332.60$-$0.17 & 16:17:29.3 & $-$50:46:12 & May 25 & \hphantom{2}79 & $-$79, $-$20 & May 24 & 130 & $-$76, $-$28 \\
 G332.94$-$0.69 & 16:21:19.0 & $-$50:54:10 & May 25 & \hphantom{2}68 & $-$79, $-$20 & May 24 & 120 & $-$76, $-$27 \\
 G332.96$-$0.68 & 16:21:22.9 & $-$50:52:59 & May 25 & \hphantom{2}73 & $-$79, $-$20 & May 24 & 130 & $-$76, $-$27 \\
 G333.03$-$0.06\makebox[0mm]{~~$^*$} & 16:18:56.7 & $-$50:23:54 & May 23 & \hphantom{2}61 & $-$70, $-$11 & May 22 & 140 & $-$61, $-$13 \\
 G333.13$-$0.44 & 16:21:03.3 & $-$50:35:50 & May 25 & \hphantom{2}80 & $-$79, $-$20 & May 24 & 140 & $-$76, $-$27 \\
 G333.13$-$0.56 & 16:21:35.7 & $-$50:40:51 & May 25 & \hphantom{2}77 & $-$79, $-$20 & May 24 & 160 & $-$76, $-$27 \\
 G333.16$-$0.10 & 16:19:42.7 & $-$50:19:53 & May 25 & \hphantom{2}54 & $-$129, $-$69 & May 24 & 130 & $-$123, $-$75 \\
 G333.18$-$0.09 & 16:19:45.6 & $-$50:18:35 & May 23 & \hphantom{2}71 & $-$119, $-$60 & May 22 & 170 & $-$109, $-$60 \\
  \hline
\end{tabular}
 \end{table*}

  \begin{table*}
 \contcaption{}
 \begin{tabular}{l@{\hskip 7mm}rcc@{ }c@{}c@{}c@{}c@{}c}
 \cthead{Source} & \multicolumn{2}{c}{Pointing centre} &\multicolumn{3}{c}{36 GHz} & \multicolumn{3}{c}{44 GHz} \\
 \cthead{name}  & & &
 \cthead{Observation} & \cthead{$1\sigma$} & \cthead{Velocity} & \cthead{Observation} & \cthead{$1\sigma$} & 
 \cthead{Velocity}\\ 
 &\cthead{$\alpha_{2000}$}&\cthead{$\delta_{2000}$} & \cthead{date}&\cthead{noise} &\cthead{range} & \cthead{date}&
 \cthead{noise} &\cthead{range}\\
&\cthead{(h:m:s)}&\cthead{(\degr:arcmin:arcsec)} & \cthead{(year 2007)} & \cthead{(mJy)}&\cthead{(\ks)} &\cthead{(year 2007)}& 
\cthead{(mJy)}&\cthead{(\ks)}\\
  \hline
 G333.23$-$0.06 & 16:19:51.3 & $-$50:15:14 & May 23 & 100 & $-$119, $-$60 & May 22 & 190 & $-$109, $-$60 \\
 G333.32$+$0.11\makebox[0mm]{~~$^*$} & 16:19:29.0 & $-$50:04:41 & May 23 & \hphantom{2}84 & $-$70, $-$10 & May 22 & 160 & $-$61, $-$13 \\
 G333.47$-$0.16\makebox[0mm]{~~$^*$} & 16:21:20.1 & $-$50:09:47 & May 23 & \hphantom{2}72 & $-$70, $-$10 & May 22 & 140 & $-$61, $-$13 \\
 G333.56$-$0.02\makebox[0mm]{~~$^*$} & 16:21:08.7 & $-$49:59:48 & May 23 & \hphantom{2}81 & $-$70, $-$10 & May 22 & 140 & $-$61, $-$13 \\
 G333.59$-$0.21 & 16:22:11.1 & $-$50:06:11 & May 25 & \hphantom{2}83 & $-$79, $-$19 & May 24 & 120 & $-$75, $-$27 \\
 G335.06$-$0.43\makebox[0mm]{~~$^*$} & 16:29:23.4 & $-$49:12:20 & May 23 & \hphantom{2}85 & $-$69, $-$9\hphantom{0} & May 22 & 190 & $-$60, $-$11 \\
 G335.59$-$0.29 & 16:30:59.3 & $-$48:43:47 & May 25 & \hphantom{2}88 & $-$77, $-$18 & May 24 & 140 & $-$74, $-$26 \\
 G335.79$+$0.17 & 16:29:45.6 & $-$48:16:16 & May 25 & \hphantom{2}84 & $-$77, $-$18 & May 24 & 130 & $-$74, $-$26 \\
 G336.41$-$0.26 & 16:34:14.3 & $-$48:06:22 & May 23 & \hphantom{2}81 & $-$117, $-$58 & May 22 & 180 & $-$107, $-$58 \\
 G337.40$-$0.40 & 16:38:49.8 & $-$47:28:18 & May 25 & \hphantom{2}75 & $-$76, $-$17 & May 24 & 110 & $-$73, $-$24 \\
 G337.92$-$0.46 & 16:41:08.3 & $-$47:07:56 & May 25 & \hphantom{2}73 & $-$76, $-$17 & May 24 & 110 & $-$73, $-$24 \\
 G338.92$+$0.55 & 16:40:31.9 & $-$45:41:53 & May 23 & 120 & $-$83, $-$24 & May 22 & 170 & $-$85, $-$36 \\
 G339.88$-$1.26\makebox[0mm]{~~$^*$} & 16:52:04.8 & $-$46:08:34 & May 23 & \hphantom{2}77 & $-$65, $-$6\hphantom{0} & May 22 & 140 & $-$57, $-$8\hphantom{0} \\
 G341.19$-$0.23 & 16:52:15.8 & $-$44:28:33 & May 25 & \hphantom{2}88 & $-$74, $-$15 & May 24 & 130 & $-$71, $-$22 \\
 G341.22$-$0.21 & 16:52:18.3 & $-$44:26:52 & May 25 & \hphantom{2}92 & $-$74, $-$15 & May 24 & 130 & $-$71, $-$22 \\
 G343.12$-$0.06\makebox[0mm]{~~$^*$} & 16:58:16.4 & $-$42:52:23 & May 23 & \hphantom{2}79 & $-$64, $-$4\hphantom{0} & May 22 & 160 & $-$55, $-$7\hphantom{0} \\
 G344.23$-$0.57 & 17:04:07.8 & $-$42:18:39 & May 25 & \hphantom{2}80 & $-$47, $+$12 & May 24 & 130 & $-$41, $+$7\hphantom{0} \\
 G345.00$-$0.22\makebox[0mm]{~~$^*$} & 17:05:09.5 & $-$41:29:04 & May 23 & \hphantom{2}99 & $-$62, $-$3\hphantom{0} & May 22 & 160 & $-$54, $-$5\hphantom{0} \\
 G345.01$+$1.79 & 16:56:49.0 & $-$40:14:20 & May 25 & \hphantom{2}76 & $-$47, $+$12 & May 24 & 110 & $-$42, $+$7\hphantom{0} \\
 G345.42$-$0.95 & 17:09:33.5 & $-$41:35:30 & May 25 & \hphantom{2}85 & $-$46, $+$13 & May 24 & 120 & $-$41, $+$8\hphantom{0} \\
 G345.50$+$0.35 & 17:04:23.5 & $-$40:44:10 & May 25 & \hphantom{2}91 & $-$46, $+$13 & May 24 & 120 & $-$41, $+$7\hphantom{0} \\
 G348.18$+$0.48 & 17:12:04.7 & $-$38:30:41 & May 25 & \hphantom{2}80 & $-$45, $+$14 & May 24 & 130 & $-$39, $+$9\hphantom{0} \\
 G349.09$+$0.11 & 17:16:25.9 & $-$37:59:23 & May 23 & 110 & $-$103, $-$57 & May 22 & 270 & $-$94, $-$56 \\
 G351.16$+$0.70 & 17:19:56.9 & $-$35:57:46 & May 25 & \hphantom{2}90 & $-$43, $+$16 & May 24 & 140 & $-$38, $+$11 \\
 G351.24$+$0.67 & 17:20:15.9 & $-$35:54:58 & May 25 & \hphantom{2}85 & $-$43, $+$16 & May 24 & 160 & $-$38, $+$11 \\
 G351.42$+$0.65 & 17:20:53.5 & $-$35:47:02 & May 25 & \hphantom{2}78 & $-$39, $+$13 & May 24 & 110 & $-$35, $+$9\hphantom{0} \\
 G351.63$-$1.26 & 17:29:18.0 & $-$36:40:09 & May 25 & \hphantom{2}70 & $-$40, $+$14 & May 24 & 130 & $-$34, $+$8\hphantom{0} \\
 G351.77$-$0.54 & 17:26:42.7 & $-$36:09:18 & May 25 & \hphantom{2}89 & $-$43, $+$17 & May 24 & 140 & $-$37, $+$11 \\
  \hline
\end{tabular}
 \end{table*}

Observations were made with the Australia Telescope Compact Array (ATCA) 2007 May 20-25
(project code C1642) using the hybrid H214C array configuration.  The most distant 
CA06 antenna (located about 4.5~km to the west of the other five antennas) was not used
in this experiment. The remaning baselines had lengths ranging from 82 to 247~m corresponding
to the synthesised beam sizes of about 7 and 6~arcsec at 36 and 44~GHz, respectively.
The targets were drawn from \citet{sly94}, \citet{val00} and \citet{ell05} and included essentially 
all sites of class~I maser emission reported in the literature at the time of observations and located 
south of declination $-$35\degr, 71 unique targets in total.  The exception was a well-known maser
in NGC~6334I(N) already imaged by \citet{kog98} and later studied in detail by \citet{bro09}. 
It is worth noting, that since the time of our observations, a large number of new class~I methanol 
masers have been found in this declination range in various single dish surveys \citep{che11,vor12,che13}.

To reduce the calibration overheads 
two transitions were observed
in separate slots, but no more than one day apart for the same source.
Each target was observed in a few cuts 2.5- to 3-min in duration spread in hour angle to improve 
$uv$-coverage (the typical total integration time is about 10 to 15~min). The targets were grouped according to their position. Each group was 
bracketed with 1.5-min observations of a nearby phase calibrator in such a way that
calibrator scans were no more than 10~minutes apart. The same calibrator sources served also as
pointing calibrators to improve antenna pointing accuracy beyond the level provided by the 
default pointing model. Following the standard practice for the reference pointing procedure,  
the solutions were obtained every hour or when the next group of targets was further away  
than 15\degr. From the statistics of these solutions, the resulting reference-pointing accuracy was 
estimated to be 4.8$\pm$2.6~arcsec. This residual pointing error affects the accuracy of flux-density 
measurements, particularly for sources of emission which are significantly offset from the pointing centre. 
The full width at half maximum (FWHM) of the primary beam was 84 and 69~arcsec at 36 and 44~GHz,
respectively. Note, the accuracy
of the absolute positions for maser emission depends primarily on the quality of the phase calibration 
and is believed to be around 0.5~arcsec (1$\sigma$) for the majority of sources. We make a specific
comment in the case when poorer accuracy of the absolute position is expected.

The project used the original Australia Telescope correlator before the new broad band backend became
available. It was configured with 1024 spectral channels across an 8-MHz bandwidth, providing a 
spectral resolution of 0.078 and 0.064~\kss at 36 and 44~GHz, respectively. The initially adopted
rest frequencies (36169.265$\pm$0.03 and 44069.410$\pm$0.01~MHz for the
$4_{-1}-3_0$~E and $7_0-6_1$~A$^+$ methanol transitions, respectively) were taken from 
\citet{mul04}. The velocity uncertainties corresponding to the rest frequency uncertainties are 
0.24 and 0.068~\kss at 36 and 44~GHz, respectively (see also section~\ref{restfrequncert} for an improved
estimate of 36169.238$\pm$0.011~MHz based on the present work).

The data reduction was undertaken in {\sc miriad} and followed the standard procedure with the exception
of bandpass calibration. For the latter, we solved for bandpass correction in the form of a third order 
polynomial, fitting independently for the real and imaginary parts of the visibility data of the bandpass calibrator 
(PKS~1253-055) using the {\sc uvlin} task. A number of spectral channels near both ends of the band (typically 
100 channels at both ends or 20 per cent of the band in total, but sometimes more if the target source 
had strong continuum emission) were excluded from the fit and further analysis.
This approach is preferable for high spectral resolution mm-wavelength observations 
(i.e. the case of low signal-to-noise ratio in the calibration data) and suits maser observations well. 
After the absolute positions had been determined, we executed a single pass of phase only self-calibration
with 1~minute solution interval. The strongest spectral feature was typically used as a reference
(sometimes weaker features with less structure had to be used). The flux-density ratio for the reference
feature before and after self-calibration is a measure of decorrelation due to atmospheric phase
variations on the time-scale shorter than the minimum temporal separation of phase calibrator
scans. We estimate that the effect of this decorrelation on the measured flux densities did not exceed a few 
per cent and was corrected by the self-calibration to first order.
The absolute flux-density scale was bootstrapped from observations of Uranus for which 
we assumed flux densities of 1.38 and 1.98~Jy at 36 and 44~GHz, respectively. 
The opacity correction has been applied using the model built into {\sc miriad} and
environmental parameters such as ambient temperature, pressure and humidity registered at the time
of observations by the weather station. We estimate the
flux density scale to be accurate to about 20 per cent. 


As is standard with interferometric observations, no real-time Doppler tracking was done. The correction
was performed off-line during imaging which was done for the full velocity range available for analysis and used 
uniform weighting. We adopted 0\farcs5 image pixels for all sources at both frequencies, which delivered sufficient 
oversampling in the image plane for adequate imaging and analysis (the synthesised beam is greater than 6~arcsec).  
Note that no additional Hanning smoothing has been done, so all image cubes
had the same frequency resolution as in the measured visibility data.
The maser emission was searched for in these velocity cubes prior to the primary beam 
correction (i.e. the noise level was constant across the field of view). We used a combination of automatic detection
in the spectral domain \citep[same procedure as used in MMB; see][]{gre09} and visual inspection of each
individual image plane. The formal detection criterion was
3 consecutive spectral channels above 3$\sigma$. However, the majority of cubes appeared to be
dynamic range limited due to the presence of strong emission and the complexity of its 
spatial distribution. We rejected all such potential detections if there were reasonable grounds to believe they
might be dynamic range artefacts. In particular, we rejected apparent detections weaker than twice the typical dynamic 
range artefacts in the field. Therefore, the ability to detect weak features is notably
reduced within the velocity range of strong emission. Another type of artefact is caused by 
spatially extended (and probably thermal) sources which cannot be properly imaged with the sparse 
$uv$-coverage attained in this experiment tailored for maser imaging. We excluded such sources
from analysis unless they could be reliably localised (i.e. their sizes did not exceed a few seconds of arc).
After the emission had been located, we applied primary beam correction at the appropriate frequency and 
extracted spectra at each location as slices along the spectral axis taken at the position of the peak. 
Given the oversampling in the image plane, this method provides
accurate flux densities for compact sources, such as masers, and is very robust to imperfections
of the point spread function caused by limited $uv$-coverage and dynamic range artefacts. For each velocity of
interest, an accurate position was determined by fitting a point source with the {\sc imfit} task into a 5$\times$5~arcsec$^2$ region centred at the 
peak pixel (see also section~\ref{results}).

The basic source-dependent parameters of observations are summarised in Table~\ref{obsdetails}.
The first three columns list the source name and the pointing direction used in observations (the same
pointing centre was used at 36 and 44-GHz). The remaining columns  
include the UT date of observations, the noise level (1$\sigma$) in the image cube before the primary beam correction (i.e. constant noise across the 
field of view) and the velocity range inspected for the 36 and 44-GHz transitions.

 
 \begin{table*}
 \caption{General overview of the results and basic associations (at arcsec scale) for each observed source. 
 The number of 6.7-GHz masers is defined as per the MMB survey. See \protect\citet{cas98} for information on OH maser detections, 
OH non-detections are mostly based on \citet{cas83} and \citet{cas87b}. References for the H$_2$O masers are \protect\citet{bre10a}, \protect\citet{bre11} and \protect\citet{zap08} for
$a$, $b$ and $c$, respectively. The asterisk means that only data with inferior quality are available, see notes on individual sources for details. 
The {\em Spitzer} dark clouds (SDC) are from the catalogue of \protect\citet{per09}, the EGOs are from the list of \protect\citet{cyg08}.
 }
 \label{srcoverview}
 \begin{tabular}{@{}lr@{}cc@{}c@{}c@{}c@{~}c@{~}c@{~}c@{~}c@{~}c@{~}c@{~}c@{~}c}
 \cthead{Source} & \multicolumn{3}{c}{Region with maser emission}& &\cthead{Pointing}&\# of&\multicolumn{2}{@{}c@{}}{other}  & &\multicolumn{5}{@{}c}{Associations with class~I masers} \\
 \cthead{name}  &\multicolumn{2}{@{}c}{centre}&radius & \cthead{velocity} &\cthead{offset} &6.7 GHz& \multicolumn{2}{@{}c@{}}{masers} & regular & \multicolumn{5}{@{}c}{at arcsec scale}\\
  \cline{11-15}
 &\cthead{$\alpha_{2000}$}&$\delta_{2000}$ & &\cthead{range}&\cthead{from centre}&masers& OH~ & H$_2$O & struct. & SDC\vphantom{$\int^a$} &
EGO &other & 8.0 $\mu$m & ~cm-$\lambda$~ \\
&\cthead{(h:m:s)}&(\degr:arcmin:arcsec) & (arcsec) &\cthead{(\ks)}& \cthead{(arcsec)} &
$<$1~arcmin & & &  & & &\cthead{4.5 $\mu$m} & & cont.\\
  \hline
 G269.15$-$1.13 & 09:03:32.2 & $-$48:28:10 & 34 & $+$7, $+$14 & 21 & 1 &  & $+^b$ & $-$ &  &  &  &  &  \\
 G270.26$+$0.84 & 09:16:40.7 & $-$47:56:16 & 11 & $+$7, $+$20 & 32 & 1 &  & $+^b$ & $-$ &  &  &  &  &  \\
 G294.98$-$1.73 & 11:39:14.1 & $-$63:29:10 & \hphantom{2}8 & $-$10, $+$3\hphantom{0} & 25 & 1 &  & $+^b$ & $-$ &  &  &  &  &  \\
 G300.97$+$1.15 & 12:34:52.3 & $-$61:39:55 & 17 & $-$48, $-$40 & \hphantom{2}7 & 1 & $+$ & $+^a$ & $-$ &  &  &  &  & $+$ \\
 G301.14$-$0.23 & 12:35:35.3 & $-$63:02:29 & \hphantom{2}4 & $-$42, $-$34 & 12 & 1 & $+$ & $+^a$ & $-$ & $-$ & $-$ & $+$ & $-$ & $+$ \\
 G305.21$+$0.21 & 13:11:10.6 & $-$62:34:45 & 39 & $-$46, $-$36 & \hphantom{2}9 & 2 & $+$ & $+^a$ & $+$ & $-$ & $-$ & $-$ & $+$ & $+$ \\
 G305.25$+$0.25 & 13:11:32.5 & $-$62:32:08 & \hphantom{2}2 & $-$39, $-$34 & \hphantom{2}5 & 1 & $-$ & $+^b$ & $-$ & $-$ & $-$ & $-$ & $+$ &  \\
 G305.37$+$0.21 & 13:12:33.1 & $-$62:33:47 & 32 & $-$43, $-$31 & 22 & 0 &  & $+^b$ & $+$ & $-$ & $-$ & $+$ & $+$ & $+$ \\
 G309.38$-$0.13 & 13:47:22.3 & $-$62:18:01 & 23 & $-$88, $-$47 & 28 & 1 & $+$ & $+^a$ & $-$ & $-$ & $+$ & $+$ & $-$ &  \\
 G316.76$-$0.01 & 14:44:56.1 & $-$59:48:02 & 22 & $-$43, $-$36 & 10 & 0 & $+$ & $+^a$ & $+$ & $-$ & $-$ & $+$ & $-$ &  \\
 G318.05$+$0.09 & 14:53:43.2 & $-$59:08:58 & 21 & $-$54, $-$48 & \hphantom{2}5 & 1 & $+$ & $+^a$ & $-$ & $+$ & $+$ & $+$ & $-$ &  \\
 G318.95$-$0.20 & 15:00:55.5 & $-$58:58:54 & 19 & $-$37, $-$27 & 22 & 1 & $+$ & $+^a$ & $+$ & $-$ & $-$ & $+$ & $+$ &  \\
 G320.29$-$0.31 & 15:10:19.1 & $-$58:25:17 & \hphantom{2}2 & $-$69, $-$60 & \hphantom{2}8 & 1 &  & $+^a$ & $-$ & $-$ & $-$ & $+$ & $-$ &  \\
 G322.16$+$0.64 & 15:18:37.0 & $-$56:38:22 & 38 & $-$60, $-$51 & 38 & 1 & $+$ & $+^a$ & $-$ & $-$ & $-$ & $+$ & $+$ & $+$ \\
 G323.74$-$0.26 & 15:31:45.6 & $-$56:30:52 & \hphantom{2}8 & $-$53, $-$47 & \hphantom{2}8 & 1 & $+$ & $+^a$ & $-$ & $-$ & $+$ & $-$ & $-$ &  \\
 G324.72$+$0.34 & 15:34:57.9 & $-$55:27:27 & 37 & $-$57, $-$44 & \hphantom{2}1 & 1 & $+$ & $+^a$ & $-$ & $+$ & $+$ & $-$ & $-$ &  \\
 G326.48$+$0.70 & 15:43:18.1 & $-$54:07:25 & 25 & $-$56, $-$29 & 18 & 2 &  &  & $-$ & $+$ & $+$ & $+$ & $-$ &  \\
 G326.64$+$0.61 & 15:44:31.0 & $-$54:05:10 & 34 & $-$66, $-$26 & 25 & 1 &  &  & $-$ & $-$ & $-$ & $+$ & $+$ &  \\
 G326.66$+$0.57 & 15:44:47.7 & $-$54:06:43 & 11 & $-$41, $-$41 & 12 & 0 & $-$ & $*$\hphantom{$^a$} & $-$ & $+$ & $-$ & $-$ & $-$ &  \\
 G326.86$-$0.68 & 15:51:13.9 & $-$54:58:05 & \hphantom{2}6 & $-$71, $-$65 & \hphantom{2}2 & 1 &  & $+^a$ & $-$ & $+$ & $+$ & $-$ & $-$ &  \\
 G327.29$-$0.58 & 15:53:09.5 & $-$54:36:57 & 19 & $-$50, $-$38 & 30 & 1 & $+$ & $+^a$ & $-$ & $-$ & $+$ & $-$ & $-$ &  \\
 G327.39$+$0.20 & 15:50:19.2 & $-$53:57:07 & 10 & $-$92, $-$85 & \hphantom{2}7 & 2 & $-$ & $+^a$ & $-$ & $-$ & $+$ & $-$ & $-$ &  \\
 G327.62$-$0.11 & 15:52:50.2 & $-$54:03:03 & \hphantom{2}4 & $-$91, $-$84 & \hphantom{2}2 & 1 & $-$ & $+^a$ & $-$ & $-$ & $-$ & $-$ & $-$ &  \\
 G328.21$-$0.59 & 15:57:59.5 & $-$54:02:14 & \hphantom{2}2 & $-$44, $-$39 & 41 & 0 & $-$ &  & $-$ & $+$ & $-$ & $-$ & $-$ &  \\
 G328.24$-$0.55 & 15:58:01.2 & $-$53:59:09 & 35 & $-$46, $-$41 & 27 & 2 & $+$ & $+^a$ & $-$ & $-$ & $-$ & $+$ & $-$ & $+$ \\
 G328.25$-$0.53 & 15:58:02.2 & $-$53:57:29 & 36 & $-$49, $-$43 & 36 & 1 & $+$ & $+^a$ & $-$ & $-$ & $-$ & $-$ & $+$ &  \\
 G328.81$+$0.63 & 15:55:48.7 & $-$52:43:03 & 36 & $-$45, $-$38 & 16 & 2 & $+$ & $+^a$ & $-$ & $+$ & $-$ & $-$ & $+$ & $+$ \\
 G329.03$-$0.20 & 16:00:30.9 & $-$53:12:34 & 25 & $-$49, $-$37 & \hphantom{2}3 & 2 & $+$ & $+^a$ & $+$ & $+$ & $+$ & $-$ & $-$ &  \\
 G329.07$-$0.31 & 16:01:09.6 & $-$53:16:08 & 30 & $-$51, $-$28 & \hphantom{2}7 & 1 & $+$ & $+^a$ & $-$ & $+$ & $+$ & $+$ & $-$ &  \\
 G329.18$-$0.31 & 16:01:46.7 & $-$53:11:38 & 36 & $-$76, $-$44 & \hphantom{2}6 & 1 & $+$ & $+^a$ & $+$ & $+$ & $+$ & $+$ & $+$ &  \\
 G329.47$+$0.50 & 15:59:39.8 & $-$52:23:35 & 41 & $-$80, $-$64 & 11 & 1 &  &  & $-$ & $-$ & $+$ & $-$ & $+$ &  \\
 G331.13$-$0.24 & 16:10:59.9 & $-$51:50:19 & 10 & $-$92, $-$81 & 21 & 1 & $+$ & $+^a$ & $-$ & $-$ & $-$ & $-$ & $-$ & $+$ \\
 G331.34$-$0.35 & 16:12:26.5 & $-$51:46:20 & \hphantom{2}0 & $-$67, $-$65 & \hphantom{2}8 & 1 & $+$ & $+^a$ & $-$ & $-$ & $+$ & $-$ & $-$ &  \\
 G331.44$-$0.19 & 16:12:11.4 & $-$51:35:09 & 22 & $-$92, $-$84 & 10 & 1 & $+$ & $+^a$ & $+$ & $-$ & $-$ & $-$ & $+$ & $+$ \\
 G332.30$-$0.09 & 16:15:45.2 & $-$50:55:54 & \hphantom{2}9 & $-$53, $-$45 & \hphantom{2}2 & 2 & $-$ & $+^a$ & $-$ & $-$ & $+$ & $+$ & $+$ & $+$ \\
 G332.60$-$0.17 & 16:17:29.4 & $-$50:46:08 & \hphantom{2}7 & $-$52, $-$45 & \hphantom{2}4 & 1 & $-$ & $+^a$ & $-$ & $-$ & $+$ & $-$ & $+$ &  \\
 G332.94$-$0.69 & 16:21:20.3 & $-$50:54:12 & 13 & $-$51, $-$47 & 12 & 1 &  &  & $-$ & $-$ & $+$ & $-$ & $+$ &  \\
 G332.96$-$0.68 & 16:21:22.5 & $-$50:52:57 & \hphantom{2}9 & $-$59, $-$46 & \hphantom{2}4 & 1 &  & $+^a$ & $-$ & $-$ & $+$ & $-$ & $-$ &  \\
 G333.03$-$0.06 & 16:18:56.7 & $-$50:23:54 & \hphantom{2}1 & $-$43, $-$39 & \hphantom{2}0 & 1 & $-$ & $+^a$ & $-$ & $-$ & $-$ & $+$ & $-$ &  \\
 G333.13$-$0.44 & 16:21:02.1 & $-$50:35:49 & 32 & $-$56, $-$45 & 11 & 4 & $+$ & $+^a$ & $-$ & $+$ & $-$ & $-$ & $+$ &  \\
 G333.13$-$0.56 & 16:21:36.1 & $-$50:40:57 & 28 & $-$64, $-$48 & \hphantom{2}7 & 2 &  & $+^a$ & $+$ & $-$ & $+$ & $-$ & $+$ &  \\
 G333.16$-$0.10 & 16:19:42.5 & $-$50:19:57 & \hphantom{2}4 & $-$93, $-$90 & \hphantom{2}5 & 1 & $-$ & $*$\hphantom{$^a$} & $-$ & $-$ & $-$ & $-$ & $+$ & $+$ \\
 G333.18$-$0.09 & 16:19:46.0 & $-$50:18:34 & \hphantom{2}5 & $-$88, $-$85 & \hphantom{2}4 & 1 & $-$ & $*$\hphantom{$^a$} & $-$ & $+$ & $+$ & $-$ & $-$ &  \\
  \hline
\end{tabular}
 \end{table*}
 
 \begin{table*}
 \contcaption{}
 \begin{tabular}{@{}lr@{}cc@{}c@{}c@{}c@{~}c@{~}c@{~}c@{~}c@{~}c@{~}c@{~}c@{~}c}
 \cthead{Source} & \multicolumn{3}{c}{Region with maser emission}& &\cthead{Pointing}&\# of&\multicolumn{2}{@{}c@{}}{other}  & &\multicolumn{5}{@{}c}{Associations with class~I masers} \\
 \cthead{name}  &\multicolumn{2}{@{}c}{centre}&radius & \cthead{velocity} &\cthead{offset} &6.7 GHz& \multicolumn{2}{@{}c@{}}{masers} & regular & \multicolumn{5}{@{}c}{at arcsec scale}\\
   \cline{11-15}
 &\cthead{$\alpha_{2000}$}&$\delta_{2000}$ & &\cthead{range}&\cthead{from centre}&masers& OH~ & H$_2$O & struct. & SDC\vphantom{$\int^a$} &
EGO &other & 8.0 $\mu$m & ~cm-$\lambda$~ \\
&\cthead{(h:m:s)}&(\degr:arcmin:arcsec) & (arcsec) &\cthead{(\ks)}& \cthead{(arcsec)} &
$<$1~arcmin & & &  & & &\cthead{4.5 $\mu$m} & & cont.\\
  \hline
 G333.23$-$0.06 & 16:19:50.8 & $-$50:15:13 & 14 & $-$107, $-$84 & \hphantom{2}5 & 2 & $+$ & $+^a$ & $-$ & $-$ & $-$ & $+$ & $-$ &  \\
 G333.32$+$0.11 & 16:19:28.3 & $-$50:04:46 & \hphantom{2}9 & $-$55, $-$44 & \hphantom{2}8 & 1 & $+$ & $+^a$ & $-$ & $-$ & $+$ & $-$ & $-$ &  \\
 G333.47$-$0.16 & 16:21:20.2 & $-$50:09:44 & 12 & $-$50, $-$39 & \hphantom{2}2 & 1 & $+$ & $+^a$ & $+$ & $-$ & $+$ & $-$ & $+$ & $+$ \\
 G333.56$-$0.02 & 16:21:08.8 & $-$49:59:48 & \hphantom{2}0 & $-$42, $-$38 & \hphantom{2}1 & 1 & $-$ & $*$\hphantom{$^a$} & $-$ & $+$ & $-$ & $-$ & $-$ &  \\
 G333.59$-$0.21 & 16:22:06.7 & $-$50:06:23 & \hphantom{2}4 & $-$52, $-$48 & 45 & 0 & $-$ & $*$\hphantom{$^a$} & $-$ & $+$ & $-$ & $-$ & $-$ &  \\
 G335.06$-$0.43 & 16:29:23.4 & $-$49:12:26 & 19 & $-$43, $-$34 & \hphantom{2}6 & 1 & $+$ & $+^a$ & $+$ & $-$ & $+$ & $-$ & $+$ &  \\
 G335.59$-$0.29 & 16:31:00.3 & $-$48:43:37 & 50 & $-$56, $-$38 & 15 & 3 & $+$ & $+^a$ & $-$ & $+$ & $+$ & $-$ & $-$ &  \\
 G335.79$+$0.17 & 16:29:46.5 & $-$48:15:50 & 23 & $-$53, $-$45 & 28 & 1 & $+$ & $+^a$ & $-$ & $-$ & $+$ & $-$ & $+$ &  \\
 G336.41$-$0.26 & 16:34:13.5 & $-$48:06:18 & \hphantom{2}5 & $-$91, $-$86 & \hphantom{2}9 & 1 &  & $*$\hphantom{$^a$} & $-$ & $-$ & $-$ & $-$ & $+$ & $+$ \\
 G337.40$-$0.40 & 16:38:48.9 & $-$47:27:55 & 25 & $-$46, $-$35 & 25 & 1 & $+$ & $+^a$ & $-$ & $-$ & $-$ & $+$ & $-$ & $+$ \\
 G337.92$-$0.46 & 16:41:08.5 & $-$47:07:44 & 52 & $-$45, $-$36 & 12 & 1 & $+$ & $+^a$ & $-$ & $-$ & $+$ & $+$ & $+$ &  \\
 G338.92$+$0.55 & 16:40:34.5 & $-$45:41:50 & 27 & $-$79, $-$56 & 28 & 2 & $+$ & $+^a$ & $-$ & $+$ & $+$ & $-$ & $-$ &  \\
 G339.88$-$1.26 & 16:52:04.3 & $-$46:08:28 & 19 & $-$33, $-$31 & \hphantom{2}9 & 1 & $+$ & $+^a$ & $-$ &  &  &  &  &  \\
 G341.19$-$0.23 & 16:52:16.4 & $-$44:28:44 & \hphantom{2}3 & $-$71, $-$15 & 13 & 0 & $-$ & $*$\hphantom{$^a$} & $+$ & $+$ & $-$ & $-$ & $-$ &  \\
 G341.22$-$0.21 & 16:52:17.4 & $-$44:27:03 & 27 & $-$47, $-$39 & 14 & 1 & $+$ & $+^a$ & $-$ & $-$ & $-$ & $-$ & $+$ &  \\
 G343.12$-$0.06 & 16:58:16.8 & $-$42:52:09 & 30 & $-$35, $-$20 & 15 & 0 & $+$ & $+^a$ & $+$ & $-$ & $-$ & $+$ & $-$ & $+$ \\
 G344.23$-$0.57 & 17:04:07.8 & $-$42:18:26 & 48 & $-$26, $-$15 & 14 & 1 & $+$ & $+^a$ & $+$ & $-$ & $+$ & $-$ & $-$ &  \\
 G345.00$-$0.22 & 17:05:10.7 & $-$41:29:13 & 26 & $-$31, $-$16 & 16 & 2 & $+$ & $+^a$ & $-$ & $+$ & $+$ & $-$ & $-$ &  \\
 G345.01$+$1.79 & 16:56:46.0 & $-$40:14:09 & 21 & $-$18, $-$10 & 36 & 2 & $+$ & $+^a$ & $+$ & $-$ & $-$ & $+$ & $+$ &  \\
 G345.42$-$0.95 & 17:09:35.6 & $-$41:35:40 & 47 & $-$25, $-$17 & 25 & 2 & $+$ & $+^a$ & $-$ & $-$ & $-$ & $+$ & $+$ &  \\
 G345.50$+$0.35 & 17:04:24.1 & $-$40:44:07 & 33 & $-$19, $-$14 & \hphantom{2}8 & 1 & $+$ & $+^a$ & $-$ & $+$ & $-$ & $+$ & $-$ &  \\
 G348.18$+$0.48 & 17:12:06.8 & $-$38:30:38 & 40 & $-$9, $-$5\hphantom{0} & 25 & 0 &  & $*$\hphantom{$^a$} & $-$ & $-$ & $-$ & $+$ & $+$ &  \\
 G349.09$+$0.11 & 17:16:24.6 & $-$37:59:43 & \hphantom{2}7 & $-$81, $-$75 & 25 & 2 & $+$ & $+^a$ & $-$ & $-$ & $-$ & $+$ & $-$ &  \\
 G351.16$+$0.70 & 17:19:56.4 & $-$35:57:54 & 30 & $-$9, $-$3\hphantom{0} & 11 & 1 & $+$ & $+^a$ & $+$ & $-$ &  & $+$ & $-$ &  \\
 G351.24$+$0.67 & 17:20:18.0 & $-$35:54:45 & \hphantom{2}6 & $-$6, $-$2\hphantom{0} & 29 & 1 & $-$ & $+^a$ & $-$ & $-$ &  & $-$ & $+$ & $+$ \\
 G351.42$+$0.65 & 17:20:52.3 & $-$35:46:42 & 32 & $-$12, $-$3\hphantom{0} & 24 & 2 & $+$ & $+^a$ & $+$ & $-$ &  & $+$ & $+$ & $+$ \\
 G351.63$-$1.26 & 17:29:17.9 & $-$36:40:29 & \hphantom{2}9 & $-$18, $-$12 & 20 & 0 & $-$ &  & $-$ &  &  &  &  &  \\
 G351.77$-$0.54 & 17:26:42.4 & $-$36:09:14 & 40 & $-$11, $+$2\hphantom{0} & \hphantom{2}5 & 1 & $+$ & $+^c$ & $+$ &  &  & $+$ & $-$ &  \\
  \hline
\end{tabular}
 \end{table*}
  
\section{Results}
\label{results}
 
A general overview of the results obtained for all the 71 observed targets is presented in Table~\ref{srcoverview}.
In agreement with the earlier results on common class~I masers \citep[e.g.,][]{kur04,vor06,cyg09,bro09}, the maps often reveal significant
spatial spread in the emission. Columns two to four of Table~\ref{srcoverview} characterise this spatial extent by listing the position and radius of the smallest circle
encompassing all the emission regardless of the frequency for each particular target. The fifth column characterises the spread in the velocity domain and 
shows the total velocity range of the class~I maser emission in both transitions.
For most targets, our observations were the first with the
spatial precision of an interferometer, and our initial target positions, taken from single dish surveys, were typically rather inaccurate and
often corresponded to the positions of some other maser species. Therefore, the pointing centres turned out to be sub-optimal 
on many occasions. In extreme cases like G333.59$-$0.21, all the detected emission happened to be outside the 50 per cent sensitivity region (primary beam
FWHM is 84 and 69~arcsec at 36 and 44~GHz, respectively). The sixth column
of Table~\ref{srcoverview} shows the magnitude of the offset between the chosen pointing centre and the centre of the maser emission region shown
in columns 2 and 3 (the actual orientation is shown in the maps, pointing centres are also listed in Table~\ref{obsdetails}). The large spatial spread of the class~I maser emission 
also has the 
potential to create confusion in cross-referencing the sources as the traditional naming convention for masers is based on their Galactic positions. 
We used the following approach to source naming throughout the paper and rounded the appropriate positions to the nearest 0\fdg01 
(see the first column of Table~\ref{srcoverview}).
For sources with only one class~II methanol maser at 6.7-GHz within 1~arcmin of any class~I maser 
emission according to the MMB survey (there are 45 such sources in our sample) we used the position of this 6.7-GHz maser. Thus, the source name
reflects the presumed location of the YSO. For 17 sources with two or more such nearby 6.7-GHz masers we used the average position of the 6.7-GHz masers, 
except for G328.24-0.55, which was treated as a source with a single 6.7-GHz maser (the other 6.7-GHz maser seems to be unrelated given the distribution 
of the class~I maser emission). For the remaining nine sources which do not have 6.7-GHz masers detected in the vicinity according to the MMB survey, we 
used the flux-weighted average position of all class~I maser emission regardless of the transition. There are only three sources in
our sample which  have prior interferometric data on common class~I masers reported in the literature, G309.38$-$0.13 \citep{vor10a}, G343.12$-$0.06 \citep{vor06},
and G351.42$+$0.65 \citep[NGC~6334F;][]{gom10}. This approach resulted in identical names as used in the previous publications for the first
two sources (and the latter is more commonly named as NGC~6334F).
The number of associated 6.7-GHz masers is given in the seventh column of Table~\ref{srcoverview}. Columns eight and nine summarise information on OH and 
H$_2$O masers if such observations have been undertaken towards our targets. For OH masers the positive detection data were taken from \citet{cas98} and the non-detections relied upon
blind surveys of \citet{cas83} and \citet{cas87b}, unless the particular target was in the primary beam of another observation (see section~\ref{notesonsrc} for details on individual sources).
As indicated in Table~\ref{srcoverview}, the H$_2$O masers were found towards every source studied in this project for which sensitive data exist. The H$_2$O 
surveys of \citet{bre10a} and \citet{bre11} cover most of our targets, although better data exist for G351.77$-$0.54 \citep{zap08}. The tenth column indicates whether the distribution
of class~I maser emission trace regular structures such as lines or arcs. The remaining columns are self-explanatory and indicate whether the class~I maser emission is associated
with {\em Spitzer} dark clouds \citep[SDC;][]{per09}, EGOs \citep{cyg08}, 4.5-$\mu$m emission sources not identified as EGOs by \citet{cyg08}, 8.0-$\mu$m filaments and cm-wavelength
continuum emission. For the latter we do not draw any conclusions from non-detections due to very inhomogeneous information available for this sample of sources
(see also section~\ref{genassoc}).

For detailed analysis of the spatial and kinematic structure of detected class~I maser emission, we fitted all spectral profiles with a sum of Gaussians. The number of 
components in the fit was typically the smallest required to represent the measured profile given the noise. However, on some occasions additional components
were fitted if warranted by the spatial distribution. In a few cases, the spectra had some (usually weak) feature which could not be fitted reliably as a sum of 
Gaussians, for example an abrupt jump in the flux density at a certain velocity. We make an explicit comment on all such individual sources where there is a 
significant systematic residual after the fit. For each Gaussian, the position fit was done under the assumption that the source is spatially unresolved. We selected a 
representative velocity channel for this fit, which corresponded to the peak unless there was significant blending of several Gaussian components. Unless 
noted otherwise, the position fits agreed within the fit uncertainty for all velocity channels dominated by the particular Gaussian component and any systematic
effects could be ignored. 

\begin{figure*}
\includegraphics[width=\linewidth]{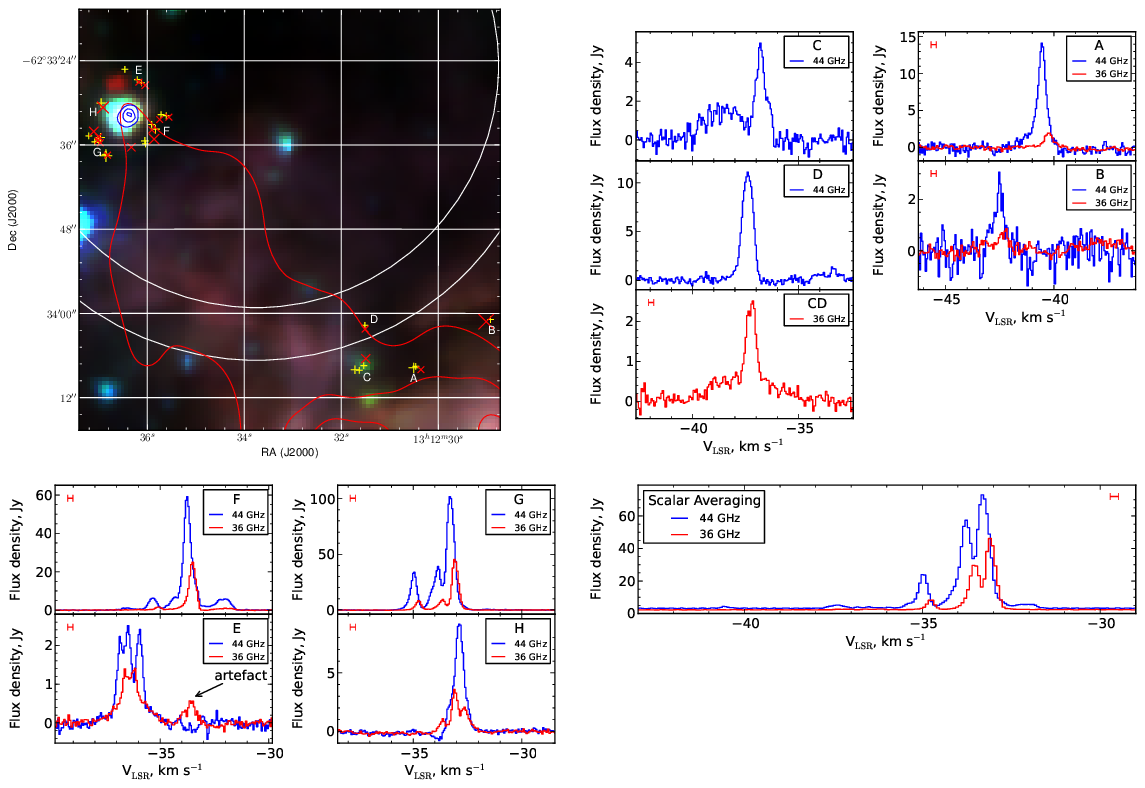}
\caption{The image and spectra for a single selected source, G305.37$+$0.21. The results for all sources are available online (where appropriate, the 6.7-GHz masers, which have not been
found in this source, are shown by squares). The background represents the
emission in  8.0, 4.5 and 3.6~$\mu$m IRAC {\em Spitzer} bands shown as red, green and blue, respectively. The class~I methanol masers are shown by crosses (36~GHz) and 
pluses (44~GHz). The red contours show the 18-cm continuum emission (MAGMO data). Blue contours show compact 3-cm continuum emission from \protect\citet{hin12}. 
White concentric circles
represent the full width at half maximum for the primary beam at 36 (larger circle) and 44~GHz (smaller circle). The horizontal error bar shown with the 36-GHz spectra
represents the velocity uncertainty due to the uncertainty of the initially adopted rest frequency of this transition. The uncertainty is comparable to the spectral 
resolution for the 44-GHz transition. The bottom right plot shows scalar-averaged visiblity spectra for both transitions to illustrate the difference between the single dish and interferometric results.}
\label{map}
\end{figure*}

The fit results for individual Gaussian components are given in Table~\ref{fitresults} (for sources other than G305.37$+$0.21 see the 
full version of the table available online). The total number of components is 848 at 36~GHz and 879 at 44~GHz (this includes
one absorption component at 36 and two such components at 44~GHz in G301.14$-$0.23, see notes on this source in section~\ref{notesonsrc}).
Following the same approach as \citet{vor10a}, the components in Table~\ref{fitresults} are grouped by a rough spatial 
location labelled with letters to assist cross-referencing between images and spectra (see Fig.~\ref{map}). We use these location labels for presentation purposes only and disregarded
them in the statistical analysis. The correspondence between individual Gaussian components and a particular location
is determined by practicality of presenting these components together in a single spectrum (individual spectral profiles were combined with appropriate weighting, if necessary). 
Typically, the components which are further apart than several 
arcseconds were assigned to different locations, unless dictated otherwise by the sidelobe structure of the synthesised beam. In some cases we were unable to 
assign locations the same way for two transitions. For example, the locations C and D of the 44-GHz emission in G305.37$+$0.21 correspond to a single
location CD of the 36-GHz emission (see Table~\ref{fitresults}). The source name, the location and the transition frequency are given in the first three columns of Table~\ref{fitresults}.
The remaining columns are self-explanatory and include the peak velocity, the positions, full width at half maximum (FWHM) and the peak flux density for each Gaussian 
component. The uncertainties are given in parentheses and expressed in units of the least significant figure. 
 
\begin{table*}
\caption{Fit results for individual Gaussian components for a single selected source, G305.37$+$0.21. The full table with details for all sources
is available online. The uncertainties are given in parentheses and expressed in units of the least significant figure.}
\label{fitresults}
\begin{tabular}{lcc@{}rlrrr}
\hline
\cthead{Source} & \cthead{Location} & \cthead{Frequency} & \cthead{LSR} & \cthead{$\alpha_{2000}$} & \cthead{$\delta_{2000}$} & \cthead{Line} & \cthead{Flux} \\
\cthead{name} & & & \cthead{velocity} &  &  & \cthead{FWHM} & \cthead{density} \\
& & \cthead{(GHz)} & \cthead{(\ks)} & \cthead{(h:m:s)} & \cthead{(\degr:arcmin:arcsec)} & \cthead{(\ks)} & \cthead{(Jy)} \\
\hline
 G305.37+0.21  & A & 36 & $-$40.21\hphantom{0}~(1) & 13:12:30.36\hphantom{00}~(4) & $-$62:34:08.0\hphantom{00}~(3) & 0.50\hphantom{00}~(3) & 1.9\hphantom{00}~(1) \\
              &   & 44 & $-$40.79\hphantom{0}~(6) & 13:12:30.51\hphantom{00}~(8) & $-$62:34:07.8\hphantom{00}~(5) & 1.10\hphantom{00}~(9) & 2.5\hphantom{00}~(4) \\
              &   &    & $-$40.511~(5) & 13:12:30.465\hphantom{0}~(6) & $-$62:34:07.55\hphantom{0}~(4) & 0.40\hphantom{00}~(2) & 11.5\hphantom{00}~(4) \\
              & B & 36 & $-$42.32\hphantom{0}~(4) & 13:12:29.0\hphantom{000}~(2) & $-$62:34:01.2\hphantom{00}~(8) & 0.65\hphantom{00}~(8) & 0.66\hphantom{0}~(8) \\
              &   & 44 & $-$42.50\hphantom{0}~(3) & 13:12:28.92\hphantom{00}~(4) & $-$62:34:00.9\hphantom{00}~(3) & 0.56\hphantom{00}~(6) & 2.3\hphantom{00}~(2) \\
              & C & 44 & $-$38.48\hphantom{0}~(6) & 13:12:31.72\hphantom{00}~(6) & $-$62:34:08.0\hphantom{00}~(4) & 2.1\hphantom{000}~(2) & 1.57\hphantom{0}~(9) \\
              &   &    & $-$36.77\hphantom{0}~(1) & 13:12:31.63\hphantom{00}~(2) & $-$62:34:08.1\hphantom{00}~(1) & 0.48\hphantom{00}~(3) & 4.5\hphantom{00}~(2) \\
              &   &    & $-$36.32\hphantom{0}~(2) & 13:12:31.54\hphantom{00}~(3) & $-$62:34:07.4\hphantom{00}~(2) & 0.21\hphantom{00}~(5) & 1.5\hphantom{00}~(3) \\
              & CD & 36 & $-$37.55\hphantom{0}~(9) & 13:12:31.50\hphantom{00}~(9) & $-$62:34:06.5\hphantom{00}~(6) & 3.4\hphantom{000}~(2) & 0.59\hphantom{0}~(4) \\
              &   &    & $-$37.20\hphantom{0}~(1) & 13:12:31.51\hphantom{00}~(4) & $-$62:34:02.2\hphantom{00}~(3) & 0.50\hphantom{00}~(3) & 1.96\hphantom{0}~(8) \\
              & D & 44 & $-$37.372~(3) & 13:12:31.51\hphantom{00}~(1) & $-$62:34:01.69\hphantom{0}~(8) & 0.596\hphantom{0}~(7) & 11.3\hphantom{00}~(1) \\
              & E & 36 & $-$36.57\hphantom{0}~(1) & 13:12:36.17\hphantom{00}~(3) & $-$62:33:27.1\hphantom{00}~(2) & 0.27\hphantom{00}~(4) & 0.68\hphantom{0}~(8) \\
              &   &    & $-$36.27\hphantom{0}~(3) & 13:12:36.0\hphantom{000}~(1) & $-$62:33:27.7\hphantom{00}~(6) & 1.59\hphantom{00}~(8) & 0.71\hphantom{0}~(6) \\
              &   &    & $-$36.18\hphantom{0}~(1) & 13:12:36.1\hphantom{000}~(1) & $-$62:33:27.3\hphantom{00}~(7) & 0.29\hphantom{00}~(4) & 0.70\hphantom{0}~(8) \\
              &   & 44 & $-$36.82\hphantom{0}~(1) & 13:12:36.21\hphantom{00}~(2) & $-$62:33:26.7\hphantom{00}~(1) & 0.31\hphantom{00}~(3) & 1.6\hphantom{00}~(1) \\
              &   &    & $-$36.45\hphantom{0}~(1) & 13:12:36.12\hphantom{00}~(5) & $-$62:33:27.3\hphantom{00}~(3) & 0.31\hphantom{00}~(3) & 1.8\hphantom{00}~(2) \\
              &   &    & $-$36.28\hphantom{0}~(6) & 13:12:36.12\hphantom{00}~(9) & $-$62:33:26.8\hphantom{00}~(5) & 1.7\hphantom{000}~(2) & 0.6\hphantom{00}~(2) \\
              &   &    & $-$35.925~(7) & 13:12:36.46\hphantom{00}~(3) & $-$62:33:25.2\hphantom{00}~(2) & 0.32\hphantom{00}~(2) & 1.8\hphantom{00}~(1) \\
              & F & 36 & $-$35.1\hphantom{00}~(1) & 13:12:35.9\hphantom{000}~(1) & $-$62:33:35.2\hphantom{00}~(7) & 2.3\hphantom{000}~(3) & 0.37\hphantom{0}~(4) \\
              &   &    & $-$35.047~(7) & 13:12:36.08\hphantom{00}~(9) & $-$62:33:36.3\hphantom{00}~(6) & 0.27\hphantom{00}~(2) & 1.32\hphantom{0}~(8) \\
              &   &    & $-$33.80\hphantom{0}~(5) & 13:12:35.74\hphantom{00}~(4) & $-$62:33:32.4\hphantom{00}~(3) & 0.57\hphantom{00}~(8) & 2.6\hphantom{00}~(2) \\
              &   &    & $-$33.469~(2) & 13:12:35.554\hphantom{0}~(3) & $-$62:33:32.04\hphantom{0}~(2) & 0.342\hphantom{0}~(3) & 23.9\hphantom{00}~(5) \\
              &   &    & $-$32.36\hphantom{0}~(6) & 13:12:35.9\hphantom{000}~(1) & $-$62:33:33.7\hphantom{00}~(7) & 0.4\hphantom{000}~(1) & 0.5\hphantom{00}~(1) \\
              &   &    & $-$31.85\hphantom{0}~(5) & 13:12:35.92\hphantom{00}~(6) & $-$62:33:33.9\hphantom{00}~(4) & 0.63\hphantom{00}~(9) & 0.94\hphantom{0}~(6) \\
              &   & 44 & $-$36.45\hphantom{0}~(2) & 13:12:36.05\hphantom{00}~(6) & $-$62:33:35.4\hphantom{00}~(4) & 0.76\hphantom{00}~(6) & 1.03\hphantom{0}~(8) \\
              &   &    & $-$35.314~(3) & 13:12:36.04\hphantom{00}~(4) & $-$62:33:35.9\hphantom{00}~(3) & 0.459\hphantom{0}~(8) & 6.35\hphantom{0}~(9) \\
              &   &    & $-$34.38\hphantom{0}~(1) & 13:12:35.83\hphantom{00}~(6) & $-$62:33:33.7\hphantom{00}~(4) & 0.51\hphantom{00}~(2) & 5.0\hphantom{00}~(1) \\
              &   &    & $-$33.746~(1) & 13:12:35.609\hphantom{0}~(5) & $-$62:33:31.82\hphantom{0}~(3) & 0.283\hphantom{0}~(3) & 35.0\hphantom{00}~(6) \\
              &   &    & $-$33.649~(5) & 13:12:35.715\hphantom{0}~(6) & $-$62:33:31.68\hphantom{0}~(4) & 0.646\hphantom{0}~(8) & 25.8\hphantom{00}~(6) \\
              &   &    & $-$32.4\hphantom{00}~(3) & 13:12:35.903\hphantom{0}~(8) & $-$62:33:32.99\hphantom{0}~(5) & 0.8\hphantom{000}~(3) & 3\hphantom{000.}~(1) \\
              &   &    & $-$31.94\hphantom{0}~(4) & 13:12:35.91\hphantom{00}~(2) & $-$62:33:33.1\hphantom{00}~(1) & 0.57\hphantom{00}~(8) & 4\hphantom{000.}~(3) \\
              & G & 36 & $-$34.99\hphantom{0}~(5) & 13:12:36.7\hphantom{000}~(2) & $-$62:33:36\hphantom{000.}~(1) & 0.27\hphantom{00}~(8) & 0.9\hphantom{00}~(2) \\
              &   &    & $-$34.718~(6) & 13:12:36.819\hphantom{0}~(4) & $-$62:33:37.47\hphantom{0}~(3) & 0.289\hphantom{0}~(8) & 8.2\hphantom{00}~(2) \\
              &  &   & $-$33.66\hphantom{0}~(1) & 13:12:37.013\hphantom{0}~(7) & $-$62:33:35.43\hphantom{0}~(5) & 0.59\hphantom{00}~(2) & 5.2\hphantom{00}~(3) \\
             &   &    & $-$33.608~(4) & 13:12:37.011\hphantom{0}~(4) & $-$62:33:35.24\hphantom{0}~(3) & 0.20\hphantom{00}~(1) & 4.3\hphantom{00}~(3) \\
              &   &    & $-$33.107~(2) & 13:12:36.968\hphantom{0}~(2) & $-$62:33:35.68\hphantom{0}~(1) & 0.212\hphantom{0}~(9) & 16\hphantom{000.}~(2) \\
              &   &    & $-$32.995~(8) & 13:12:37.01\hphantom{00}~(2) & $-$62:33:34.8\hphantom{00}~(2) & 0.336\hphantom{0}~(6) & 35\hphantom{000.}~(2) \\
              &   &    & $-$32.49\hphantom{0}~(3) & 13:12:37.10\hphantom{00}~(8) & $-$62:33:34.1\hphantom{00}~(7) & 0.38\hphantom{00}~(6) & 0.66\hphantom{0}~(6) \\
              &   & 44 & $-$35.6\hphantom{00}~(1) & 13:12:36.85\hphantom{00}~(9) & $-$62:33:37.5\hphantom{00}~(6) & 0.4\hphantom{000}~(3) & 0.6\hphantom{00}~(3) \\
              &   &    & $-$34.99\hphantom{0}~(2) & 13:12:36.87\hphantom{00}~(2) & $-$62:33:37.4\hphantom{00}~(1) & 0.53\hphantom{00}~(5) & 12\hphantom{000.}~(3) \\
              &   &    & $-$34.930~(3) & 13:12:36.850\hphantom{0}~(7) & $-$62:33:37.33\hphantom{0}~(4) & 0.26\hphantom{00}~(1) & 22\hphantom{000.}~(3) \\
              &   &    & $-$34.05\hphantom{0}~(2) & 13:12:37.082\hphantom{0}~(6) & $-$62:33:35.56\hphantom{0}~(4) & 0.42\hphantom{00}~(3) & 18.7\hphantom{00}~(8) \\
              &   &    & $-$33.808~(3) & 13:12:37.023\hphantom{0}~(2) & $-$62:33:35.26\hphantom{0}~(2) & 0.259\hphantom{0}~(7) & 30\hphantom{000.}~(2) \\
              &   &    & $-$33.262~(2) & 13:12:36.999\hphantom{0}~(2) & $-$62:33:35.49\hphantom{0}~(2) & 0.404\hphantom{0}~(4) & 100\hphantom{000.}~(4) \\
              &   &    & $-$33.0\hphantom{00}~(2) & 13:12:37.20\hphantom{00}~(2) & $-$62:33:34.7\hphantom{00}~(1) & 0.7\hphantom{000}~(1) & 6\hphantom{000.}~(2) \\
              &   &    & $-$31.57\hphantom{0}~(8) & 13:12:36.96\hphantom{00}~(6) & $-$62:33:35.0\hphantom{00}~(5) & 0.6\hphantom{000}~(2) & 0.6\hphantom{00}~(1) \\
              & H & 36 & $-$32.61\hphantom{0}~(1) & 13:12:36.9\hphantom{000}~(1) & $-$62:33:30.7\hphantom{00}~(8) & 0.56\hphantom{00}~(3) & 1.87\hphantom{0}~(5) \\
              &   & 44 & $-$32.847~(5) & 13:12:36.95\hphantom{00}~(8) & $-$62:33:30.0\hphantom{00}~(4) & 0.518\hphantom{0}~(9) & 9.0\hphantom{00}~(3) \\
\hline
\end{tabular}
\end{table*}

The spatial distribution of the class~I maser emission for each source is shown in Figure~\ref{map} along with the spectra for each location marked in the image (for sources other than G305.37$+$0.21 see the 
full version of the figure available online). For the single selected source G305.37$+$0.21 we also show the scalar-averaged visibility spectra at 36 and 44~GHz to illustrate the 
amount of additional information provided by interferometric data in comparison to the typical single dish experiment as well as the biases of the latter when the spatial distribution is as wide as
the primary beam. To declutter the image
plots we showed Gaussian components co-located in position and velocity within 3$\sigma$ by a single symbol (crosses for the 36-GHz and pluses for the 44-GHz masers). The position uncertainty
does not exceed the symbol size. Some symbols were enlarged to ensure this. The wide concentric circles (or arcs) in Fig.~\ref{map} represent the full width at half 
maximum (FWHM) of the primary beam in both transitions. Centred at the pointing position, these circles therefore enclose the 50 per cent sensitivity region 
for each target. With the exception of six 
sources for which no {\em Spitzer} data were available, the three-colour
background in Fig.~\ref{map}  shows the emission in the 8.0, 4.5 and 3.6-$\mu$m IRAC {\em Spitzer} bands as red, green and blue, respectively. A survey of ground-state hydroxyl 
masers, designated MAGMO \citep[][and further data in preparation]{gre12b}
provided 18-cm radio continuum measurements for the majority of our targets (except for eight sources specifically commented on
in the following section) with a typical 1$\sigma$ noise level of about 5$-$10~mJy. If detected within the image boundaries, this continuum emission (referred to as MAGMO data)
is shown by contours in Fig.~\ref{map},
unless more appropriate data are available for the particular source (see the notes on individual sources for details). Note, the radio
emission from H{\sc ii} regions is usually optically thick and weak at low frequency. Therefore, observations at higher frequencies are required to establish reliably whether an H{\sc ii} region
is present in a particular source. However, such sensitive observations are scarce at the moment for class~I maser targets. 
The positions of the class~II methanol masers at 6.7-GHz measured as part of the Methanol
Multibeam survey \citep[MMB;][]{cas10,cas11a,gre12a} are shown by squares in Fig.~\ref{map}, if such masers were found. The absolute radial velocity in the spectra depends on the 
adopted value of rest frequency. The
velocity uncertainty associated with the uncertainty of the rest frequency is shown by horizontal error bars for all 36-GHz spectra (Fig.~\ref{map}; this uncertainty
is comparable to the spectral resolution at 44~GHz). 
However, in subsequent analysis (section~\ref{restfrequncert} and~\ref{veldistribution}) we argue that the best estimate rest frequency at 36~GHz should be lower than currently
adopted, and the plotted 36-GHz spectra should be shifted to lower velocity by approximately the width of the error bar shown; peaks of many 36- and 44-GHz features
then become aligned.

\subsection{Notes on individual sources}
\label{notesonsrc}

\subsubsection{G269.15$-$1.13}
This source is located in the Vela Molecular Ridge cloud~B \citep[e.g.,][]{mur91} in the area devoid of
radio continuum emission stronger than a few mJy \citep[e.g. 3-cm measurement of ][18-cm MAGMO data]{wal98}. 
However, outside the field shown in Fig.~\ref{map}, there is an H{\sc ii} region about 30~arcsec west
of the class~I maser location~C \citep{cas87a}. This area is well outside the half-power point of the primary beam and may
contain additional masers (e.g. associated with an H{\sc ii} region) which our data are not sensitive to. The rest of the class~I
maser emission is found to cluster around the position of the 6.7-GHz maser \citep[Fig.~\protect\ref{map};][]{cas09,gre12a}. 
There is also an H$_2$O maser near this position \citep{bre11}. Note,
the uncertainties of the profile fit (listed in Table~\ref{fitresults}) may be underestimated for the 36-GHz maser at~A.
Guided by the spatial distribution, we had to fit this profile with one more Gaussian component than would be
required by the spectral profile and this destabilised the fit in the spectral domain. Partially resolved thermal emission
may have contributed to 36-GHz profiles of~E and~B.



\subsubsection{G270.26$+$0.84}

This target is associated with the Vela Molecular Ridge cloud~A \citep[e.g.,][]{mur91}.
The class~I masers are scattered around the position of the 6.7-GHz maser \citep{cas09,gre12a} near the half-power point of the primary beam.
They are red-shifted in velocity with respect to the 6.7-GHz maser, which peaks at around 4~\ks. The largest velocity offset (over 15~\ks) is at the location G detected 
at 36~GHz only. \citet{cas89} detected a strong H$_2$O maser towards this source albeit without accurate position measurement (note that \citet{bre11} measured the 
accurate position but their spectrum suffers from ringing artefacts). The water maser peaks at 8~\kss and is also
red-shifted with respect to the 6.7-GHz maser up to velocities of about 30~\ks. No 18-cm continuum has been detected towards this source in the MAGMO project
(1$\sigma$ rms is 0.8~mJy). 


\subsubsection{G294.98$-$1.73}

This class~I maser source was found towards an IRAS source \citep{sly94}, which is associated 
with a weak ($\sim$~1~mJy at 3cm) H{\sc ii} region \citep{wal98}.
The pointing centre lies just beyond the northern edge of the field shown in Fig.~\ref{map},
more than 20~arcsec from the actual position of 
class~I masers. This offset explains the rather marginal appearance of the 95-GHz spectrum \citep{val00} which was registered at the same pointing
position and with a narrower beam. A highly variable 6.7-GHz maser was found near the class~I masers in the MMB project \citep{gre12a}. The class~I
masers at A and B are blue-shifted with respect to the 6.7-GHz maser emission, while masers at C, detected at 36~GHz only, are red-shifted.
No 18-cm continuum was found within the boundaries of Fig.~\ref{map} (MAGMO continuum rms is 0.7~mJy). 


\subsubsection{G300.97$+$1.15}

The class~I masers are spread over an elongated 20$\times$40~arcsec$^2$ region located approximately 15~arcsec south of the extended H{\sc ii}
region revealed by the MAGMO data at 18-cm (see Fig.~\ref{map}). The H{\sc ii} region has a compact core situated near its edge. This
core is associated with the 6.7-GHz maser which is red-shifted with respect to all detected class~I masers \citep{cas09,gre12a}.
The 2MASS data revealed a chain of compact 2-$\mu$m sources roughly tracing the shape of the leading edge of the H{\sc ii} region a few 
arcseconds south of it \citep{mot07}. There is also a weak filament between two 2-$\mu$m sources possibly tracing shocked gas. One of these 
sources (located near B), is associated with a 10.4-$\mu$m source and also detected at 21$\mu$m (MSX). Therefore, this source is likely to
be the site of induced star formation. The class~I masers are most likely associated with the new generation of YSOs, although the masers
at B, the maser location closest to the H{\sc ii} region, could be related to the ionization front.



\subsubsection{G301.14$-$0.23}

This is a reasonably well studied source which harbours a molecular outflow deeply embedded
in a dense molecular cloud \citep{hen00}. The {\em Spitzer} IRAC images reveal four strong infrared sources labelled MIR1 to MIR4
by \citet{hen00} in addition to a number of weak ones. Two infrared sources have associated radio continuum emission, 
presumably from H{\sc ii} regions 
\citep[see MAGMO data in Fig.~\protect\ref{map} and][]{wal98}. There is an extended 4.5-$\mu$m halo around these two sources and
a weaker source which lies immediately to the north in the area where \citet{hen00} detected nebulous 2.12-$\mu$m H$_2$ emission.
Therefore, the presence of shocked gas is likely in this source despite the fact that it is not listed as an EGO by \citet{cyg08}, 
perhaps due to the effects of strong sources. The class~I masers at A are situated near the edge of the northern H{\sc ii}
region, which is also at the rim of the H$_2$ emission \citep[c.f.][]{hen00} presumably tracing the outflow. The maser at B is
associated with the southern H{\sc ii} region and has a remarkable absorption feature in the spectrum. This absorption feature
is unresolved at the ATCA resolution and was fitted with Gaussians and presented in Table~\ref{fitresults} the same way as the maser 
componets. However, we have excluded the absorption features from the statistical analysis presented in the following sections. This 
location also has a weak ($\sim$1~Jy) 6.7-GHz maser \citep{cas09,gre12a} which
spans the velocity range from $-$41 to $-$37~\ks, comparable to the velocities of the absorption feature
\citep[c.f. Fig.~\protect\ref{fitresults} and][]{gre12a}, and a co-located OH maser which shows an unusually wide spread of
velocity components from $-$64 to $-$33~\kss \citep{cas87b,cas98}.


\subsubsection{G305.21$+$0.21}


This is a well studied region of high-mass star formation (see \citet{wal07} and references therein). There are two 6.7-GHz masers in the region, the 
stronger 305.208$+$0.206 and the weaker 305.202$+$0.208 \citep{gre12a}, which were referred to as G305A and G305B, respectively by \citet{wal07}.
The latter site (G305B) is associated with a strong infrared source (see Fig.~\ref{map}) but no detected radio continuum emission, while the
former (G305A) is a deeply embedded hot core detected in a number of molecular tracers which has either no associated infrared source or a very 
weak one \citep[e.g.][]{wal07}.  The source has two prominent H{\sc ii} regions, one located approximately 30~arcsec  to the south-east of G305A just outside 
the region shown in Fig.~\ref{map} \citep[referred to as G305HII(SE) by][]{wal07} and 
the other located 15~arcsec to the west of G305B \citep[referred to as G305HII by][]{wal07}. None of the 6.7-GHz masers appear to be directly 
associated with any of these H{\sc ii} regions \citep{phi98}. The contours in Fig.~\ref{map} represent reprocessed ATCA archival data at 6~cm (project code C865, 6A array
configuration). The emission at 36 and 44~GHz is scattered around the large area exceeding the FWHM of the primary beam. The locations A$-$E, G and L seem to be
associated with G305A and, except for A and B which are situated in the vicinity of the 6.7-GHz maser, are arranged in an elliptic arc 
approximately 10$\times$7~arcsec$^2$ across 
possibly tracing an outflow cavity. The strongest masers at both 36 and 44~GHz are at location~F, 
approximately 3~arcsec to the east of the 6.7-GHz maser position at 
G305B. This region is active in a number of rare class~I methanol maser transitions which require higher than average temperatures and densities to form 
\citep[see the discussion of this source by][]{vor10b}. It is worth noting that in addition to the 25-GHz methanol masers 
at F \citep{vor05b,vor10b}, \citet{wal07} found 25-GHz maser emission at D as well as the broad and 
possibly quasi-thermal emission at B which may be a counterpart
to the broad 36- and 44-GHz features seen in the spectrum (see Fig.~\ref{map}).
Detected at 44-GHz only, the maser emission at J is seen projected onto the continuum source G305HII and may, therefore, be associated with the interface 
between this H{\sc ii} region and the molecular gas in the cold cloud identified by \citet{wal06} in the area to the south-west of G305B and G305HII. The maser
emission at K, H and I may also originate from this cloud or its interface. However, given the angular separation of at least 10~arcsec, it seems unlikely that the emission
at K and H is caused by interaction with G305HII.

\subsubsection{G305.25$+$0.25}

This region also belongs to the G305 complex and has widespread star formation activity \citep[e.g.][]{lon07}. The class~I methanol
masers are distributed very tightly around a 6.7-GHz maser \citep[Fig.~\protect\ref{map};][]{cas09,gre12a}. All class~I maser emission is
blue-shifted with respect to the class~II maser velocities. An infrared overlay reveals a nearby filament most likely tracing PAH emission. There is also a nearby infrared source
with 4.5-$\mu$m excess, but it is too compact to be considered as an EGO.  No 18-cm continuum has been detected in the MAGMO project (rms is 1~mJy).
Note, the uncertainties of the fit are likely to be underestimated for the most red-shifted Gaussian component of~A at both 36 and 44~GHz
(at $-$35.0 and $-$35.68~\ks, respectively).



\subsubsection{G305.37$+$0.21}
There are two clearly separated sites of class~I maser emission: the south-western locations A$-$D situated near the boundary of an extended H{\sc ii} region traced by the 18-cm emission contours and
the north-eastern locations E$-$H encircling a compact infrared source at the tip of the ridge of the 18-cm emission (Fig.~\ref{map}). We also overlay an image of the 15~mJy 3-cm continuum source located
in the north-eastern site obtained by \citet{hin12} using the spatial filtering of the data to enhance the compact structure. The circular distribution of the class~I maser emission  supports 
the idea of possible association of such masers with expanding H{\sc ii} regions \citep{vor10b,cyg12}.
\citet{vor07} reported detection of a strong class~I methanol
maser in the 25~GHz series (J=5 transition of the series was observed) with two components, which both seem to be associated with the location G, according to the measured position and velocity.
No 6.7-GHz methanol maser has been detected in this source in the MMB survey. 


\subsubsection{G309.38$-$0.13}

This source has been studied in detail at 36 and 44~GHz by \citet{vor10a}, who revealed a maser feature blue-shifted by over 30~\kss from the systemic velocity. In this paper,
we present an independent observation of this source, which was carried out a few days after the original dataset had been taken. In particular, the new observation was made at
a different receiver tuning which allowed us to better register the high-velocity feature at both 36 and 44~GHz (the original 44-GHz dataset did not cover the velocities corresponding to the
high-velocity feature at all and the 36-GHz dataset had them near the edge of the band). The two datasets provide an excellent cross-check on the reliability of the measurements.
The comparison of Table~\ref{fitresults} and Figure~\ref{map} with table~2 and figure~2 from \citet{vor10a} shows that they are in a good agreement (note,
the coordinate grid in Figure~2 of \citet{vor10a} has a rounding error in the labels due to a software error, the plot itself reflects correct relative positions). The main difference,
apart from the detecton of a weak 44-GHz counterpart of the high-velocity feature (also labelled~E for consistency with the earlier paper), is the detection of a few additional weak
components in the new dataset, including a 36-GHz counterpart at the location~A at $-$51.0~\kss and a new 2-Jy feature at 44~GHz near $-$50.2~\kss labelled
J \citep[both were most likely missed due to confusion with dynamic range artefacts caused by other strong emission in][]{vor10a}. The new data confirmed the presence of the 44-GHz emission 
towards the location~I and also revealed a marginal 44-GHz component at the same location near $-$88~\ks, i.e. near the velocities of the high-velocity feature (the spectrum is 
labelled I$^m$ in Fig.~\ref{map}). The broad component of H is the only one with more than 3$\sigma$ disagreement with the previous results. Finer structure  
of this component is also evident in the spectrum (Fig.~\ref{map}). This minor discrepancy could be due to the presence of a partially resolved thermal emission.
The 6.7-GHz methanol maser in the region peaks at $-$49.6~\ks. It is located just a few seconds of arc south of locations B and E. There is also an EGO located in the area indicating
presence of the shocked gas \citep[classified as a likely outflow candidate according to][]{cyg08}. No 18-cm radio-continuum emission has been found in this region with an rms of 2~mJy (MAGMO data).


\subsubsection{G316.76$-$0.01}
The field is dominated by an infrared source which has  a 4.5-$\mu$m excess but  was not classified as an EGO by \citet{cyg08}, perhaps due to its regular shape (Fig.~\ref{map}). 
The MMB survey revealed no 6.7-GHz methanol maser in this source \citep{gre12a}.
The class~I maser emission is spread over a large area with separation of up to 0\farcm5 from the infrared source. The most offset emission components (at F, D and E) are weak and 
detected at 36~GHz only. The emission profiles at locations F and D are broad and possibly correspond to quasi-thermal emission from the parent molecular cloud. The maser emission at 
C (this location contains the strongest peak at both 36 and 44~GHz) and B forms an arc structure around the eastern side of the infrared source. Both the H$_2$O and
hydroxyl masers are located near~C \citep{bre10a,cas98} and may be part of the same arc. No 18-cm continuum emission was detected towards this source (MAGMO, with rms of 5~mJy).

\subsubsection{G318.05$+$0.09}

Most of the class~I masers are distributed over a large area about 30~arcsec in extent to the south-east from the 6.7-GHz maser position 
\citep[Fig.~\protect\ref{map};][]{cas09,gre12a}. The exception is the maser location B, which lies approximately 20~arcsec to the south-west. The {\em Spitzer} image
reveals a number of infrared sources in the general area occupied by class~I masers and an elongated 4.5-$\mu$m feature, possibly an outflow emanating from the
YSO near the 6.7-GHz maser, which seems to be associated with the class~I masers at C and G \citep[Fig.~\protect\ref{map}; likely outflow candidate by][]{cyg08}.
Location~A is situated a few arcseconds to the north of this structure and also likely to be associated with the same outflow.  
The 36-GHz emission at D$-$F is located futher to the south-east and does not follow the direction traced by the 4.5-$\mu$m feature. Therefore, either 
the outflow is bent or all locations except~B trace an arc-like structure at the rim of the dark cloud identified by \citet{per09} approximately 0\farcm5 to the south of
class~I masers. Diffuse 4.5-$\mu$m emission extends
to the west and south-west of the 6.7-GHz maser. There is also a brighter knot in the immediate vicinity of B.  
No 18-cm continuum emission has been detected in this source (MAGMO, with rms 1~mJy).



\subsubsection{G318.95$-$0.20}

The distribution of class~I masers in this source is elongated in the north-west to the south-east direction symmetrically with respect to the 
position of the 6.7-GHz maser, which is associated with the strongest infrared source \citep[Fig.~\protect\ref{map};][]{gre12a}. 
This morphology suggests the presence of an outflow in the region. Assuming a distance of 10.6~kpc \citep{gre11}, this outflow is at least
1.1~pc long. It is worth noting that the 36-GHz spectrum of C (likely in the terminal area of the outflow) shows a pronounced red-shifted wing 
spanning several \kss (see Fig.~\ref{map}), although the spatial resolution of our data was insufficient to detect a velocity gradient in this feature.
The baseline of the 36-GHz spectrum of B contained weak (a few 100~mJy) undulations raising the detection threshold a few times. The cause of this is not entirely
clear and could indicate the presence of resolved, probably thermal, emission with complex kinematics.
No radio continuum emission has been detected in this source (the 18-cm MAGMO image has rms of 4~mJy; the 3-cm measurement of \citet{ell96} has a 5$\sigma$ limit of 0.82~mJy).


\subsubsection{G320.29$-$0.31}

The class~I masers in this source are tightly clustered around a rather weak 6.7-GHz maser \citep[Fig.~\protect\ref{map}; 0.8~Jy at 6.7~GHz according to][]{gre12a}.
They are located at the interface between a dark area in the {\em Spitzer} image, presumably a molecular cloud, and two infrared sources with the 4.5-$\mu$m excess 
\citep[neither qualified as an EGO according to][]{cyg08}. There is a stronger infrared source and an extended PDR at the northern tip of the molecular cloud,
approximately 10~arcsec north from the masers (Fig.~\ref{map}). The spectral fit required broad pedestals at both frequencies to fit the data within the noise
(peaks are 1.5 and 3~Jy at 36 and 44~GHz, respectively). However, we were unable to position these features and therefore do not list them in Table~\ref{fitresults}.
No 18-cm continuum has been detected in this source (MAGMO, with rms 6~mJy).


\subsubsection{G322.16$+$0.64}

This region has a complex distribution in the infrared emission (see Fig.~\ref{map}). It contains 3 well-separated sites of class~I methanol maser 
emission which could be independent sources. The most abundant site is around the position of the 6.7-GHz maser \citep{cas09,gre12a} and an infrared
source with a strong 4.5-$\mu$m excess (locations A$-$D and F in Fig.~\ref{map}). The latter may actually be a cluster of compact sources without 
any significant extended component and, therefore, is not listed as an EGO by \citet{cyg08}. Some emission at A, B and C is located near the edge
of the 4.5-$\mu$m source. There is a hint of 36-GHz emission at A not shown in the spectra due to poor signal-to-noise ratio. It is worth noting that the
hydroxyl maser (co-located with the 6.7-GHz methanol maser within the measurement uncertainty) is rather weak ($<$1~Jy) and blue-shifted with respect 
to the class~I methanol masers \citep{cas98}. The 6.7-GHz methanol maser is strong ($>$200~Jy) and has a wide velocity range 
\citep[from $-$66 to $-$51~\ks;][]{gre12a} covering that of both class~I methanol and hydroxyl masers.

The second class~I maser site is
represented by the location E near the pointing centre in the south-east corner of the field shown in Fig.~\ref{map}. This location is near the edge
of an H{\sc ii} region with some class~I methanol masers projected onto it. There is also an associated water maser 322.165$+$0.625, but no 6.7-GHz methanol maser 
at this position \citep{bre10a,gre12a}.
The remaining site is represented by locations G and H. Situated in the north of the
field, this area is devoid of strong infrared emission and presumably contains a molecular cloud \citep[although not listed as a dark cloud by][]{per09}. 
The location H is near the interface of this cloud traced by the 8.0-$\mu$m PDR emission. The PDR emission shows some symmetry and has a dark area
along the perceived symmetry axis which goes through G and H (Fig.~\ref{map}). The continuation of this line further to the south-east
goes through the H{\sc ii} region at E suggesting that a physical connection between these two sites cannot be completely excluded, although this is
pure speculation.
It is also worth noting that the blue and red wings
of the peak at G may correspond to slightly different positions (although the separation is below the uncertainty of our measurement).


\subsubsection{G323.74$-$0.26}

The class~I masers are scattered around a pronounced EGO \citep[designated as a possible outflow candidate by][]{cyg08} and very strong 6.7-GHz methanol 
maser \citep[over 3000~Jy;][]{phi98,gre12a}. There are at least three infrared sources embedded in the EGO \citep{wal01} with the two
dominant ones clearly visible in the {\em Spitzer} overlay (Fig.~\ref{map}). Note, there are two components of the 6.7-GHz maser \citep[see e.g.][]{phi98}
which the MMB project treated as a single source due to their close proximity (about 3~arcsec). The stronger 6.7-GHz maser is associated with the western 
infrared source and an OH maser showing a wide velocity spread from $-$62~\kss to $-$39~\kss \citep{cas98}. The latter (along with the EGO) suggests 
a strong outflow activity in the region.
The eastern infrared source is associated with the weaker 6.7-GHz maser and the class~I maser emission at C. The rest of the class~I maser sites
(A, B and D) are in close proximity to the edge of the EGO.
There is an extended H{\sc ii} region associated with the pronounced PDR in the north of the field shown in Fig.~\ref{map}, but no radio continuum
emission in the vicinity of class~I masers \citep[below 0.4~mJy at 18~cm according to MAGMO data and 0.2~mJy at 3~cm according to][]{phi98}.



\subsubsection{G324.72$+$0.34}

The class~I masers are spread over a large area comparable to the ATCA field of view. The distribution is approximately symmetric with respect to the position
of the 6.7-GHz methanol and mainline OH masers
\citep[Fig.~\ref{map};][]{cas98,cas09,gre12a}, which are located at a tip of an EGO \citep[likely outflow candidate according to][]{cyg08}. 
There is a dark cloud located to the north-west of the 6.7-GHz maser position \citep{per09}, so the class~I masers at F may be associated with its interface.
The broader spectral feature of F and the masers at E are located in the immediate vicinity of the EGO. The masers at A, B and C are the most offset from it. 
Assuming the distance to the source is 11.1~kpc \citep{gre11}, this implies a linear separation in excess of 2~pc for the most offset location B. The
EGO has a faint tail of emission which curves towards A suggesting a possible physical association. The masers at C and B are likely to be independent sources.
No 18-cm continuum emission was detected (MAGMO, rms 1~mJy).


\subsubsection{G326.48$+$0.70}

This region has two class~II methanol masers at 6.7-GHz, 326.475+0.703 and 326.476+0.695 \citep{cas09,gre12a}. The former is a strong maser (flux density in
excess of 100~Jy) located in an area devoid of strong infrared sources in the north-west corner of the field shown in Fig.~\ref{map}. It is estimated to 
be at the far kinematic distance of 11.8~kpc by \citet{gre11}. In contrast, the latter 6.7-GHz maser is co-located with a strong infrared source (Fig.~\ref{map})
and assumed to be at the near kinematic distance of 2.9~kpc. This drastic difference in assumed distances makes identification of the class~I masers
with a particular 6.7-GHz maser a non-trivial task because the region of influence of 326.476$+$0.695 extends virtually across the whole field of view shown
in Fig.~\ref{map}. In particular, location F is most likely attributed to 326.476$+$0.695, despite a slightly greater angular separation than from
the other 6.7-GHz maser. The distribution of the class~I masers in the vicinity of 326.475$+$0.703 (locations A$-$C, G and H) is rather compact, which is
consistent with the large assumed distance (although the Galactic latitude implies a large offset from the Galactic plane in this case). 
The maser at~C (detected at 44~GHz only) could not be positioned and fitted reliably due to confusion with other strong
emission (mainly~A). We only denote its approximate location in~Fig.~\ref{map}.
There are two EGOs in the region \citep[both are likely outflow candidates according to][]{cyg08}, one is several
arcsec to the east of A and the other is at the position of the bulk of the class~I maser emission at this location (obscured by the maser symbols in Fig.~\ref{map}).
There is also a compact infrared source with 4.5-$\mu$m excess near G and a more diffuse 4.5-$\mu$m source several seconds of arc north of F (see Fig.~\ref{map}), 
neither classified as EGOs. It is worth noting that \citet{per09} identified a dark cloud immediately to the north of G. However, this could be a foreground
object.
\citet{gar06} reported the most sensitive continuum measurement available to date towards this region, but detected no radio continuum emission
in the field shown in Fig.~\ref{map} (rms below 3~mJy at 20 and 13~cm and below 0.3~mJy at 6 and 3~cm).

   
\subsubsection{G326.64$+$0.61}

The class~I masers are spread over a large area exceeding 1~arcmin in angular extent. Most of the emission (locations A, B, D, E, G and J) is tightly 
clustered immediately to the north-west of the 6.7-GHz maser (326.641$+$0.611; see Fig.~\ref{map}) located near the edge of the prominent 8.0-$\mu$m filament 
(barely captured in the corner of the field shown in Fig.~\ref{map}). The remaining emission is further away, with the locations C, F and H being near 
the half power offset of the primary beam at 44~GHz and the most offset locations I and K being well outside the 50 per cent sensitivity region at both 
frequencies. Therefore, our data were sensitive only to the strongest emission in this area. 
There is a marginal detection of a high-velocity feature at K
(labelled K$^*$ in the spectra in Fig.~\ref{map}) peaking near $-$63~\kss (other emission detected at the same location is confined to the velocities 
from $-$31 to $-$25~\ks). The area occupied by the class~I maser emission is largely devoid of infrared sources, with the
exception of weak infrared sources near A, F and K (Fig.~\ref{map}).
No 18-cm continuum emission was detected in the MAGMO project within the field presented in Fig.~\ref{map} (rms noise is 12~mJy).



\subsubsection{G326.66$+$0.57}

There are two sites of class~I maser emission in this source separated
by about 20~arcsec. Labelled as locations~A and B, they were detected only at 36 and 44~GHz, respectively. Both masers are quite faint, so no self-calibration has been 
performed as part of the reduction process. This had no effect on the presented data because there was only one emission component at each frequency and, therefore, no
need to determine relative positions. The masers are located in an area devoid of strong infrared sources and presumably associated with
a dark cloud \citep[see Fig.~\protect\ref{map} and the catalogue of][]{per09}. 
The MAGMO \citep{gre12b} 18-cm continuum data revealed a weak (around 30~mJy or 2$\sigma$) source at the northern side of Fig.~\ref{map} near the 
8.0-$\mu$m emission which belongs to a long filament surrounding the whole dark cloud. The continuum
source could be the result of imaging with inadequate $uv$-coverage of an extended 0.1-Jy H{\sc ii} region located further north (outside the field of view).
The MMB survey revealed no 6.7-GHz methanol maser in this source \citep{gre12a}. \citet{cas98} found a weak OH maser 
about 1\farcm5 south-east of the class~I masers peaking at $-$40.8~\ks, which implies no emission
in the immediate vicinity of the methanol masers (1$\sigma$ rms is 0.1~Jy) given the primary beam size. 
However, the class~I masers are located near the edge of the field of view for the water maser observations 
of \citet{bre10a}. Therefore, it only excludes
associations with a strong water maser (tens of Jy). This source may be an example of a very young star-forming region prior to the appearance
of other masers, most notably the class~II methanol maser at 6.7~GHz.


\subsubsection{G326.86$-$0.68}

This compact source is located at the western edge of a dark cloud \citep{per09}, the interface of which might be directly associated with class~I masers at B and C.
The 6.7-GHz maser is also located nearby and is red-shifted with respect to all class~I masers \citep[Fig.~\protect\ref{map};][]{cas09,gre12a}. 
There is an EGO \citep[likely outflow candidate according to][]{cyg08} 
several arcsec further west, which may be associated with the class~I maser at D and some of the emission represented by A (the 36-GHz emission at A 
corresponds to both locations A and B at 44~GHz). An additional 1.5-Jy feature near $-$66.68~\kss may also be present, but 
has been excluded from Table~\ref{fitresults} due to confusion with the peak at A. No 18-cm continuum emission has been detected in the MAGMO project (rms is 0.5~mJy).


\subsubsection{G327.29$-$0.58}

This is a relatively well studied region with a prominent hot core, infrared dark cloud \citep{leu13a,wyr06} and an EGO 
\citep[Fig.~\protect\ref{map}; possible outflow candidate according to][]{cyg08}. There are also indications that triggered star formation is taking place
in this region \citep{min09}. The position of the hot core \citep{wyr06} suggests association with the class~I masers at B,C and G
as well as a rather weak (a few Jy) 6.7-GHz methanol maser and an infrared source \citep[see Fig.~\protect\ref{map} and Table~\protect\ref{fitresults};][]{cas09,gre12a}. 
The observations of \citet{wyr06} 
registered this hot core in thermal methanol as an elongated 26$\times$7~arcsec$^2$ source extending up to the class~I maser site at A. 
The EGO is located approximately the same angular distance further north-east on the same line (Fig.~\ref{map}). The class~I masers at E show a
close position association with the EGO. Detected at 36-GHz only, the maser site H is located nearby.
The masers at F, I and D also form a line, which is parallel but offset about 10~arcsec to the north. It seems likely that
class~I masers is this source trace an outflow emanating from the hot core in the north-east direction. The EGO and the masers at H and E
could trace the terminal surface of the outflow. The hot core and the EGO correspond to the 450-$\mu$m sources SMM1 and SMM4, respectively \citep{min09}.
The dark cloud is located north of the EGO position \citep{per09}.
No 18-cm emission has been detected in the MAGMO project within the boundaries shown in Fig.~\ref{map} with rather high rms of 18~mJy, perhaps due to
the emission of H{\sc ii} regions located outside the field. \citet{wyr08} detected 3-mm continuum emission co-located with the hot core.
It is worth noting that the gaussian decomposition of the 44-GHz spectral profile at A does not approximate well the small dip around $-$46.1~\kss
between the peak and the red-shifted component.


\subsubsection{G327.39$+$0.20}

This field has two 6.7-GHz masers, 327.392$+$0.199 and 327.395$+$0.197, each associated with its own cluster of class~I maser emission (locations B and A,
respectively) and infrared source
\citep[see Fig.~\protect\ref{map};][]{cas09,gre12a}. The infrared source associated with the western 6.7-GHz maser (327.392$+$0.199, location B) is weaker
but has an exended 4.5-$\mu$m halo and was classified as an EGO \citep[possible outflow candidate according to][]{cyg08}.
This 6.7-GHz maser has a notably wider velocity spread (from $-$87 to $-$79~\ks) than the eastern 6.7-GHz maser \citep[from $-$90 to $-$88; see][]{gre12a}.
Similarly, the associated class~I maser emission at B spans an almost 3 times wider velocity range than the maser at A, which suggests that an outflow
might indeed be present. Due to the presence of a resolved (and most likely thermal) source, the position of the broad component in B may have
an uncertainty at 36~GHz exceeding the values quoted in Table~\ref{fitresults}. No 18-cm continuum emission has been detected towards this source
(MAGMO project, rms is 0.3~mJy).


\subsubsection{G327.62$-$0.11}

The distribution of class~I maser emission in this source is rather compact (Fig.~\ref{map}). The strongest emission is at A located in the immediate vicinity
of a rather weak 6.7-GHz maser \citep[$<$3~Jy;][]{cas09,gre12a}, which is associated with the dominant infrared source in the {\em Spitzer} overlay (Fig.~\ref{map}). 
The other site of class~I maser emission (B, several arcsec to the south of A) contains a marginal spike and a broad component which may be quasi-thermal. It is
worth noting that \citet{gre11} attributed this source to the far kinematic distance of 8.9~kpc, although the presence of a quasi-thermal emission 
would favour the near kinematic distance. No 18-cm continuum emission has been detected towards this source (MAGMO project, rms is 0.4~mJy).


\subsubsection{G328.21$-$0.59}

This is a rather compact source of class~I maser emission located in an area devoid of strong infrared (see Fig.~\ref{map}) and radio continuum sources 
(the 1$\sigma$ upper limit at 18-cm is 2~mJy according to MAGMO data), and most likely associated with a dark cloud~\citep{per09}. All the detected emission
lies near the half-power point of the primary beam at 36~GHz. Therefore, the map may be missing maser emission further away from the pointing
centre. The MMB project
detected no 6.7-GHz maser in this source \citep{gre12a} and there seems to be no OH maser either (being about 3~arcmin offset from the following target, 
G328.24$-$0.55, this source is well within the 18-cm primary beam of \citet{cas98} observations which implies non-detection with noise rms of 0.1~Jy). 
This source may be a good example of the youngest stage of star formation, before the 6.7-GHz maser appears.



\subsubsection{G328.24$-$0.55}

Most of the class~I maser emission is located near the 6.7-GHz maser associated with a compact H{\sc ii} region 
\citep[see Fig.~\protect\ref{map} and][]{phi98}. 
The exception is the 36-GHz maser at C, which is most likely an unrelated source located well  outside the 50 per cent sensitivity region. As an independent
confirmation of this detection, there is also a hint at the position and velocity of this feature in the 
data on the following source, G328.25$-$0.53. But it does not satisfy the
formal detection criterion and is further away from the pointing centre in that dataset. Therefore, the maser at C is described only in relation to this source. 
This is the only pair of targets in our sample which is sufficiently close to get the same feature detectable in two separate
pointings. 

The 36-GHz emission is quite weak at the remaining locations A and B. Therefore, no self-calibration was done at this frequency.
The mid-infrared image is dominated by the strong 8.0~$\mu$m cometary-shaped PDR located several seconds of arc north of the masers at A (see Fig.~\ref{map}).
An association of some of these masers with the PDR interface cannot be excluded. No radio continuum emission has been detected towards this PDR
(data of \citet{phi98} imply a 1$\sigma$ limit of 0.2~mJy at 3~cm; MAGMO data yielded no 18-cm detection in the whole field shown in Fig.~\ref{map} with
rms of 2~mJy). The contours in Fig.~\ref{map} represent 3-cm continuum data \citep{phi98}. There are a number of sources with 
emission excess at 4.5~$\mu$m near class~I masers at both A and B. However, none of these sources have qualified as an EGO, perhaps due to their size \citep{cyg08}.



\subsubsection{G328.25$-$0.53}

The class~I maser emission is scattered across a large area. Most of the emission is at or beyond the half-power point of the primary beam with the exception
of location A near the 6.7-GHz maser (south-western corner in Fig.~\ref{map}, near the pointing centre). The locations C and F are in the
area with diffuse 8.0~$\mu$m emission, while B, D and E are situated in the area devoid of strong infrared sources, presumably at the edge of a 
dark cloud. The 6.7-GHz maser is associated
with a hydroxyl maser \citep{cas98} and a strong infrared source, but no 18-cm continuum emission has been detected in the MAGMO project 
(1$\sigma$ rms is 2~mJy). There is also an EGO to the north of the dominant infrared source \citep[likely outflow candidate according to][]{cyg08}. 


\subsubsection{G328.81$+$0.63}

The class~I maser emission is spread over a large area comparable to the field of view. There are two very close and strong 6.7-GHz masers in this source 
\citep[2\farcs5 separation; in excess of 300~Jy;][]{cas97,wal98,gre12a} seen projected onto an H{\sc ii} region \citep[$\sim$200~mJy at 3~cm;][]{cas97,wal98} at 
the northern side of a complex infrared source (see Fig.~\ref{map}). The weak 36-GHz emission at J is found nearby.
The infrared source has been classified as a possible outflow candidate by \citet{cyg08}, although the 4.5-$\mu$m excess is not
immediately obvious in the image. The southern diffuse part of the infrared source looks more prominent at 8.0~$\mu$m and may actually represent a PDR.
The class~I masers at B, C and D detected at 44~GHz only seem to be associated with the southern interface of this PDR (see Fig.~\ref{map}) and the dark
cloud identified by \citet{per09} in the area located further south. The class~I masers at A and H as well as the broad line and possibly quasi-thermal
36-GHz emission at I are also likely to be associated with this dark cloud. The remaining class~I maser sites E, F and G are arranged in a line towards
the northern direction and may trace an outflow propagating into a less dense environment away from the dense cloud.


\subsubsection{G329.03$-$0.20}

This region has two 6.7-GHz methanol masers, 329.029$-$0.205 and 329.031$-$0.198, in the south-east and the north-west, respectively. The former is quite strong and
has a flux density in excess of 100~Jy. This source, along with the 
two following sources (G329.07$-$0.31 and G329.18$-$0.31), are associated with a continous filamentary infrared dark cloud \citep[see Fig.~17 of][]{ell06}
and likely to have similar ages. This
association strongly favours the near kinematic distance, although the ambiguity resolution of \citet{gre11}, based on the absence of H{\sc i} self-absorption,
implied the far kinematic distance of 12~kpc for both 6.7-GHz masers in this region.
Each 6.7-GHz maser has its own class~I maser cluster. The locations A$-$D and I correspond to the north-western 
6.7-GHz maser and trace a curved structure (with B,C and D) near the northern edge of the image in Fig.~\ref{map}, which resembles a bow-shock. The remaining
class~I maser sites E$-$H correspond to the south-eastern 6.7-GHz maser. The locations F$-$H trace a linear structure with the 6.7-GHz maser situated between F and G,
while E lies approximately 10~arcsec to the west and forms a separate arc structure (Fig.~\ref{map}). The two 6.7-GHz masers have distinct velocity ranges:
from $-$48.5~\kss to $-$41.5~\kss for the north-west maser and $-$41.5~\kss to $-$33.5~\kss for the south-east maser \citep{gre12a}. While the north-western class~I masers 
seem to share a similar velocity range with the associated 6.7-GHz maser, the other south-eastern group tends to cover a wider velocity range (from $-$45~\kss 
to $-$38~\ks) encompassing that of both 6.7-GHz masers. The OH masers associated with the 6.7-GHz masers show a similar behaviour \citep{cas98}.
Embedded inside a dark cloud \citep{per09}, this region is devoid of strong infrared sources. There is a faint 4.5-$\mu$m source 
\citep[classified as an EGO; likely outflow candidate according to][]{cyg08}, which is located a few arcsec south of the class~I masers at A and
the 6.7-GHz masers. The location I is seen projected onto this EGO suggesting a physical association.
No 18-cm continuum has been detected towards this source (MAGMO project, rms is 0.2~mJy). 


\subsubsection{G329.07$-$0.31}

The class~I maser emission is scattered across a large area over 1~arcmin in extent. The IRAC overlay is dominated by two strong infrared sources. The western 
infrared source is associated with the 6.7-GHz methanol and OH masers \citep{cas95, cas98, gre12a}. There might be some extended 4.5-$\mu$m emission near 
both infrared sources \citep[possible outflow candidates according to][]{cyg08}, although imaging artefacts cannot be completely excluded. 
In particular, one of these suspected EGOs (near the eastern source) may be associated with the class~I masers at the location~C. 
There is also a compact 4.5-$\mu$m source near the maser location D. The 36-GHz maser at D shows a large radial 
velocity spread and a broad wing of emission towards the higher velocities (see Fig.~\ref{map} and Table~\ref{fitresults}). An unexpectedly
large number of Gaussian components was required for the profile fit (see also Table~\ref{fitresults}). However, the profile was represented within the noise.
Perhaps, location~D is the site of interaction
with the dense molecular cloud located between D, E and A \citep{sim06, per09}. 
The fit overestimates the measured profile for the maser at~E at the boundaries of components at the level of 2$-$3$\sigma$. 
No 18-cm continuum has been detected towards this source in the MAGMO 
project (1$\sigma$ rms is 0.3~mJy). It is worth noting that \citet{gre11} attributed this source to the far kinematic distance of 11.8~kpc. However,
a large angular spread of class~I maser spots results in linear offsets up to 2~pc and favours the near kinematic distance.  
As commented on the previous source, this source is associated with a continuous infrared dark filament \citep[see Fig.~17 of][]{ell06} along with G329.03$-$0.20
and G329.18$-$0.31, which would also suggest the near kinematic distance. 




\subsubsection{G329.18$-$0.31}

The class~I methanol masers are spread around a large area comparable to the field of view. They trace an elongated dark cloud, which has a number
of infrared sources embedded in it \citep[Fig.~\protect\ref{map};][]{per09}. The 6.7-GHz methanol maser \citep{cas09,gre12a}  and the associated OH maser 
\citep{cas98} are co-located with one of these sources enshrouded by a very pronounced EGO \citep[likely outflow candidate according to][]{cyg08}.
The class~I masers at E are seen projected onto this EGO and, therefore, likely to be in close physical association. The emission at this location
along with C, D and F trace a regular arc, with the 6.7-GHz maser located close to its vertex. Thus the structure may represent the walls of an outflow
propagating at an angle to the plane of the sky. The 6.7-GHz maser and the class~I masers at G, B and H form a line along the dark cloud 
(Fig.~\ref{map}). This line may trace the other arm of the outflow which perhaps does not have as much material to interact with and extends
further from the YSO. It is worth noting that there is a knot of the 4.5-$\mu$m emission near G elongated in the same direction. Located
near the northern tip of this knot is a marginal high-velocity feature, G, only at 44-GHz, which is blue-shifted
by 20.9~\kss with respect to the middle of the velocity range of the 6.7-GHz maser (Table~\ref{fitresults}). Both these facts corroborate the idea
that the outflow (most likely the approaching arm) is present in the area. The remaining class~I maser sites A, I$-$K are likely to be associated with
the external interface of the dark cloud. No 18-cm continuum has been detected in this region (MAGMO project, rms is 3~mJy). As discussed 
for the previous two sources (G329.03$-$0.20 and G329.07$-$0.31), this source is also associated with the same filamentary infrared dark cloud
\citep[see Fig.~17 of][]{ell06}. In contrast to the previous two sources, \citet{gre11} attributed this source to the near kinematic distance of 3.3~kpc
(which is consistent with the spread of maser spots and association with an IRDC). It is worth noting that the assignment of the previous two sources to 
the far kinematic distance was based on the absence of a detectable self absorption signal, which always carries a greater uncertainty. The assignment of 
this source also carried uncertainty as it was complicated by the presence of a local trough in the H{\sc i} emission at the same velocity as the 
absorption signal.


\subsubsection{G329.47$+$0.50}

The bulk of class~I maser emission (locations B, C, E$-$G and I) is tightly clustered around the position of the 6.7-GHz maser \citep[Fig.~\ref{map};][]{cas09,gre12a} associated
with an EGO \citep[likely outflow candidate according to][]{cyg08}. The maser at I (detected at 36-GHz only) shows very broad wings detectable over the range of velocities
exceeding 15~\ks.  The remaining class~I masers are well separated from the 6.7-GHz maser: A, D and H in the north, K in the south and J in the south-west, which is associated
with the infrared source (Fig.~\ref{map}). Note, that \citet{gre11} attributed this source to the far kinematic 
distance of 10.8~kpc. Although this estimate is entirely
consistent with the compactness of the main cluster of masers, it implies projected linear separations between 1.2 and 2~pc for the offset sites. Therefore,
the offset sites may represent unrelated sources. No cm-wavelength radio continuum has been detected towards this 
source \citep[rms of 0.4~mJy at 18~cm according to MAGMO data and 1~mJy at 3-cm according to][]{wal98}.


\subsubsection{G331.13$-$0.24}

This source has two prominent H{\sc ii} regions \citep[see MAGMO data in Fig.~\protect\ref{map} and][]{phi98}. The 6.7-GHz methanol and OH masers are located
near the edge of the north-eastern H{\sc ii} region \citep{phi98,cas11a,cas98} bordering an EGO \citep[likely outflow candidate according to][]{cyg08}. The class~I
methanol masers are scattered around this H{\sc ii} region, with the location A situated near its southern edge. This location has one of a few known class~I 
methanol masers at 9.9-GHz, which requires higher temperatures and densities to form than
the common masers at 44 and 36~GHz \citep{vor10b}. Its velocity ($-$91.16~\ks) agrees with that of the peak 36- and 44-GHz component at A if the spectral
resolution and the rest frequency uncertainty of the two experiments are taken into account. None of the class~I maser locations seem to be close to the EGO.
\citet{deb03} found a single knot of 2.12~$\mu$m H$_2$ emission in the area a few seconds of arc north of B. There is an outflow in the source 
traced by the SiO emission \citep{deb09} which is likely to be the cause of the class~I maser emission at least in B and C. However, the tight geometry
of the region does not allow us to draw reliable associations given the spatial resolution of the available data. The distribution of the class~I maser emission
suggests that the outflow axis may be close to the line of sight (this has also been considered by \citet{deb09} as a possible explanation for significant
spatial overlap of the blue-shifted and red-shifted SiO emission) with perhaps some inclination in the north-eastern direction.



\subsubsection{G331.34$-$0.35}

This is a very compact class~I maser, which is associated with an EGO \citep[possible outflow candidate according to][]{cyg08} and located at the southern
edge of a strong infrared source associated with the 6.7-GHz and OH masers \cite[see Fig.~\ref{map};][]{cas96,cas98,cas11a}. There is a pronounced 8.0-$\mu$m
filament, thinner to the south-east of the infrared source (Fig.~\ref{map}). At least some of this emission may trace the interface region of the dark cloud 
located just over 10~arcsec south of the masers \citep{per09}. No 18-cm continuum emission has been detected within the boundaries 
shown in Fig.~\ref{map} (MAGMO project, rms is 3~mJy). Note, the profile fit systematically overestimates the peak at 44~GHz at the level of 2$-$3$\sigma$. 



\subsubsection{G331.44$-$0.19}

Most of the class~I maser emission (locations A$-$C and E) lies in a straight line extending south-west from the position of a 
continuum source, presumably an ultra-compact H{\sc ii} region (see Fig.~\ref{map}, contours show 18-cm MAGMO data). The 6.7-GHz methanol \citep{cas96,cas11a},
OH \citep{cas98} and class~I masers at D are seen projected
onto the edge of the H{\sc ii} region, a few arcsec north of the line traced by A$-$C and E and may also be associated with a stronger infrared source located
a few arcsec west of the H{\sc ii} region. The maser at F is situated about 20~arcsec north of the 6.7-GHz maser and may be an unrelated source.


\subsubsection{G332.30$-$0.09}

This region has two 6.7-GHz masers, 332.295$-$0.094 and 332.296$-$0.094 which are only 3~arcsec apart \citep{cas09,cas11a}. The eastern 6.7-GHz maser (332.296$-$0.094), newly 
discovered in the MMB project, is weak (a few Jy) and variable. It is associated with the northern edge of a well developed H{\sc ii} region with a pronounced PDR
traced by the 8.0-$\mu$m emission (Fig.~\ref{map}, contours show MAGMO 18-cm data). The majority of class~I masers at A trace the same edge of the H{\sc ii} region.
The western 6.7-GHz maser (332.295$-$0.094) is located at the south-eastern tip of an EGO \citep[possible outflow candidate according to][]{cyg08}. The class~I masers at B
as well as the $-$46.74~\kss 44-GHz feature of A are associated with this EGO and located at the opposite sides. The remaining class~I maser site C is located near the edge
of another H{\sc ii} region located about 10~arcsec west of the 6.7-GHz masers. There is also an infrared source with 4.5-$\mu$m excess at C although it is not listed
as an EGO by \citet{cyg08}.


\subsubsection{G332.60$-$0.17}

The class~I masers at A (see Fig.~\ref{map}) and the 6.7-GHz maser \citep{cas96,cas11a} are associated 
with a faint EGO \citep[likely outflow candidate according to][]{cyg08}.
The location B, offset by about 10~arcsec to the north from the EGO position, corresponds to broad and possibly quasi-thermal lines at both 36 and 44~GHz.
The source is located in a darker area on the {\em Spitzer} image surrounded by the 8.0-$\mu$m emission, but not formally associated with any dark cloud
identified by \citet{per09}. No 18-cm continuum has been detected towards this source (MAGMO data, rms is 0.9~mJy).


\subsubsection{G332.94$-$0.69}

This region has an EGO \citep[likely outflow candidate according to][]{cyg08} extending to the south-eastern direction from an infrared source associated with a
weak ($<$2~Jy) 6.7-GHz maser discovered in the MMB project \citep{cas11a}. The class~I masers at A are associated with the EGO. The other class~I maser sites
are offset (a few arcsec offset for B and over 15~arcsec for C) from the EGO and seen at the edge of a darker area in the {\em Spitzer} image enclosed by
the 8.0-$\mu$m emission and the EGO. Although this area is not listed as a dark cloud by \citet{per09}, it probably harbours a molecular cloud. No 18-cm
continuum emission has been detected in this source (MAGMO data, rms is 2~mJy).


\subsubsection{G332.96$-$0.68}

Most of the class~I masers in this source are associated with a pronounced EGO \citep[likely outflow candidate according to][]{cyg08}. Associated with the 6.7-GHz maser 
\citep{cas09,cas11a} is an infrared point source embedded in this arc-shaped EGO (see Fig.~\ref{map}). 
Location~B corresponds to the southern arm of this EGO, while location~A corresponds to the northern arm and the edge of
the source, and has a noticeably blue-shifted component at 36~GHz.
\citet{per09} identified two dark clouds, located towards the western and the south-east direction from the EGO. Perhaps these
clouds are a part of a common structure where this star-forming region is embedded. No 18-cm continuum has been detected within the boundaries shown in Fig.~\ref{map}
(MAGMO data, rms is 3~mJy).



\subsubsection{G333.03$-$0.06}

This very compact source is located between a strong infrared source and a more extended region with 4.5-$\mu$m excess (Fig.~\ref{map}). However,
the latter is not listed as an EGO by \citet{cyg08}, perhaps due to the presence of the strong source. A rather weak ($<$2~Jy) 6.7-GHz maser is co-located with the
majority of the class~I masers within the measurement uncertainty and shares a similar velocity range. No 18-cm continuum emission has been detected
towards this source (MAGMO data, rms is 1~mJy).

\subsubsection{G333.13$-$0.44}

This is the only target in our sample which contains four 6.7-GHz methanol masers: a close pair of weak masers (333.126$-$0.440 and 333.128$-$0.440, 
separated by 7~arcsec, both weaker than several Jy) near the
centre of the image shown in Fig.~\ref{map}, 333.121$-$0.434 in the west and another weak maser, 333.135$-$0.431, in the north \citep{cas11a,cas97}.
The distribution of class~I maser emission suggests rather straightforward identification of each class~I maser site with a particular 6.7-GHz maser. However,
it must be noted that ambiguity cannot be completely excluded for class~I masers corresponding to the close pair of the 6.7-GHz masers. 
The dominant infrared source in Fig.~\ref{map} is associated with the south-western 6.7-GHz maser of the pair, 333.126$-$0.440. The class~I masers
at A are at the position of the 6.7-GHz maser within the measurement uncertainty and scattered within several arcsec of it. 
Note, decomposition of the spectral profile into the Gaussian components leaves systematic residuals near minima between individual peaks. The effect
is seen at both frequencies, but is much more pronounced at 44~GHz.
The class~I maser at C is located 20~arcsec south of A and may be an unrelated source. The profile looks very similar to the peak of A, but is
too strong to be an artefact. The north-eastern 
6.7-GHz maser of the pair, 333.128$-$0.440, is located near an infrared source (showing 3.6-$\mu$m excess), but in an area devoid of extended 
infrared emission in the {\em Spitzer} overlay (Fig.~\ref{map}). The associated class~I masers (locations D$-$F and H) are spread around 10$\times$10~arcsec$^2$ area
to the north-east, further away from the pair of 6.7-GHz masers. All this class~I maser emission is blue-shifted with respect to the associated 6.7-GHz maser.
The northern 6.7-GHz maser, 333.135$-$0.431 is associated with the continuum source and an extended infrared source which appears green in the three-colour
map (Fig.~\ref{map}) suggesting an excess at 4.5$\mu$m. However, this source is not listed as an EGO by \citet{cyg08}. Unlike the other three sites, this one
harbours an OH maser \citep{cas98}. The class~I methanol maser at G (only 44-GHz masers are detected) are located nearby, but appear not to be directly 
associated with the infrared source (about 10~arcsec offset from the 6.7-GHz maser). The 44-GHz maser is blue-shifted with respect to the northern 6.7-GHz maser. 
This location is beyond the half-power point of the primary beam at 44-GHz. Therefore, additional class~I masers related to the same YSO could have been missed.
The western 6.7-GHz maser, 333.121$-$0.434, is related to the class~I masers at B located within a few arcsec from it towards a north-western direction. 
At a comparable offset in the
south-eastern direction lies a weak continuum source which may represent a working surface of an outflow. This site (B) is located near the edge of a dark
cloud \citep{per09} with the rim traced by a long 8.0-$\mu$m filament. \citet{mot07} also reported a radio continuum emission extending from a prominent
H{\sc ii} region at the northern edge of Fig.~\ref{map} along the filament towards B \citep[see also][]{urq07}. 
Although we cannot exclude that the class~I masers at B are directly associated
with the 8.0-$\mu$m filament, most likely they are related to an outflow in a separate high-mass star-forming region whose formation has been triggered
by interaction of the dark cloud with its surroundings. 


\subsubsection{G333.13$-$0.56}

This region has two close (separated by 6 arcsec) 6.7-GHz masers, 333.128$-$0.560 at the south-west and 333.130$-$0.560 at the north-east \citep{cas09,cas11a}.
The latter is located a few arcsec from the infrared source which has been classified 
as an EGO and a possible outflow candidate by \citet{cyg08} despite being 
rather compact (see Fig.~\ref{map}). Due to the close proximity of the 6.7-GHz masers, identification with the class~I maser emission may be ambiguous, especially
for location~D which is equidistant. The class~I masers at~B are clustered around the EGO with some components seen projected onto it. Some of this emission forms a curved
structure with the north-east 6.7-GHz maser (333.130$-$0.560) and class~I masers at~D located along the line of symmetry suggesting a possible association. Note, the 
decomposition into Gaussian components of the 36-GHz emission at this location (Table~\ref{fitresults}) may have underestimated uncertainties, although it
represents the profile (Fig.~\ref{map}) within the noise. Most of the class~I maser emission in this region is located at or near the edge (E and some emission at B) of 
extended 8.0$\mu$m emission (Fig.~\ref{map}). The masers at~C are well separated (about 30~arcsec) from both 6.7-GHz masers and may be an unrelated source. \citet{gar04}
found a massive dense core in the area using 1.2-mm emission. Despite the low spatial resolution, this core seems to be located between C and the other class~I masers,
perhaps centred in the region with a darker 8.0$\mu$m background (see Fig.~\ref{map}). The majority of class~I maser emission in this source may be associated
with the interface of this core. No 18-cm continuum has been found towards this source (MAGMO data, rms is 6~mJy).


\subsubsection{G333.16$-$0.10}

The infrared {\em Spitzer} image of this region is dominated by a strong source associated with an H{\sc ii} region (18-cm peak is about 30~mJy according to MAGMO data; Fig.~\ref{map}).
The stronger class~I masers at B and the 6.7-GHz maser \citep{cas96,cas11a} are located near the edge of the H{\sc ii} region, the weaker masers at A are a few arcsec
south of it. The only reported limit on the water maser emission in this region is a relatively weak one imposed by the HOPS survey \citep[complete at $\sim$10~Jy;][]{wal11}.


\subsubsection{G333.18$-$0.09}

This source has a pronounced EGO \citep[likely outflow candidate according to][]{cyg08} with embedded infrared source associated with the 6.7-GHz methanol maser
\citep[Fig.~\protect\ref{map};][]{cas96,cas11a}. The class~I maser emission at~B and the $-$85.89-\kss 44-GHz feature of~A trace the edge of the EGO, which is presumably
the interface with a dark cloud located towards the north-east \citep{per09}. Note, the emission at~B seems to be slightly extended suggesting a second component at a 
very close velocity and position. However, our present data do not allow us to separate these components. The remaining masers at the location~A are situated near the
6.7-GHz maser and most likely associated with the EGO. No 18-cm continuum has been detected towards this source (MAGMO data, rms is 2~mJy). 
The only reported limit on the water maser emission in this region is a relatively weak one imposed by the HOPS survey \citep[complete at $\sim$10~Jy;][]{wal11}.


\subsubsection{G333.23$-$0.06}

This source has two weak (both $<$2~Jy) 6.7-GHz masers separated by just over 5~arcsec \citep{cas96,cas11a}, 333.234$-$0.060 in the north-west and 333.234$-$0.062 in the south-east.
The former is associated with an OH maser and a strong infrared source in the {\em Spitzer} overlay \citep[Fig.~\protect\ref{map};][]{cas98}. The whole complex is located in 
a darker area on the infrared overlay surrounded by the 8.0$\mu$m emission suggesting the region is most likely embedded in a molecular cloud.
The class~I maser emission is
scattered around the 6.7-GHz masers making the association with a particular 6.7-GHz maser ambiguous. We attribute emission at B and~C to the south-eastern 
6.7-GHz maser (333.234$-$0.062) based purely on the smallest angular separation. Some emission at~B traces an arc-like structure, suggesting a possible association with an outflow cavity.
The remaining sites of the class~I maser are attributed to the north-western 6.7-GHz maser (333.234$-$0.060). The locations D and F are near the edge of the infrared
source and the remaining locations A and E are separated from other class~I maser emission by more than 10~arcsec. The class~I masers at~A include the strongest 
component (almost 300~Jy at 44~GHz) and have a very compact spatial distribution. Another noteworthy feature is a marginal high-velocity component at $-$106.4~\kss detected
at 36~GHz only. It is blue-shifted by about 20~\kss with respect to the middle of the velocity range of the north-western 6.7-GHz maser.
No 18-cm continuum has been detected towards this source (MAGMO data, rms is 2~mJy).




\subsubsection{G333.32$+$0.11}

The {\em Spitzer} infrared image of this region (see Fig.~\ref{map}) is dominated by a strong source with an EGO at the north-western end 
\citep[possible outflow candidate according to][]{cyg08}. The 6.7-GHz maser \citep{cas96,cas11a} and the 44-GHz masers at~A are associated with the 
infrared source. The class~I masers represented by location~B include components seen projected at the EGO and near its edge. The rest
of the class~I maser emission lies several seconds of arc offset from the infrared source. No 18-cm continuum has been detected towards this source (MAGMO data, rms is 1~mJy).


\subsubsection{G333.47$-$0.16}

The striking feature of this source is a curved distribution of the class~I maser masers at A, B, D and F mimicking a bow-shock. The 6.7-GHz methanol
and OH masers are located near the apex of this structure \citep{cas98,cas11a}. This is also the location of an 18-cm continuum source 
with the peak flux density of about 30~mJy (see Fig.~\ref{map}). Note, that the MAGMO data have rather high rms of 11~mJy for this source. 
From comparison with the 4-cm observation of \citet{wal98}, one may speculate that at least some of this emission corresponds to a working surface of
a jet. However, given the quality of the available data, more sensitive continuum observations are required to establish the true nature of 
this source.  The mid-infrared image reveals a pronounced EGO in this source 
\citep[likely outflow candidate according to][]{cyg08} elongated roughly in the north-south direction. The class~I masers at C, G, E and H as well
as some spectral features represented by A trace this EGO. This source is a good demonstration of the mutual complementarity of the two
class~I methanol maser transitions at 36 and 44~GHz: the southern end of bow-shock structure is mainly traced by the 36-GHz masers while
the southern end of the EGO is only traced by the 44-GHz masers. If only
the 44-GHz data were available, this source would most likely be interpreted as an outflow along the EGO axis. The full data suggest that the outflow is
most likely inclined towards the north-eastern direction and the EGO traces a side wall of this outflow.


\subsubsection{G333.56$-$0.02}

This is a very compact class~I methanol maser source associated with the 6.7-GHz methanol maser \citep{cas11a} and
located at the edge of the dark cloud \citep{per09} surrounded by the 8.0-$\mu$m PDR emission (Fig.~\ref{map}).
The MAGMO data revealed no 18-cm continuum emission within the field of view shown in Fig.~\ref{map} with the rms of 7~mJy.
The only reported limit on the water maser emission in this region is a relatively weak one imposed by the HOPS survey \citep[complete at $\sim$10~Jy;][]{wal11}.
This source may be at the earliest stage of its evolution.


\subsubsection{G333.59$-$0.21}

This source was found by \citet{sly94} towards an IRAS source associated with the prominent H{\sc ii} region G333.6$-$0.2 enclosed by a strong PDR ring 
traced by the 8.0-$\mu$m emission (seen in the north-eastern corner of the field in Fig.~\ref{map}). However, our interferometric observations revealed that all
class~I maser emission is located 45~arcsec away which explains only marginal 95-GHz detection by \citet{val00}. The MMB project revealed no
6.7-GHz methanol maser in the area \citep{cas11a}. Also this location is within the primary beam of the OH measurement of \citet{cas98} which targeted masers
near the H{\sc ii} region. This implies the flux density limit of 0.1~Jy for OH masers (1$\sigma$) in the vicinity of the class~I masers. No H$_2$O maser
emission has been reported by \citet{bre10a} at this location (the offset corresponds roughly to the half power point of their measurement implying an rms of 0.3~Jy).
The class~I masers are located at the edge of the dark cloud \citep{per09} and may represent the earliest stage of the evolution. The separation from the ionization front
traced by the 8.0-$\mu$m suggests that the maser emission is unlikely to be directly related to this front. However, the interaction of this front with the 
molecular gas seen as a dark cloud could have induced star formation which in turn gave rise to class~I masers.


\subsubsection{G335.06$-$0.43}

This source has an EGO \citep[likely outflow candidate according to][]{cyg08} distributed symmetrically about the position of the 6.7-GHz and OH masers
\citep[see Fig.~\ref{map};][]{cas98,cas09,cas11a}. The class~I methanol masers at A, B and D appear to trace this EGO, and two other locations C and E are 
20~arcsec offset to the west and east, respectively. The masers at A appear to form a curved structure which resembles the morphology of the 4.5-$\mu$m
emission in the northern side of the EGO. Most likely masers at A, B and D are tracing an outflow.
No 18-cm continuum emission has been detected in the MAGMO project (rms is 0.5~mJy).

\subsubsection{G335.59$-$0.29}

This is a complex with 3 nearby 6.7-GHz maser sites, a close pair comprising 335.585$-$0.290 and 335.585$-$0.289, and the third source
335.585$-$0.285, about 15~arcsec away. The close pair and the third source are both associated with the OH and H$_2$O masers \citep{cas98,bre10a}.
\citet{gar02b} found no 6-cm continuum within the field shown in Fig.~\ref{map} (3$\sigma$ limit is 0.2~mJy). 
The mid-infrared overlay is dominated by a single strong source near the close pair of 6.7-GHz masers. The area surrounding the 
mid-infrared source and the 6.7-GHz masers has a number of sites with diffuse 4.5-$\mu$m emisison which qualify as an EGO
\citep[likely outflow candidate according to][]{cyg08}. Most class~I methanol masers
are scattered around the EGO and 6.7-GHz masers over the area approximately 30~arcsec in extent. The exception is the 44-GHz maser at C, which is
located well outside the 50 per cent sensitivity region and may be a separate source (assuming the distance of 3.3~kpc, the projected linear separation
from the nearest 6.7-GHz maser is about 0.9~pc). 
This source has recently been the object of a detailed investigation by \citet{per13}. The 6.7-GHz methanol masers in this source are associated with 
a 550~M$_\odot$ dust clump and lie towards the centre of an infared dark cloud from which a number of filaments appear to emanate 
\citep[in particular, class~I masers at G and D may be associated with the edge of this cloud, see also][]{per09}.
Mopra HCO$+$~(1$\rightarrow$0) observations show blue-shifted self-absorption features within both the filaments and the body of the cloud and this 
has been interpreted as evidence for large-scale infall towards the central high-mass core.
The morphology of the class~I maser distribution seems to have a preferential direction from 
south-east to north-west, roughly along the line traced by the 6.7-GHz masers and the axis of the EGO, which likely indicates the presence of an 
outflow in this source. It is worth noting that the SiO (2$\rightarrow$1) data corroborate the presence of a bipolar outflow \citep{gar02b}.



\subsubsection{G335.79$+$0.17}

The {\em Spitzer} infrared image (Fig.~\ref{map}) reveals a very pronounced EGO \citep[likely outflow candidate according to][]{cyg08} and an 8.0-$\mu$m 
filament. A rather strong 6.7-GHz maser ($>$100~Jy) is located at the southern tip of the EGO \citep{cas96,cas11a} and coincides with the mainline OH 
maser \citep{cas98}. The class~I maser emission at~C, D and G
is seen projected at the EGO or located near its edge. Note, the uncertainties of all parameters of the $-$48.933-\kss 44-GHz Gaussian component of~C
(Table~\ref{fitresults}) are underestimated due to confusion with the peak.
The masers at~A, B, E and F are likely to be related to the 8.0-$\mu$m filament. The 44-GHz emission represented by~F contains a broad component which 
is most likely a compact part of a resolved thermal source. Therefore, the corresponding fit parameters have large systematic uncertainties.
The remaining maser at~H, which has been detected only at 36~GHz, is slightly ($\sim$5~arcsec) offset from the EGO. 
There is no 18-cm measurement for this source in the MAGMO data.



\subsubsection{G336.41$-$0.26}

The class~I maser emission in this source is confined to two sites located about 10~arcsec apart. Both sites are near the edges of infrared sources
surrounded by extended 8.0-$\mu$m structure (Fig.~\ref{map}), perhaps an embedded cluster of young sources. The 6.7-GHz maser is located
within this structure at a clearly distinct position \citep{cas09,cas11a}. The infrared source at~A is associated with a weak (9~mJy at 3~cm) H{\sc ii} region
\citep[Fig.~\ref{map}; see also][]{phi98}. The contours in Fig.~\ref{map} represent 3-cm continuum based on 
archival data (project code C379, observations made on the 5th of July 1995). \citet{per09} identified a compact dark cloud to the south of the 8.0-$\mu$m source.
However, no class~I maser emission appears to be associated with it.
\citet{phi98} indicated that additional weaker continuum emission may 
be present in the area, including the location of the 6.7-GHz maser. They also suspected that an additional weak 6.7-GHz maser might be present near the strong
infrared source about 20~arcsec south-west of class~I masers. The only reported limit on the water maser emission in this region is a relatively weak one 
imposed by the HOPS survey \citep[complete at $\sim$10~Jy;][]{wal11}.



\subsubsection{G337.40$-$0.40}

Most of the class~I maser emission is located in the vicinity of the strong radio continuum and mid-infrared source (see Fig.~\ref{map}), including
a tight cluster near the continuum peak, which lies a few 
seconds of arc south of the 6.7-GHz methanol \citep{wal98,cas09,cas11a} and ground-state OH masers \citep{cas98} but coincident with 6035-MHz OH
masers \citep{cas01}.
There is extended infrared emission with a 4.5-$\mu$m excess around this source. However this site is not listed as an EGO
by \citet{cyg08}, perhaps due to the effects of the strong source. The radio-continuum emission was interpreted by \citet{guz12} 
as an ionized jet which is driving a high-velocity molecular outflow detected between $-$85 and 41~\ks.
The location~A is most likely associated with this outflow. It has multiple 44-GHz spectral features and a weak 
continuous 36-GHz emission across the velocity interval over 10~\ks, roughly symmetric with respect
to the 6.7-GHz maser \citep[c.f. spectra of][]{wal98,cas09,cas11a} and the ambient gas velocity \citep[assumed to be $-$40.7~\ks;][]{guz12}. 
The 36-GHz feature at $-$36.53~\kss is located at a spatially distinct position but overlaps in velocity with the 36-GHz emission emanating from 
the location~A. Therefore, its spectrum is presented separately (location E in Fig.~\ref{map}). The 44-GHz spectrum can be adequately repesented
by a single plot (location A) despite the spatial spread covering both locations A and E at 36~GHz.
A weaker
extended radio-continuum source, presumably an evolved H{\sc ii} region,  is associated with the PDR located in south-east corner of the 
image (Fig.~\ref{map}) and traced by the strong 8.0~$\mu$m emission. Its relation to the jet is unknown at present. Two sites of weak 36-GHz masers
(at C and B) are offset by over 30~arcsec, and may be unrelated.



\subsubsection{G337.92$-$0.46}

The class~I maser emission is spread over a large area exceeding the nominal field of view of our experiment. The {\em Spitzer} image reveals
a long curved 8.0-$\mu$m filament spanning across the whole area shown in Fig.~\ref{map} and a pronounced EGO in the south-eastern corner 
\citep[likely outflow candidate according to][]{cyg08}. The filament contains a strong infrared source at the northern edge of the field which is
associated with an H{\sc ii} region \citep[Fig.~\protect\ref{map};][]{wal98}. Some weaker and probably partially 
resolved 18-cm emission has also been detected at other parts of the 
filament (MAGMO data, see Fig.~\ref{map}). \citet{per09} identified a large dark cloud north-east of the filament. Perhaps, this filament traces
a PDR at the interface of the molecular cloud.
The class~I masers at A and E (note, the 44-GHz maser at~A is quite strong exceeding 300~Jy) are located
near the edge of the EGO and are presumably associated with it. The masers at B$-$D are in the same general area, but clearly separated from the EGO.
The remaining class~I masers (at F$-$I) are all located outside the 50 per cent sensitivity region at 44~GHz and trace the western edge of the 8.0-$\mu$m 
filament being several arcsec further west. A 6.7-GHz maser is located close to~G and associated with a weak 4.5-$\mu$m source 
\citep[Fig.~\ref{map};][]{wal98,cas09,cas11a} which is not listed as an EGO by \citet{cyg08}.
\citet{cas98} reported two OH masers in the area, the weaker one is associated with the 6.7-GHz maser and the stronger one is located near the EGO and the class~I 
methanol masers at~E. Therefore, the class~I masers clustered around the EGO (locations A$-$E) may correspond to a separate star forming region which
happens to have no 6.7-GHz methanol maser.


\subsubsection{G338.92$+$0.55}

This is a complex source with class~I maser emission scattered over a large area about 1~arcmin in extent and two 6.7-GHz methanol masers, 
338.920$+$0.550 in the south and 338.925$+$0.557 in the north (Fig.~\ref{map}). The southern 6.7-GHz maser has an order
of magnitude higher flux density and is associated with a strong infrared source \citep{cas09,cas11a}. The northern 6.7-GHz maser has an
associated OH maser \citep{cas98}. \citet{cyg08} identified two
EGOs in the region (both are classified as possible outflow candidates), one extending to the south from the position of the southern 6.7-GHz maser
(338.920$+$0.550) and the other a few seconds of arc south of the northern 6.7-GHz maser (338.925$+$0.557). Note, there is also a 4.5-$\mu$m source
exactly at the position of the northern 6.7-GHz maser which is not classified as an EGO by \citet{cyg08}.
The class~I methanol masers at
B$-$D are situated near the southern EGO and the infrared source associated with the 6.7-GHz maser. The northern EGO appears to be related to the
class~I masers at H and the location~I is only a few seconds of arc north-west of the norhtern 6.7-GHz maser and the 4.5-$\mu$m source. The other
class~I maser sites do not show obvious associations and those at G and J could even be unrelated sources. \citet{per09} identified two 
rather extended dark clouds to the east of the region. The class~I masers at G and J, and potentially also masers at E and F, could be 
associated with the interface of the corresponding molecular cloud. It is worth noting that both 36 and 44~GHz masers are quite strong at a number of locations 
in this source (the peak flux density at G is about 400 and 200~Jy at 44 and 36~GHz, respectively). Despite the strength, the 44-GHz spectral profile of~G is 
fitted very well with a single
Gaussian (see Table~\ref{fitresults}) with only marginal systematic error in the tail (the actual profile seems 
a bit narrower than the fit). In contrast, it was found to be difficult to fit the slightly weaker double-peaked 44-GHz profile of~A within the noise. The actual
profile is a bit narrower than the fit and there is a couple of Jy systematic error in the peak fluxes.
Another noteworthy feature of this source is a large velocity spread (about 20~\ks) of both 36- and 44-GHz masers at F (blue-shifted with respect to the middle 
of the 6.7-GHz maser velocity range by about 15~\ks). The water maser associated with the nothern 6.7-GHz methanol maser 
(located about 12~arcsec away from F) has a high velocity emission extending to $-$86~\kss \citep{bre10a}.
No radio continuum emission has been detected towards this source \citep[MAGMO rms is 2~mJy; 3- and 6-cm rms is about 0.2~mJy according to][]{urq07}.



\subsubsection{G339.88$-$1.26}

This region is outside the GLIMPSE coverage, but was studied in the near infrared by \citet{wal99}. It harbours one of the strongest 
known 6.7-GHz masers \citep[$>$1000~Jy; see e.g.][]{cas09,cas11a} which is located roughly on the line
joining two sites of the class~I maser emission separated by about 40~arcsec (Fig.~\ref{map}). The direction of this line
roughly corresponds to the direction of the dark lane at 2~$\mu$m pointed out by \citet{wal99}.
The 6.7-GHz maser is seen projected onto an
H{\sc ii} region \citep{ell96} associated with the near-infrared source \citep{wal99}, and has nearby OH 
and H$_2$O masers \citep{cas98,bre10a}. Contours in Fig.~\ref{map} represent the 3-cm continuum measurement of \citet{ell96}.


\subsubsection{G341.19$-$0.23}

The spatial distribution of class~I maser emission is very compact in this source (only a few seconds of arc across), but it has 
an extremely wide velocity spread at 36~GHz, with almost continuous emission covering over 50~\kss (Fig.~\ref{map}). This spectrum looks ordinary 
at 44~GHz, although the extreme 36-GHz velocity components are outside the velocity coverage at 44~GHz. The class~I maser emission forms an arc-like structure 
with the lower velocities tending to be at the southern end. It is seen projected onto an infrared dark cloud \citep[see the catalogue of][]{per09} 
and is located approximately 0\farcm5 from the edge of a bubble seen in IRAC images \citep[labelled S24 in][]{chu06}.
No 6.7-GHz maser was detected in the MMB project at the location of class~I masers \citep{cas11a}. Observations of the OH maser towards another source, G341.22$-$0.21, by \citet{cas98} included
this source in the field of view implying no hydroxyl maser emission stronger than 0.1~Jy (1$\sigma$). However, this source is at the edge of the field of view for the 22-GHz 
measurement of \citet{bre10a}, so the only reported limit on the water maser emission in this region is a relatively weak one imposed by the HOPS survey \citep[complete at $\sim$10~Jy;][]{wal11}.
The MAGMO data yielded no 18-cm continuum within the field of view shown
in Fig.~\ref{map} with rms noise level of 2~mJy. This source appears to be a good example of a very young star-forming region prior to the phase when the 6.7-GHz
methanol maser is switched on. Perhaps an outflow has just turned on in this source and has not yet had time to propagate into the surrounding
cloud. 


\subsubsection{G341.22$-$0.21}

Most of the class~I maser emission is scattered around the strongest infrared source projected onto the edge of a small bubble about 10~arcsec in
diameter which is traced by the 8.0-$\mu$m emission (shown in red in Fig.~\ref{map}). This complex is located 0\farcm5 north from the much larger bubble 
discussed in relation to G341.19$-$0.23 \citep[labelled S24 in][]{chu06}. No 18-cm radio continuum emission has been detected in the MAGMO project 
(rms noise level is 2~mJy) within the shown boundaries of the image, including the area inside the small bubble. 
The MMB project detected a strong ($>$100~Jy) 6.7-GHz methanol maser near the peak of the dominant infrared source with the emission spanning the velocity range
from $-$50 to $-$35~\kss \citep{cas11a}. There are both hydroxyl \citep{cas98} and water \citep{bre10a} masers near this location with similar velocities.


\subsubsection{G343.12$-$0.06}

Also known as IRAS~16547-4247, this region comprises a triple cm-wavelength continuum source interpreted as a thermal jet with two internal working 
surfaces and a jet-driven molecular outflow \citep{gar03,bro03}. The 2.12-$\mu$m molecular hydrogen emission mapped by \citet{bro03} is shown
in Fig.~\ref{map} by magenta contours. The triple continuum source is shown by red contours \citep[12-mm data from][]{vor06}.  We did not have a convincing 
detection of this source in the MAGMO data given the 1$\sigma$ rms noise level of about 2~mJy, which is comparable to the peak flux density expected from
the measurement of \citet{gar03}.  
No 6.7-GHz methanol maser has been detected towards this source in the MMB survey \citep{cas11a}, but an OH maser coincides with the continuum core \citep{cas98}.

This is a well known site of class~I methanol maser emission, which can perhaps be called an archetypal class~I maser source with outflow association \citep{vor06}. 
In this paper we present the first interferometric observation of this region at 36 and 44~GHz. The previous study of \citet{vor06} mapped the region at 84 and 95-GHz, which
are high excitation energy counterparts of the 36- and 44-GHz methanol transitions, respectively, and are expected to show similar observational properties. 
\citet{vor06} found these common class~I methanol masers to be scattered over a large area tracing well the southern part of the molecular outflow
(locations A-C in Fig.~\ref{map}). A number of rare masers found in this source were shown to be confined to a single location~B  associated with the 
strongest knot of the 2.12-$\mu$m molecular hydrogen emission (see Fig.~\ref{map}). In general, there is a good agreement between this study and the results 
of \citet{vor06} for the common locations A to~F despite different transitions, as expected.  However, there is a mismatch between radial 
velocities of corresponding Gaussian components of up to 0.4~\kss (but typically about 0.1~\ks), which notably exceeds the formal fit uncertainty in most cases. At the same time, the position
fit usually agrees well even within the formal 3$\sigma$ fit uncertainty (there are additional systematics affecting the absolute position which strictly speaking should be taken into account 
to compare separate measurements at different frequencies). Although in some cases the velocity mismatch could be explained by the rest frequency uncertainty (1$\sigma$ uncertainty is 
0.04 and 0.03~\kss for 84 and 95-GHz transitions, respectively), it most likely suggests that a finer structure with different trends on the excitation energy is present. A higher
spatial resolution observation is required to confirm this. Having a wider field of view and observing transitions which are easier to excite, the present study has 
revealed class~I maser emission
in a few additional locations in the region, labelled G to J, as well as additional weaker components for the known locations A to~F.

The new class~I masers at~H and~J are located only a few seconds of arc from the northern part of the molecular outflow (Fig.~\ref{map}). Sensitive observations along the whole
span of the outflow \citep[about 2 arcmin;][]{bro03} are likely to reveal further sites of the class~I masers in this source. The infrared image reveals an EGO 
\citep[called likely outflow candidate by][]{cyg08} covering the area around
the southern lobe of the triple continuum source and towards the central source (Fig.~\ref{map}). There is also a region with an excess of 4.5-$\mu$m emission between the
central source and the northern lobe which is not very pronounced due to the presence of a strong nearby infrared source. The area of shocked gas traced by the 4.5-$\mu$m 
emission is largely devoid of class~I masers (location D is the exception).  


\subsubsection{G344.23$-$0.57}

The class~I masers in this source are scattered over a large area exceeding 1\farcm5 across. The distribution hints at two linear structures formed by class~I maser emission, which 
intersect near~H (Fig.~\ref{map}). The north-east
to south-west line comprises class~I masers at~B, C, H, I and L. Masers at A, D and G are loosely associated with the same direction without a perfect alignment. The other
line lies approximately at 70\degr{ }angle connecting class~I masers at E, F, H and K. The masers at~G may also be loosely associated with this direction. The morphology
suggests that either one or two outflows originating near G, H or~I are likely to be present in this source. The 6.7-GHz maser is located near~I  and
has a velocity spread exceeding 20~\kss\citep{cas95,wal98, cas11a}. Interestingly, the spectral profiles for both 36- and 44-GHz masers at~I show a 
pronounced blue wing covering the whole velocity range spanned by class~I masers in this source which exceeds 10~\ks. The same general area hosts
the mainline OH maser emission which tends to be blue-shifted with respect to the class~I maser emission \citep{cas98,arg00} and a weak (a few~Jy) water maser 
spanning a large velocity range over 60~\kss \citep{bre10a}. \citet{bre10a} also reported an additional weak ($<$1~Jy) H$_2$O maser located between E and~K, on the 
line traced by class~I masers. It is worth noting that \citet{cyg08} identified 
two EGOs in this source, one likely outflow candidate is in this exact region, near class~I masers at G and~H, and the other EGO is listed in the possible outflow category 
and located at~B. Located near the northern edge of the image shown in Fig.~\ref{map}, the class~I masers at~J are not part of the linear structures described above and may be an 
unrelated site. No 18-cm radio continuum emission has been detected towards this source (MAGMO project, rms is 3~mJy). Note, the 44-GHz profile of~C is fitted with a systematic
residual in the wings of the main peak, comparable to the noise level. The fit also does not represent well the dip between the main peak and the blue-shifted satellite, overestimating 
the flux density at these velocities by up to 0.5~Jy. Although fitted within the noise, the Gaussian decomposition for the 36-GHz profile corresponding to the 
location~I (Table~\ref{fitresults}) is 
unlikely to represent well the blue wing due to artefacts caused by confusion with~G.



\subsubsection{G345.00$-$0.22}

The region has two 6.7-GHz masers, 345.003$-$0.223 and 345.003$-$0.224, which are separated by only 3~arcsec and do not overlap in velocity \citep{cas09,cas10}. 
Both masers are quite strong, with the western maser (345.003$-$0.223) being the strongest, exceeding 200~Jy \citep[see also][]{phi98}.
The class~I methanol masers are scattered around them and cannot be unambiguously attributed to a particular 6.7-GHz maser (see Fig.~\ref{map}), neither using
the offset in position, nor the velocity. 
The {\em Spitzer} overlay is dominated by two strong infrared sources (note, the 8.0-$\mu$m image suggests the presence of a third source in this cluster), the stronger
of which is associated with the 6.7-GHz masers (Fig.~\ref{map}).
The infrared sources are enshrouded by diffuse 4.5-$\mu$m emission presented by \citet{cyg08} as two EGOs, both in the possible outflow category.
The class~I masers at~C, and possibly the strongest site~A as well, may be associated with these EGOs. The masers at~B are likely to be associated with an elongated 
filamentary dark cloud which is located immediately to the east of the infrared sources \citep{per09}. The remaining sites D and~E have unclear associations. However,
it is worth mentioning that they are located in the region with the overall darker 8.0-$\mu$m background although not formally listed as a dark cloud.
There is no 18-cm measurement for this source in the MAGMO data, but \citet{phi98} reports the detection of the 3-cm continuum with the peak flux density of about 
0.18~Jy associated with the weaker eastern 6.7-GHz maser (345.003$-$0.224). The same site also has an OH maser \citep{cas98,arg00} and a nearby water 
maser  \citep{bre10a} detected. The latter exhibited highly variable high-velocity features with about 100~\kss velocity span \citep{bre10a} suggesting outflow activity.


\subsubsection{G345.01$+$1.79}

This source is renowned for a large number of detected methanol masers of both classes along with thermal molecular emission of other species
\citep[see][and references therein]{sal06, ell12, val98}. There are two 6.7-GHz masers in the region separated by 19~arcsec, 
345.010$+$1.792 in the south and 345.012$+$1.797 in the north \citep[see Fig.~\protect\ref{map};][]{cas09,cas10}. The southern 6.7-GHz maser is associated
with an ultra-compact H{\sc ii} region (Fig.~\ref{map}), OH maser \citep{cas98}, and is quite strong \citep[$>$200~Jy;][]{cas10}. 
It is this site which harbours the rare class~II masers \citep{cra01}. This 6.7-GHz maser has largely more negative velocities than  
the class~I maser emission in the region with the notable exception of the 44-GHz maser at~F peaking at $-$17.34~\ks. The class~I masers have velocities
similar to that of the thermal gas \citep[compare Fig.~\protect\ref{map} with][]{cra01, sal06}.
The {\em Spitzer}
mid-infrared overlay shown in Fig.~\ref{map} is dominated by two infrared sources associated with each of the 6.7-GHz masers and an extended
feature to the west of the southern 6.7-GHz maser which is more prominent at 8.0$\mu$m. The latter may be part of a shell-like PDR structure with diffuse 
edges. The MAGMO data indicate the presence of resolved 18-cm continuum emission inside the shell (Fig.~\ref{map}), which corroborates the idea that
the 8.0-$\mu$m feature corresponds to a PDR around an old H{\sc ii} region. The two infrared sources associated with the 6.7-GHz masers are both seen 
embedded in extended 4.5-$\mu$m emission sources, neither of which qualified as an EGO according to \citet{cyg08} perhaps due to rather regular
shape and confusion effects. Contrary to the conclusions of \citet{val98}, most of the class~I maser emission is found offset from the position of the 
southern 6.7-GHz maser. The exception is the 44-GHz maser at~F which can be associated with either the southern possible EGO 
or the PDR emission. The class~I masers at A and~B are located in the vicinity of the northern 6.7-GHz maser and the 4.5-$\mu$m emission source. 
The maser site~C, which is the strongest in this source with peak flux density of almost 100~Jy at 44-GHz (note, the fit presented in
Table~\ref{fitresults} systematically underestimates the red wing of this profile for a number of consecutive spectral channels
at the level of 1$\sigma$), and~E might both be associated
with the diffuse 8.0-$\mu$m PDR emission (the extension to E is too weak to see in Fig.~\ref{map}). The remaining class~I masers at D and~G are located
in the vicinity of a chain of weak deeply embedded infrared sources (too faint to see in Fig,~\ref{map}) and may be related
to the stars forming in this part of the molecular cloud. It is worth noting that \citet{bre10a} found a water maser (additional to two water masers near each of the 
6.7-GHz masers) located approximately half way between D and~G and the northern 6.7-GHz maser. This region of the image is outside the 50 per cent sensitivity area
and therefore additional class~I methanol masers may be missing.



\subsubsection{G345.42$-$0.95}

The class~I masers in this source were first detected by \citet{sly94} towards the water maser position quoted by \citet{bat80}, which had
only an arcmin accuracy. Therefore, this was essentially a serendipitous detection given the actual distribution of the class~I masers shown in Fig.~\ref{map} 
which are all located at or beyond the half-power point of the primary beam at 36~GHz. The regon has two well-separated 6.7-GHz masers,
345.424$-$0.951 in the north-east corner of Fig.~\ref{map} and 345.407$-$0.952 in the south \citep{cas09,cas10}. The latter has an associated OH
maser \citep{cas98} and an H{\sc ii} region (Fig.~\ref{map}). The contours in Fig.~\ref{map} show 3-cm data from the ATCA archive 
(observations done in 1994, on July, 30th; project code C380). The peak is about 240~mJy, but these continuum data do not have adequate $uv$-coverage to capture extended 
structures which are likely to be present in this region, in particular in the area towards the strong infrared cluster located to the west of the southern 6.7-GHz 
methanol maser and a shell-like feature seen at 8.0$\mu$m \citep[Fig.~\protect\ref{map}; this region is also known as RCW~117, see][]{rog60}.
\citet{bre10a} detected a number of water masers in the area, which included counterparts of both 6.7-GHz methanol
masers, another location about 15~arcsec to the north-east of the southern 6.7-GHz methanol maser (in the vicinity of~A and 
a compact 4.5-$\mu$m source), and a number of masers associated with the infrared cluster and the shell. 
The class~I maser emission is found in three well-separated locations. With only modest offsets, the masers at A and C are likely to be related to the southern and northern 6.7-GHz 
masers, respectively. The strongest site~B is seen projected onto the 8.0-$\mu$m shell. It is worth noting that there is also a weaker (about 20~mJy at 3-cm) 
continuum source in the vicinity of class~I masers at~C. \citet{per09} identified an extended dark cloud north-west of the infrared cluster which does not seem
to be directly related to any of the detected class~I maser sites. However, the water maser referred to as G345.406$-$0.942 by \citet{bre10a}, which, with the
caution about variability, could be the original water maser observed by \citet{bat80}, is likely to be associated with an interface of this cloud.


\subsubsection{G345.50$+$0.35}

The {\em Spitzer} overlay is dominated by a strong infrared source (Fig.~\ref{map}) which is associated with an 
OH \citep{cas98} and a strong (about 300~Jy) 6.7-GHz methanol maser \citep{cas09,cas10}. No radio continuum 
emission has been detected towards the infrared source \citep[rms is about 0.5$-$0.7~mJy;][]{gar06}. There is however a nearby
extended H{\sc ii} region located just outside the region shown in Fig.~\ref{map}. The strong infrared emission in the south-western corner
of Fig.~\ref{map} originates in the associated PDR of this H{\sc ii} region. The region has a prominent 4.5-$\mu$m filament extending
from the dominant infrared source associated with the 6.7-GHz maser towards C \citep["outflow-only" EGO according to][]{cyg08}. Extended
4.5-$\mu$m emission is also found in other parts of the field, particularly in the vicinity of the infrared source. 
There is also a hint at a second 4.5-$\mu$m filament in the direction towards E (see Fig.~\ref{map}). No class~I maser emission was found 
directly associated with either of these filaments. 
However, the masers at A and~B may be associated with the other sources of 4.5-$\mu$m emission mentioned above.
It is worth noting that the class~I masers at~B may also be related to the ionization front of the H{\sc ii} region and its associated PDR. 
\citet{per09} identified a large dark cloud south-west of this target. The class~I masers at~A and~E may be associated with the interfaces of
this cloud. The weak 44-GHz maser at~D has a small angular separation from the 6.7-GHz methanol maser and is seen projected onto the dominant
infrared source. 
\citet{gre11} remark that, kinematically, a distance allocation of 10.8~kpc seemed likely. However this location would make the 6.7-GHz maser
the most luminous in the far portion of the 3~kpc arm, and evidence from H{\sc i} self-absorption (and previous literature) favoured a near
distance (close to 1~kpc, in the Carina-Sagittarius arm ). We, too, suggest that the near distance is more likely, based on the
class~I emission: sites~C and~E and the 6.7-GHz maser appear to be physically associated (linked by 4.5-$\mu$m filaments) and
their separation of 0.2~pc (at the near distance) seems far more likely than the improbably large value of 2~pc that would be implied
by a distance exceeding 10~kpc.



\subsubsection{G348.18$+$0.48}

The 44-GHz emission was first reported in this source by \citet{sly94}, who named it 348.18$-$0.49, with the sign error in latitude. 
They observed this source at the single dish position of the H$_2$O maser found by \citet{bra82}. Due to the absence of better data, 
the same position was used as the pointing centre in our observations. \citet{bra82} found this source targetting the OH maser
reported by \citet{tur70}. However, as discussed by \citet{cas83}, the latter might have been a spurious detection, although an
absorption feature centred at $-$8~\kss (similar to velocities of class~I masers)  was definitely present at both 1665 and 1667~MHz. 
The MMB survey found no 6.7-GHz methanol maser towards this source and, therefore, 
there was no 18-cm measurement for this source in the MAGMO project. The class~I methanol masers are located near the H{\sc ii}
region RCW~120, a perfect bubble with a well defined ionization front traced by the 8.0~$\mu$m emission \citep[red stripe in Fig.~\protect\ref{map}; c.f. condensation~1 of][]{deh09}.
The maser emission at A$-$C and F seems to trace the ionization front, while other locations are situated further away, on the opposite side of a massive
140$-$250~M$_\odot$ core \citep{deh09}. As discussed by \citet{deh09}, the area is the site of ongoing induced star-formation \citep[see also][]{tho12}. 
Therefore, the latter locations
of maser emission may be caused by the second generation stars. The infrared map reveals a number of sources with the excess of
4.5-$\mu$m emission, but most of them are compact (see Fig.~\ref{map}). \citet{cyg08} found an EGO which is a possible outflow candidate
located just over a minute of arc to the south-east from the most eastern class~I maser locations A to C. 


\subsubsection{G349.09$+$0.11}

This source has two 6.7-GHz masers, 349.092$+$0.105 and 349.092$+$0.106, separated by just 2~arcsec \citep{cas09,cas10}. 
The north-western 6.7-GHz maser (349.092$+$0.106) is weaker and has an associated OH maser \citep{cas98}.
Both 6.7-GHz masers are seen projected onto an infrared source with 4.5-$\mu$m excess (see Fig.~\ref{map}) which did not 
qualify as an EGO according to \citet{cyg08}, perhaps due to its small size and regular shape. The class~I maser emission is found in two
sites within several arcseconds of the likely EGO. The weaker maser at~B may be directly associated with the edge of the 4.5-$\mu$m 
source. Due to the close separation of the 6.7-GHz masers it is not possible to attribute each class~I maser site to either particular 6.7-GHz maser.
No 18-cm continuum emission has been detected towards this source (MAGMO project; rms is 6~mJy).



\subsubsection{G351.16$+$0.70}

This source belongs to the NGC~6334 complex \citep[source~V; see, e.g.,][]{has08,rus13}. The {\em Spitzer} overlay reveals a 
nebulosity with a bipolar appearance \citep[Fig.~\protect\ref{map}; see also][]{has08} which would normally be
interpreted as an outflow given the strong 4.5-$\mu$m emission in the lobes \citep[note, the source is outside the region inspected by][]{cyg08}.
\citet{has07} argued that there are at least two high-mass YSOs near the centre of the symmetry of this outflow, and the 
western and eastern nebulae may actually be powered by different sources. It is worth noting that the 6.7-GHz maser is offset by a few
seconds of arc from the strongest infrared source \citep[Fig.~\protect\ref{map};][]{cas09,cas10}. An associated OH maser 
(offset by 1\farcs8 from the position of the 6.7-GHz methanol maser) is located closer to the strongest infrared core and has a wide velocity
range of 17~\kss \citep{cas98}. \citet{bre10a} found two water masers in the area, one near the OH and the 6.7-GHz methanol masers and the
other located a few seconds of arc offset, at the northern edge of the infrared nebulosity.
Most of the class~I masers (sites A$-$C and F) are concentrated towards the western edge of Fig.~\ref{map},
near the half-power point of the primary beam at 44-GHz. The masers at B and~C are located in the general direction of the western outflow
lobe (and some faint 4.5-$\mu$m emission is present in the immediate vicinity of B). All these sites, including A and F, are likely to be associated
with the outflow propagating in the westward direction. Note, the parameters of Gaussian components representing the peak 
of~A and its blue wing at both 36 and 44~GHz are likely to be more uncertain than the formal errors shown in Table~\ref{fitresults} due to
complexity of the spectra. There are also systematic residuals at both frequencies in the profile fits for the masers at~B 
(systematic offsets below 2$\sigma$ at the velocities corresponding to the blue wing at 44~GHz and underestimation of the
peak flux density at a 6$\sigma$ level at 36~GHz). The site~D detected only at 44~GHz is located at the edge of the eastern lobe. The 
remaining site~E, also detected only at 44-GHz, is located several seconds of arc to the south 
of the centre of the bipolar structure. The infrared maps \citep[Fig.~\protect\ref{map} and, e.g.,][]{has08} suggest somewhat higher
levels of extinction towards the south, south-west and north-west of the outflow. The class~I masers appear to reside in the 
corresponding molecular clouds.
No 18-cm continuum emission has been detected within the boundaries of the field shown in Fig.~\ref{map} in the MAGMO project 
(with a rather high rms of 39~mJy due to confusing sources). However, the whole region is known to have diffuse radio-continuum
emission \citep[see, e.g.,][]{mun07} which has been either resolved or otherwise not imaged adequately with the $uv$-coverage of the MAGMO
experiment tailored for the maser observations.



\subsubsection{G351.24$+$0.67}

This is another source from the NGC~6334 complex \citep[source~IV, which is also known as source~A; see, e.g.,][]{has08,mun07}. The class~I methanol
masers are clustered around the 6.7-GHz maser, which does not appear to be associated with any mid-infrared source
\citep[Fig.~\protect\ref{map};][]{cas10}. 
The class~II maser site 351.242$+$0.670 (or 351.243$+$0.671) is discussed extensively by \citet{cas08}. It is notable for its strong water
maser which is dominated by a blue-shifted outflow extending more than 100~\kss from the systemic velocity. \citet{cas08} noted
the non-detection of OH, with upper limit 0.4~Jy.
\citet{bre10a} summarise additional data on nearby water masers,  one associated 
with the infrared source located to the south of the 6.7-GHz maser, and another associated with an extended H{\sc ii} region located 
in the eastern part of Fig.~\ref{map}. It is worth noting that the class~I masers at~C are seen projected onto the edge of this H{\sc ii} region
and may be associated with the ionization shock and, therefore, unrelated to the YSO traced by the 6.7-GHz maser.  Although the strongest 44-GHz
emission feature of~B was fitted within the noise (by the sum of Gaussian components shown in Table~\ref{fitresults}), the red wing is 
systematically underestimated  at the velocities greater than $-$3.2~\ks.



\subsubsection{G351.42$+$0.65}

This is perhaps one of the most famous class~II masers, also known as NGC~6334F, with the 6.7-GHz flux density in excess of 3000~Jy
and a number of rare class~II masers detected \citep[see, e.g.,][]{ell96, cas97, cra01, cra04}. It belongs to the southern source (labelled~I)
of the twin cores I/I(N) in the NGC~6334 complex \citep[see, e.g.,][]{mcc00}. The northern core I(N), which is renowned for the class~I maser emission,
is almost 2~arcmin offset from this site. Described in detail elsewhere \citep[see, e.g.,][]{kog98, bro09}, the northern core has not been studied in 
our survey. The MMB project presented the 6.7-GHz masers in this source as two distinct sites, 351.417$+$0.645 and 351.417$+$0.646, only
a few arcseconds apart \citep[Fig.~\protect\ref{map};][]{cas09,cas10}. The south-eastern maser (351.417$+$0.645), which is 
the stronger of the two, is seen projected onto a cometary H{\sc ii} region associated with the dominant infrared source on the
{\em Spitzer} overlay (Fig.~\ref{map}). It also has associated OH masers in the ground state \citep{cas98} and excited state \citep{cas11b}.
The water maser reported by \citet{bre10a} is located 
closer to the north-western 6.7-GHz maser (351.417$+$0.646) and shows a large velocity spread of components exceeding 100~\ks.
The class~I masers at~A and the broad-line 36-GHz emission at~E, which likely corresponds to
a partially resolved quasi-thermal source, are located in the vicinity of the strongest infrared source. Due to their spatial position, it seems unlikely that
the masers at~A are caused directly by expansion of the H{\sc ii} region, however the emission at~E may have such an 
origin (although shocks related to the outflow traced by the water maser seem more likely). The class~I
masers at~C are located near a weaker infrared source about 20~arcsec north-east from the 6.7-GHz masers. There is also faint extended 4.5-$\mu$m
emission in the area \citep[note, this source is outside the region inspected for EGOs by][]{cyg08}. The two strongest class~I maser sites, B and D, are located
outside the half-power point of the 36-GHz primary beam and, therefore, potentially have a high flux density uncertainty. Note, the sum of the
Gaussian components listed in Table~\ref{fitresults} for the site~B at 36~GHz systematically underestimates both peaks at the level of 2$-$3$\sigma$.
The contours in Fig.~\ref{map} represent the 3-cm  continuum measurement of \citet{ell96}. 

It is worth noting that this source was observed at 44~GHz with the Very Large Array (VLA)  by \citet{gom10} about 1.5 years prior to our observations. 
Although their data were not analysed to the same level of detail (in particular, no spectra were published), the comparison of their map with the map 
shown in Fig.~\ref{map} reveals that the 44-GHz masers at~A and~C  are found in both datasets. The strongest masers at~B and~D were not registered by \citet{gom10}, perhaps
due to large offsets from the pointing centre. Instead, \citet{gom10} reported three additional 44-GHz maser sites, the two most nothern locations as well as the most 
southern site in their Fig.~1. The latter seems to be present in our data as a 0.8-Jy maser \citep[which is in agreement with the results of][]{gom10}, but it did not qualify as
a detection due to confusion with dynamic range artefacts. However, there is no indication of any kind in our data that the 44-GHz emission could be present at the location
of the two most northern maser sites of \citet{gom10}. The peak radial velocities and velocity ranges reported by \citet{gom10} for these masers suggests that they might
be artefacts caused by the 80-Jy maser at~B, which is located just outside the field presented by \citet{gom10}.



\subsubsection{G351.63$-$1.26}

This region has not been studied well and, in particular, there are neither {\em Spitzer} IRAC data, nor MAGMO 18-cm measurement 
available for this source. This region does not seem to be searched for the H$_2$O maser emission.
The MMB survey detected no 6.7-GHz maser \citep{cas10}, and likewise no mainline OH maser was found \citep{cas74}.
Therefore, the class~I methanol maser at 44~GHz reported by \citet{sly94} was the first maser species found in this source.
The contours shown in Fig.~\ref{map} correspond to the 3-cm data of \citet{wal98}. Reanalysing the data of \citet{wal98} extracted
from the ATCA archive we noticed deficiencies in the adopted calibration procedure which are inconsistent with the current best
practices of ATCA observing. Therefore, the fine-scale structure of the continuum shown in Fig.~\ref{map} is not to be trusted, although
the absolute position seems to be accurate to within a few arcsec. The H{\sc ii} region has also been observed by \citet{cas72} with
arcmin resolution (labelled 351.6$-$1.3). These observations suggest that the H{\sc ii} region is quite extended and most likely very
evolved. 
There are two sites of class~I maser emission separated by just over 15~arcsec, both of which seem to be well separated from the 
H{\sc ii} region.  Without a more accurate continuum measurement showing the diffuse component it is hard to argue whether the 
shocks associated with the expansion of the H{\sc ii} region are directly responsible for the masers. However, with the caution about
limited information available on the source, it is likely to represent a very evolved stage of high-mass star formation where the OH and
the 6.7-GHz methanol masers have already disappeared. 
Note, there is a hint at position shift across the blue wing of the 44-GHz profile of~B (at 2$\sigma$ level) suggesting that
an additional component is likely to be present.



\subsubsection{G351.77$-$0.54}

Also known as IRAS~17233$-$3606, this is a well studied region of high-mass star formation, which harbours multiple
distinct outflow systems \citep[e.g.,][and references therein]{zap08, leu09, leu13b}. The infrared map reveals a 
tight cluster of sources along with three remarkable 4.5-$\mu$m features: an EGO, a faint jet in the south-east to north-west
direction, and a bow-shock about 10~arcsec north of the EGO \citep[Fig.~\protect\ref{map}; see also Fig.~3 of][]{leu09}. Note,
this region is outside the area of the Galactic plane inspected for EGOs by \citet{cyg08}. The 
thinner eastern part of the 4.5-$\mu$m bow-shock feature and the thicker western part, which has not been detected in the
2.12-$\mu$m molecular hydrogen emission, trace working surfaces of two different outflows \citep[labelled OF2 and OF3 by][]{leu09}
propagating approximately in the north-south direction. The other outflow investigated by \citet{leu09} and labelled OF1 is propagating
in the similar direction to that of the 4.5-$\mu$m jet, but lies about 10~arcsec to the south. However, inspection of the 2.12-$\mu$m 
H$_2$ image of \citet{leu09} suggests that these are most likely two independent outflows. The radio continuum image at low frequencies is
dominated by a well known H{\sc ii} region (Fig.~\ref{map}) of large angular size (10~arcsec) which is associated with one of the infrared sources \citep[labelled IRAC~9 by][]{zap08}.
Note, \citet{zap08} presented much more sensitive continuum data with a higher spatial resolution than that attained in both the MAGMO 
data shown in Fig.~\ref{map} and the observations of \citet{wal98}. This fact allowed them to detect additional more compact continuum sources in the field,
an H{\sc ii} region associated with the infrared source near the southern boundary of Fig.~\ref{map} \citep[labelled IRAC~3 by][]{zap08} and
a number of weaker sources near the southern rim of the EGO, including the source previously reported by \citet{hug93}.
When considering the angular sizes and separations, it is important to emphasize that the site is likely to be very near, with a distnace of 1~kpc
adopted by \citet{zap08}.

This region also has remarkable maser properties revealed by high angular resolution observations, particularly that of hydroxyl \citep{arg00,fis05} 
and water masers \citep{for90,zap08}. The former are found to trace the OF2 outflow with a clear
separation between the blue-shifted and red-shifted components \citep{fis05,leu09}, while the latter form a ring centred approximately 4~arcsec
to the west of the presumed location of the outflow source \citep{zap08}, which seems to be related to the OF1 outflow \citep{leu09}. 
Located somewhat in the middle is the 6.7-GHz methanol maser \citep[it is quite strong, $>$200~Jy;][]{cas09,cas10}. 
Note, although the separations between positions of various  maser species is the same magnitude as the typical 
uncertainty of the absolute position measurement, it is most likely real (especially the separation between the OH and H$_2$O masers, which is 
well above 5$\sigma$). On the larger scale, all these masers are located near the southern edge of the EGO, in the same area where the 
weak continuum sources mentioned above are located \citep[Fig.~\protect\ref{map};][]{zap08,cas10}.

The class~I methanol masers are spread across a large area up to the half-power point of the primary beam (Fig.~\ref{map}). The remarkable
feature of this distribution is an approximately 10~arcsec-long elliptical arc (only a couple of arcsec along the minor axis) formed by class~I masers 
at~B and~D near the southern and western edges of
the EGO. The masers at~B are tightly clustered around the position of the 6.7-GHz maser and have a large spread of radial velocities (about 10~\ks; Fig.~\ref{map})
which seem to correspond to the blue-shifted OH emission. 
Even taking the uncertainty of the absolute position into account, these class~I masers tend to be located towards the western part of the OH outflow
(but near the central region) and the eastern part of the H$_2$O maser ring. They can be associated with any of the three outflows (OF1$-$OF3), but given the velocity
trend, at least some of these class~I masers seem to be related to the OF2 outflow. Note, the 36-GHz spectrum of~B has an abrupt dip of the flux density
near $-$6.76~\ks, which is not represented well by the sum of the fitted Gaussians listed in Table~\ref{fitresults}. A similar drop of flux density, which has also left 
a systematic residual, is present in the 44-GHz spectrum of D~near $-$2.88~\ks. The fit underestimates the flux density around this velocity by about 0.8~Jy (or 6$\sigma$).
The class~I masers at~D form the bulk of the structure discernible as an elliptical arc. There is a clear velocity gradient along the structure (4.3~\kss over 7~arcsec),
with more negative velocities being in the northern part. The northern end of the structure is just a few arcsec south of the 4.5-$\mu$m jet (Fig.~\ref{map}). The arc extends
well outside the maser zone populated with the OH and H$_2$O masers \citep[compare Fig.~\protect\ref{map} with Fig.~3 of][]{zap08}. The curvature would suggest
an association with the OF1 outflow (the arc runs across the flow direction), although association with the OF3 outflow (the arc is along the flow direction) cannot be
excluded. The class~I masers at~A are located at the northern side of the EGO (Fig.~\ref{map}) and seem to be associated with the OF1 outflow. It is worth 
noting that \citet{leu09} detected a knot of 2.12-$\mu$m H$_2$ emission in the vicinity of~A which suggests the presence of shocked gas. The profile decomposition
into Gaussian components left systematic residual at the level of 2$-$3$\sigma$ around $-$8.0~\ks, the local minimum in the spectrum at the blue side of the peak.
The other class~I masers are well offset from the EGO (C, E and H are at the east and north-east, and G and~F are in the south-west), not in a general direction
of  any of the considered outflows (Fig.~\ref{map}).



\section{Discussion}
\label{discuss}

The Gaussian components presented in Table~\ref{fitresults} are not directly suitable for detailed comparison of 36- and 44-GHz masers. The complexity
of the spectra means that the number of Gaussians used in the fit may vary from one transition to another. To circumvent this, we grouped the Gaussian
components co-located in both position and velocity within 3$\sigma$. 
For each group we calculated the peak 
velocity, the peak flux density and the effective full width at half maximum of the appropriate sum of Gaussians and propagated individual 
uncertainties of the fit to these cumulative figures. These group figures were used in the statistical analysis presented in this section. It is
worth noting that more than 85 per cent of the groups appeared to be simple and contained just one Gaussian component. For these simple groups
the parameters used for statistical analysis (e.g. peak flux density) and their uncertainties are equivalent to that of a single Gaussian component.
In total, across all 71 targets, we extracted 817 groups at 44~GHz and 740 groups at 36~GHz, with only 292 (or about 23 per cent of the total number) groups having
emission in both transitions. The matching criterion between two frequencies included the systematic shift, which is discussed in the following section,
as well as the alignment of both position and velocity (with the shift) to within 3$\sigma$.

\subsection{Rest frequency uncertainty}
\label{restfrequncert}

\begin{figure}
\includegraphics[width=1\linewidth]{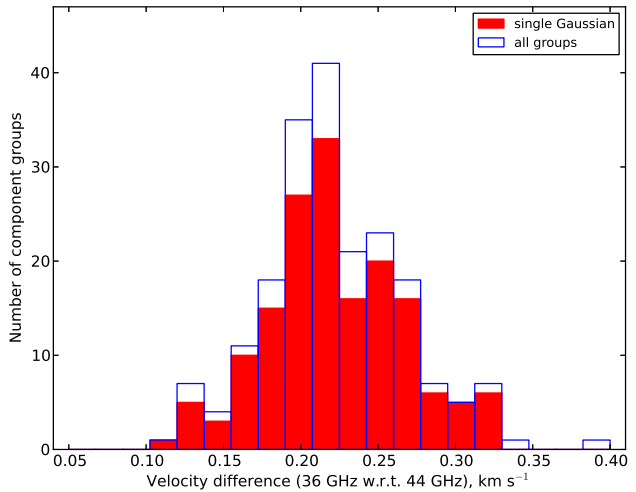}
\caption{The histogram of the velocity difference (peak V$_{LSR}$ of the 36-GHz maser minus that of the 44-GHz maser) 
distribution calculated across all matching groups of Gaussian components (blue bars) and across simple groups 
containing just one Gaussian component at each frequency (filled red bars).}
\label{restfreqplot}
\end{figure}

The velocity alignment of spectra corresponding to the two different transitions requires accurate rest frequencies. In contrast to the 
44-GHz transition, the laboratory measurement of the rest frequency for the 36-GHz transition has a large 
uncertainty equivalent to 0.24~\kss \citep{mul04}, which exceeds the spectral resolution 
of this experiment. Indeed, there appears to be a clear systematic red shift of the 36-GHz Gaussian components with respect to
the associated 44-GHz components in Table~\ref{fitresults} and Fig.~\ref{map}. We have previously used high spectral resolution maser data to refine
the rest frequencies of other methanol transitions \citep{vor06,vor11}. However, identification of the same physical component at 36 and 44~GHz
is a more difficult task due to the complexity of the spectral profiles and the lack of clear one to one correspondence 
between the fitted Gaussian components. We adopted an iterative approach progressively refining the systematic 
offset between the radial velocities corresponding to two different molecular transitions used in the match criterion. 
During each iteration we compared the velocity distributions at 36 and 44~GHz and updated the systematic offset to bring them into line.
As a starting point we used the
average velocity offset calculated over all single component groups matching in position (within 3$\sigma$) regardless of their velocity. 
Such iterations quickly converged, with the final velocity difference distribution being shown in Fig.~\ref{restfreqplot}. 
Only those
component groups which had sufficiently accurate peak velocities (an uncertainty below 0.035~\kss which is approximately half the spectral resolution) were
considered for this analysis. In particular, this cutoff excluded broad spectral features which might have some contribution from partially resolved thermal emission and,
therefore, would be very sensitive to the actual $uv$-coverage which was different at 36 and 44~GHz.
In total, 204 component groups contributed to the histogram shown in Fig.~\ref{restfreqplot}. For comparison, we also show the histogram made using a subset 
containing 163 simple groups with just one Gaussian component for each molecular transition. The results were similar in both cases corroborating the
approach to deal with complex spectral profiles represented by a number of Gaussian components. Therefore, we used the full sample of 204 groups to refine the 
rest frequency. The weighted (by the reciprocal of the square of uncertainties) mean velocity difference between 36- and 44-GHz masers was found to 
be 0.22$\pm$0.03~\ks, equivalent to 26.5$\pm$3.6~kHz for the 36-GHz transition. 
Taking into account the rest frequency uncertainty for the 44-GHz transition (it is used as a reference), this weighted mean 
corresponds to a refined rest frequency value of 36169.238$\pm$0.011~MHz. The new value agrees with the laboratory 
measurement from the catalogue of \citet{mul04} within their 1$\sigma$ uncertainty. It is worth noting that velocity alignment of different methanol transitions observed
with high spectral resolution has previously been used to constrain variations of fundamental constants \citep[for details see][and references therein]{lev11,ell11}. 
However, the lack of clear one to one correspondence between individual components and general complexity of the spatial distribution suggests that 
the 36- and 44-GHz transitions are not very well suited for this purpose.



\subsection{Separation from YSOs}
\label{sepysosect}

As mentioned earlier, all but nine of our targets have at least one class~II methanol maser at 6.7-GHz within 1~arcmin according to 
the MMB survey (see Table~\ref{srcoverview}). The 1~arcmin cutoff used throughout this study was based on the distribution of the 
measured angular offsets between
the class~I maser emission and the 6.7-GHz masers computed using a sample of sufficiently simple sources which have just one 6.7-GHz 
maser within a wider area of 2~arcmin in size (which exceeds the field of view of this experiment). We do not show this distribution here as its shape
is qualitiatively similar to the shape of the linear offset distribution discussed below. The numbers fall off quickly as the angular distance
increases. Over 70~per cent of the class~I maser emission components appear to be within 30~arcsec of the 6.7-GHz maser position
(note, the fraction is even larger if the associations are limited to 1~arcmin). There is a secondary maximum around 1\farcm5 which is entirely 
attributed to false associations from the analysis of individual sources contributing to this part of the distribution.  
The 1~arcmin cutoff separates these two populations and is conservative enough to keep all components which may be associated with the
corresponding 6.7-GHz maser. At 3.5~kpc,
which is roughly the distance where the majority of known class~I methanol masers are located (the median is 3.2~kpc for our target sample),
this cutoff corresponds to the spread of 1~pc usually quoted for the class~I methanol masers \citep[e.g.,][]{kur04}. It is
worth noting that the distribution does not change qualitatively if a minimum flux density selection is imposed on class~I maser components. 
In addition, we rarely pointed directly at the 6.7-GHz maser positions. Therefore, the limited field of view of our experiment and the 
sensitivity tapering away from the pointing centre seem unlikely to cause any significant selection effect.

\begin{figure}
\includegraphics[width=1\linewidth]{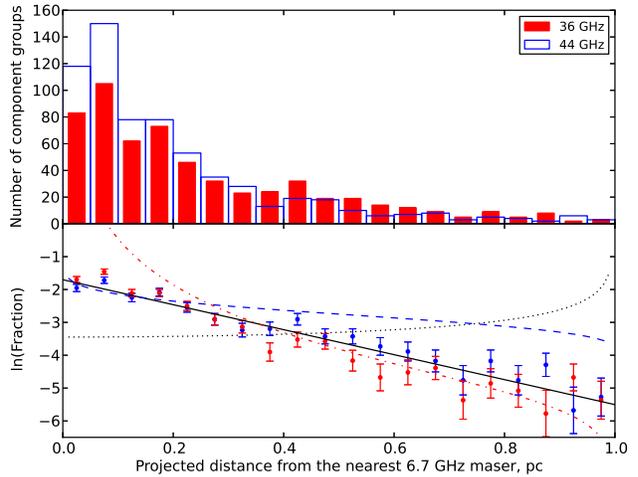}
\caption{The distribution of the projected linear separations between the class~I maser emission and the nearest 6.7-GHz methanol maser in the linear (top)
and logarithmic scale (bottom). The solid line represents the linear least-square fit to the combined 36- and 44-GHz data which corresponds to an exponential
decay in the numbers of the class~I maser emission sites with a scale of 263$\pm$15~milliparsec. The dotted curve represents the model where the masers are located on a thin shell with
the radius of 1~pc. The blue dashed curve shows the expected behaviour 
in the case of uniform spatial distribution of class~I maser sites within a sphere with the radius of 1~pc. The red dash-dotted curve corresponds to 
the inverse square law decrease in the density of the class~I maser sites with the distance from the centre within the same sphere. Note, the normalisation of the latter distribution is
arbitrary.}
\label{lindisthist}
\end{figure}

Pumped by infrared radition, the class~II methanol masers at 6.7-GHz reside in the immediate vicinity of high-mass YSOs, so their
positions can be regarded as the exact YSO location at the spatial resolution of ATCA experiments.
In contrast, the 36- and 44-GHz masers belong to class~I and are pumped by collisions with molecular hydrogen. They trace shocks,
perhaps caused by a range of phenomena. Therefore, the projected linear separation between these maser classes is indicative of
how far the shocks have propagated into the parent molecular cloud and, ultimately, is a measure of the region of influence 
for such high-mass YSOs. The resulting distribution over projected linear separations is shown in the top panel of Fig.~\ref{lindisthist}. 
The histogram was obtained using all groups of Gaussian components representing class~I maser emission which had 
a 6.7-GHz maser within 1~arcmin. If more than one such 6.7-GHz masers existed (15 targets in our list have two 6.7-GHz masers within 1~arcmin and
two targets, G335.59$-$0.29 and G333.13$-$0.44, have three and four 6.7-GHz masers, respectively), the nearest maser was deemed to be the association 
for this particular group. It is worth noting that
constraining the analysis to targets with just one 6.7-GHz maser in the vicinity yielded qualitatively the same distributon. 
The distances to the associated 6.7-GHz maser were used to convert the angular offsets into linear 
separations. We assumed the same distances as in a number of recent studies based on the MMB sample of the 6.7-GHz masers \citep[e.g.,][]{ell13},
i.e. mainly near kinematic distances unless a better estimate was available for a particular source (see \citet{gre11} for the ambiguity resolution 
details and the published list of distances).

The distribution is consistent with an exponential decay in the numbers of the class~I maser sites which is best
illustrated by the bottom panel of Fig.~\ref{lindisthist} where the fraction of the total number of the maser component groups in each transition is shown 
in the logarithmic scale. The error bars correspond to the number of component groups in each bin of the histogram. Interestingly, the distribution
is the same for the 36- and 44-GHz masers within the uncertainties imposed by the current size of the sample (note, the normalisation excluded 
component groups corresponding to the offsets over 1~pc, which are largely
considered to be false associations). The best linear fit to the combined data on both transitions corresponds to the decay scale of 263$\pm$15~mpc 
(the solid line in Fig.~\ref{lindisthist}). The second bin of the histogram manifests a single significant (more than 3$\sigma$) deviation 
from the exponential decay trend for the 36-GHz masers (the 44-GHz masers show the same tendency, but the numbers are formally within 3$\sigma$). This implies a slight 
excess of the numbers of class~I masers in the immediate vicinity of the 6.7-GHz masers but not exactly at their location.
Zooming into this distribution at finer resolution in linear separations  (the numbers allow a meaningful study at small offsets) suggests
the excess of class~I masers takes place at 0.03$-$0.1~pc. Perhaps, there is a second population responsible for this excess, for example masers associated
with outflows aligned roughly with the line of sight. Although we cannot completely exclude the possibility of bias caused by the preference towards 
the near kinematic distance during
the ambiguity resolution (distant sources are expected to have smaller angular extent and would cause a bias if assigned to the near kinematic distance), it seems 
unlikely due to the number of sources included and the steep fall off of the distribution.
It is worth noting that \citet{kur04} studied the distribution of the projected linear offsets between 
the 44-GHz methanol masers and geometric centres of compact H{\sc ii} regions. Despite smaller statistics, their distribution \citep[see Fig.~29 in][]{kur04} shows 
similar characteristics to the distribution found in our study, including the presence of a peak at small but non-zero offset.

The distribution in Fig.~\ref{lindisthist} is obtained in terms of projected linear separations and, therefore, corresponds to the actual spatial distribution integrated
along the cylindrical annuli. In order to achieve typical observed brightnesses, the modelling of class~I methanol masers usually assumes methanol column densities of
about 3$\times$10$^{16}$~cm$^{-2}$, hydrogen number densities of $10^6-10^7$~cm$^{-3}$ and an abundance of methanol as high as 10$^{-5}$ relative to 
hydrogen due to the effects of weak shocks \citep[e.g.,][]{cra92,vor06}. This range of parameters implies the size of the maser clump (or a region with effective
velocity coherence) of the order of a milliparsec. Therefore, when dealing with large-scale projected offsets of the order of 1~pc, it seems more appropriate to consider 
a three-dimensional distribution of individual masers rather than a two-dimensional distribution in the plane of the sky where each maser site corresponds to the combined
effect of the whole column of gas. Note, in the unlikely case that the latter is realised, the histogram shown in Fig.~\ref{lindisthist} would be a direct measurement of 
this two-dimensional distribution. Otherwise, one has to convert the measured projected distribution into a three-dimensional one implying some symmetry considerations
to regularise the problem. Assuming a spherically symmetrical distribution of methanol masers with a spatial density of $\rho(r)$ up to a radius 
of $R$ (i.e. $\rho(r>R)=0$), one can compute the distribution with respect to the projected separation $d$ as
\begin{equation}
\label{pdfconv}
p(d) = \frac2\pi\int\limits_d^R\frac{\rho(r)\;dr}{\strut\sqrt{r^2-d^2}}\mbox{.}
\end{equation}
To analyse Fig.~\ref{lindisthist}, we consider three special cases with analytical solutions: case one, a uniform distribution on a thin shell of radius $R$;
case two, a uniform spatial distribution ($\rho(r)=R^{-1}=const$); and case three, 
an inverse square law ($\rho(r)\sim r^{-2}$). Note, normalisation can only be achieved in the third case by assuming a cutoff at some minimal radius. 
Substituting $\rho(r)=\delta(r-R)$, where $\delta$ is the Dirac delta function, into (\ref{pdfconv}), it follows immediately
that case one implies rising $p(d)$ at $d<R$, contrary to the observed distribution (see Fig.~\ref{lindisthist}, this case is represented by the dotted curve). 
The statistics represented by Fig.~\ref{lindisthist}
are built across all sources in our sample. Therefore, we would like to emphasize that the simplistic considerations of the thin shell in case one
imply a standard radius of the shell for all sources, which is unlikely to be the case. Incorporating some distribution across a range of shell radii into equation (\ref{pdfconv})
would mimic to some extent the other two cases considered below. In addition, one could speculate that detectability of the maser depends on its position in the shell due to
different orientations with respect to the observer. However, we will not consider these options in the current study keeping the number of degrees of freedom to a 
minimum. It is worth noting that we did not find any systematic trend in the maser flux density with offset from the associated 6.7-GHz maser 
(and also no relationship between the luminosity of the class~I maser and the linear separation from the 6.7-GHz maser).

For the case two of a uniform spatial distribution, (\ref{pdfconv}) transforms to
\begin{equation}
\label{uniform}
p(d)=\frac 2\pi\ln\left(\frac{R+\sqrt{R^2-d^2}}d\right)\mbox{.}
\end{equation}
Its logarithm is shown by the blue dashed curve in the lower panel of Fig.~\ref{lindisthist} for $R=1$~pc. Although it matches the
fitted slope at distances below 0.2~pc, $\ln(p(d))$ flattens significantly at larger projected linear separations. The inverse square law, case three, results in 
\begin{equation}
\label{sqlaw}
p(d)= C \frac{\sqrt{1 - \left(d/R\right)^2}}{d^2}\mbox{,}
 \end{equation}
where $C$ is an arbitrary normalisation constant which is a function of the inner cutoff radius. The logarithm of (\ref{sqlaw}) is shown in the bottom panel of Fig.~\ref{lindisthist} by the red dash-dotted
curve (for $R=1$~pc). The resulting slope matches well the observed slope at the projected offsets beyond 0.3~pc. However, this
model is unable to explain the exponential decay across the whole range of offsets. Most likely a combination of cases two and three is realised 
with the spatial distribution being close to uniform up to 0.2$-$0.3~pc offsets from the YSO with an inverse square law fall off afterwards. This might
reflect the large-scale distribution of molecular gas around the driving YSO. In its immediate vicinity, we expect excess residual material of the parent
molecular cloud from which the YSO formed. On the other hand,  the gas might be too dense for the shock to propagate far beyond a certain distance 
without disocciating methanol or there could be a smaller chance to hit a sufficiently dense clump. 

\subsection{Velocity distribution}
\label{veldistribution}

\begin{figure}
\includegraphics[width=1\linewidth]{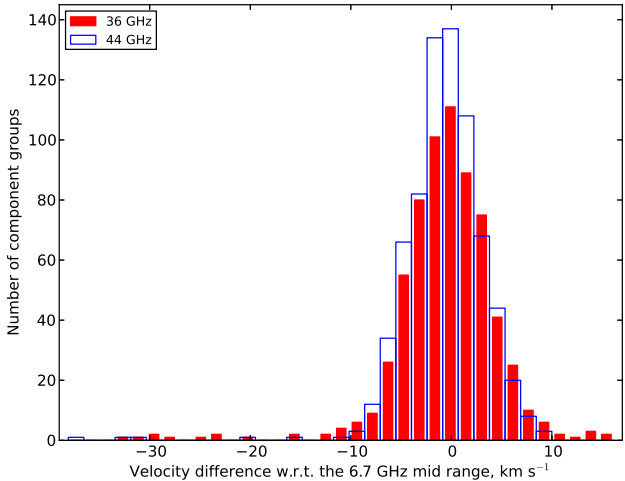}
\caption{The distribution of velocity offsets between class~I methanol maser emission and the middle of the velocity range of the associated 6.7-GHz methanol maser.}
\label{velhist}
\end{figure}

Previous studies have suggested that class~I methanol masers have radial velocities close (within a few \ks) to that of the quiescent gas \citep[e.g.][]{vor06,vor10b}. 
This is consistent with the expectation that only a small fraction of molecular gas is accelerated to shock velocities while most of the gas is only mildly 
disturbed \citep[e.g.][]{gar02a}. A proxy for the quiescent gas velocity is the middle of the velocity range of the 6.7-GHz methanol maser. It is often used for 
estimates of kinematic distance and has proven to be much better than the velocities of individual maser peaks \citep[e.g.][and the MMB project]{ell13}. Our data 
allow detailed comparison of the class I and class~II maser velocities. Using the same identification criterion between individual groups of  Gaussian components
describing class~I maser emission and the 6.7-GHz maser as in the previous section, we have constructed a histogram of the relative velocities (Fig.~\ref{velhist}).
Apart from a small number of blue-shifted high-velocity components (with offsets below $-$15~\ks), the distribution is well approximated by a Gaussian with
the same mean of $-$0.57~\kss for both 36- and 44-GHz masers (uncertainties are 0.06 and 0.07~\ks, respectively), and 
the standard deviations are 3.65$\pm$0.05~\kss and 3.32$\pm$0.07~\ks, respectively. Note, the mean for the distribution of the 36-GHz masers and its uncertainty 
have been corrected for the systematic shift due to the rest frequency uncertainty (see section~\ref{restfrequncert}). The agreement between the means 
of the 36- and 44-GHz velocity distributions is an independent check of the rest frequency correction without a need to make any assumptions about the
individual associations of the 36- and 44-GHz features, and has the advantage of a five times larger sample.
Both histograms for the 36- and 44-GHz masers peak around zero
(and are symmetric) confirming that the middle of the 6.7-GHz maser velocity interval is a good estimate for the quiescent molecular gas velocity. However, 
there is a small but statistically significant blue-shift asymmetry. Perhaps, the chance of detecting a class~I methanol maser on the observer's side of the
cloud (associated with the approaching portion of the outflow, and therefore blue-shifted) is higher. 
Assuming that high-mass star formation typically occurs towards the edges of molecular clouds or filaments, it is reasonable to expect that a shock would 
encounter different density gradients depending on the direction of propagation causing 
systematic differences in appearance (e.g. the strength or the number of maser spots distinguishable at a given resolution).




\subsection{High-velocity features}
\label{hvfeaturesect}

The velocity histogram shown in Fig.~\ref{velhist} suggests the presence of a small number of high-velocity components, which are largely blue-shifted
and detected more easily at 36~GHz. The largest velocity offsets (over 30~\ks) correspond to G309.38$-$0.13, a source described in detail by \citet{vor10a}
based on the preliminary 36- and 44-GHz data. In this paper we present an independent dataset obtained with a different receiver tuning, which allowed
us to register the 36-GHz high-velocity feature better and also to detect it at 44~GHz (see section~\ref{notesonsrc}). In addition, the new data revealed a marginal
44-GHz component corresponding to the location~I (see the map of G309.38$-$0.13 in Fig.~\ref{map}), which is several arcsec offset from the location
of other blue-shifted emission \citep[labelled~E following the notation introduced by][]{vor10a}, but has a similar radial velocity. This lateral offset suggests 
that the two locations are probably related to the opposite sidewalls of the same outflow cavity. With a velocity offset of about $-$38~\kss with respect to the middle of the 6.7-GHz 
maser velocity range, this marginal component corresponds to the most extreme non-zero bin of the histogram (Fig.~\ref{velhist}).

Including G309.38$-$0.13, there are seven sources in our sample which contain emission at either 36 or 44~GHz beyond 15~\kss (just over 4$\sigma$ of the velocity distribution
shown in Fig.~\ref{velhist}) from the middle of the 6.7-GHz maser velocity range. Only two of them, G270.26$+$0.84 and G329.07$-$0.31, show red-shifted emission 
(by just over 15~\ks) and the rest have blue-shifted emission. Among the sources with blue-shifted high-velocity components, three sources, G326.64$+$0.61, G329.18$-$0.31 and 
G333.23$-$0.06, have a marginal detection only. The blue-shifted emission in the remaining source, G338.92$+$0.55, is just over 15~\kss from the estimated systemic velocity, but 
along with the other maser emission in the immediate vicinity (see location~F in G338.92$+$0.55) spans a wide velocity range of about 20~\ks. This is also the only
source in addition to G309.38$-$0.13 where the high-velocity emission was detected in both transitions (and in this particular source the high-velocity emission is stronger at 44~GHz). 
With the exception of G329.18$-$0.31 and the new marginal component 
in G309.38$-$0.13 (location~I), which were detected at 44~GHz only, and two sources with high-velocity emission detected in both transitions, all other 
sources were detected only at 36~GHz. With the caution about low number statistics, there seems to be a tendency
for high-velocity components to be stronger (and more easily detectable) at 36~GHz. This might be related to this transition being inverted in a broader
range of conditions \citep[also suggested by its widespread distribution in the Galactic Centre, see][]{yus13}.
Note, in the case of G326.64$+$0.61, the source of the high-velocity emission is located
at a significant spatial offset from the pointing centre, and therefore corresponds to a larger primary beam attenuation factor at 44~GHz.
It is also worth noting that the widest velocity spread (over 50~\ks) was found in the 36-GHz maser in G341.19$-$0.23. There is no 6.7-GHz maser in this source
and therefore it did not contribute to the histogram in Fig.~\ref{velhist}. This source has a very compact arc-shaped distribution of the maser emission suggesting
association with a sidewall of an outflow cavity. A similar arc-shaped morphology was found at the location of the high-velocity features in G329.07$-$0.31 and
G338.92$+$0.55.

As discussed by \citet{vor10a}, the detection of high-velocity features was somewhat unexpected for methanol masers \citep[see also][]{bac90,kal10}. Although shocks
are considered essential to achieve high methanol abundance in the gas phase as the efficiency of pure gas phase reactions is quite low \citep[e.g.,][]{har95}, methanol 
is expected to survive sputtering and desorption of grain mantles only if such shocks are relatively mild \citep[i.e. shock velocities not greatly exceeding 10~\ks;][]{gar02a}.
Therefore, the velocity offsets in excess of 30~\kss led \citet{vor10a} to suggest that interaction with a moving parcel of gas must have been taking place. The exact mechanism
causing this prior acceleration is still unknown. However, the presence of emission at a wide range of velocities at nearby spatial locations (e.g. see G341.19$-$0.23 in Fig.~\ref{map})
and the morphology of G309.38$-$0.13 (the new marginal high-velocity feature at~I, which is spatially offset from the high-velocity feature at~E) suggest that this velocity gradient could
arise in a cascade of wakes accompanying a complex shock interface between an outflow and entrained ambient material.

It is worth noting that G309.38$-$0.13 is the only source where the high-velocity feature was found at a nearby location to the 6.7-GHz methanol maser (a few seconds of arc
offset). For all other sources with a 6.7-GHz maser, the angular offset is typically in the range of 10$-$20~arcsec with the exception of G326.64$+$0.61 where this offset is about 1~arcmin 
(and so the high-velocity feature may actually be unrelated to the YSO traced by the 6.7-GHz maser in this source). These sources have smaller relative velocities of the high-velocity
features than in G309.38$-$0.13 and may either contain more inclined outflows or outflows unrelated to the YSO traced by the 6.7-GHz maser.





\subsection{Evolutionary stages}
\label{evolstagesect}

\begin{figure}
\includegraphics[width=1\linewidth]{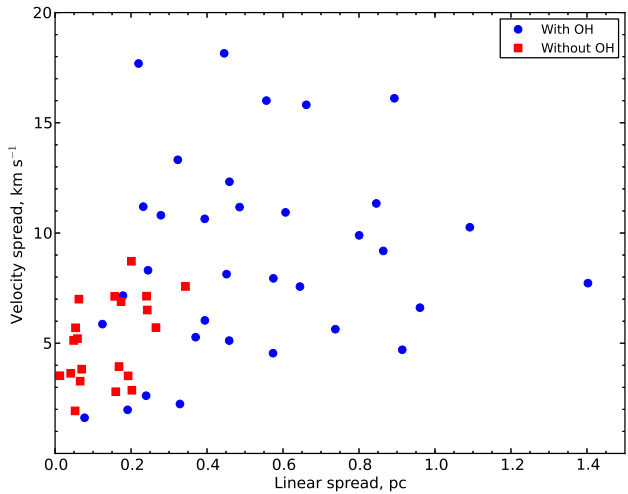}
\caption{The linear spatial spread of class~I maser emission (i.e. an estimate of the source size) and corresponding radial velocity spread for sources with (blue circles) and without (red squares) 
associated mainline hydroxyl maser.  
High-velocity features (offset by more than 15~\kss from the middle of the velocity range of the associated 6.7-GHz maser, see section~\protect\ref{hvfeaturesect}) were excluded from
the analysis as were four sources with ambiguous OH association.}
\label{scatterstats}
\end{figure}

The question whether different maser species trace 
distinct evolutionary stages has recently gained much attention \citep[see, e.g.,][and references therein]{bre10b}. However, the place of 
class~I masers in this picture is poorly understood. Initially, these masers were thought to appear prior to the onset of the 6.7-GHz (and other
class~II) masers \citep{ell07}. This conclusion was based on the analysis of the infrared colours of GLIMPSE point sources associated with the
6.7-GHz masers \citep{ell06}, where sources with the class~I methanol masers appeared to have redder colours, and the expectation that the
outflows, which were thought to be largely responsible for the class~I maser emission, occurred at the earliest stages of evolution. 
Focusing on other maser species, \citet{bre10b} further refined the evolutionary timeline for them, but, in the absence of new
observational data, inherited the same conclusions regarding the class~I methanol masers. 

The controversy emerged from detailed 
studies of selected sources \citep[e.g.,][]{vor06,vor10b} and notable overlap between the class~I methanol and OH masers \citep[e.g.,][]{vor12}, which are
usually considered a signature of somewhat evolved stages \citep[e.g.,][]{for89}. \citet{vor10b} also suggested that the class~I masers may be caused by other phenomena, 
which are not limited to the outflow scenario, e.g. by expanding H{\sc ii} regions (this conclusion is further reinforced by the present study). This finding undermines another
implicit assumption of the maser-based evolutionary sequence that each maser species occur only once during the course of evolution. Therefore, there could be
both populations of young and evolved sources which have class~I masers. 

High angular resolution data allow us to study the spatial spread of class~I maser emission, which could be another metric indicative of the evolutionary stage. The YSOs
are likely to expand their zone of influence with age as shocks have time to propagate further into the parent molecular cloud. Therefore, one might expect a larger
spatial spread of the class~I maser emission for more evolved sources, particularly those where the YSO is associated with an OH maser. The main difficulty arises again
from the complex morphology as multiple YSOs and unrelated class~I maser emission may be present in the same field.
We used the same approach as in sections~\ref{sepysosect} and~\ref{veldistribution} to identify a group of Gaussian components aligned in position and velocity
within 3$\sigma$ with a particular 6.7-GHz maser. Then, we estimated the largest linear separation from the 6.7-GHz maser (calculated as in section~\ref{sepysosect})
and the velocity spread across all groups attributed to the same 6.7-GHz maser. These linear and velocity spreads are shown in Fig.~\ref{scatterstats} with different
symbols depending on the OH association status of the corresponding 6.7-GHz maser. In total, the sample includes 37 6.7-GHz masers with OH association and
20 without (we did not have OH data available for the remaining sources). We have also explicitly excluded from the analysis four sources with potentially 
ambigous associations. Three of them, G345.00$-$0.22, G328.81$+$0.63 and G351.42$+$0.65, have very close 6.7-GHz masers, which have
different OH association cases. The fourth source, G337.92$-$0.46, has an OH maser in the field which is not associated with any 6.7-GHz maser and may
represent a separate YSO at a later stage of evolution. The high-velocity features (defined as offset by more than 15~\kss from the middle of the velocity range of the 
associated 6.7-GHz maser) were excluded from the velocity spread calculation. Fig.~\ref{scatterstats} suggests that when the YSO has an associated hydroxyl maser
\citep[and therefore expected to be more evolved, see][]{for89}, the corresponding class~I masers indeed have larger spread of emission both spatially and in the velocity domain. 
The latter probably reflects the increase of the velocity dispersion in the zone of influence as it expands with age. The difference in spatial spreads of sources with and without 
associated OH masers also stands out in the distribution of linear separations of individual groups of Gaussian components (i.e. if these two populations of sources are shown 
separately in Fig.~\ref{lindisthist}). However, the analysis described in this section allows us to avoid potential biases due to different number of maser sites (i.e. different number
of groups of components contributing to the distribution) which might be present in these two cases.
It is worth noting that we found no relationship between either of these spread figures and the luminosities of either the class~II or the class~I masers (both the peak and the mean
flux densities calculated across all corresponding groups of the Gaussian components were also inspected).

For younger sources, it still remains unclear whether there is a population of class~I methanol masers pre-dating the stage with class~II maser acitivity
(see \citet{vor10b} and \citet{che11} for detailed discussion of these issues). A positive identification of class~I maser emission with a dark cloud, especially
if it is devoid of other masers, would be a good example that such a population exists. It is worth noting that \citet{bay12} investigated association of the whole
population of class~I methanol masers with dark clouds using published low spatial resolution data. Despite the high
association rate (over 70\%) reported by \citet{bay12}, this result only means that such an association cannot be excluded, a stronger 
conclusion is precluded by the complexity of high-mass star forming regions and a weak association tolerance of 2~arcmin in their study. Our data provide the means
to search for associations at the arcsec scale. In total, among 64 sources in our sample where comparison data exist,
we found some class~I emission to be associated
with a dark cloud identified by \citet{per09} in 19 sources ($\sim$30 per cent). 

\subsection{Class~I masers without nearby 6.7-GHz masers}
\label{no6src}

The current knowledge of sources which have class~I maser emission but no 6.7-GHz (or other class~II) masers is quite limited due to 
selection biases in most surveys conducted to date. These surveys \citep[e.g.,][]{sly94} largely targeted known 6.7-GHz masers directly or  
regions of high-mass star formation selected by other means (e.g. by presence of other maser species), where the 6.7-GHz masers are also often found. 
More recent searches with different selection
criteria \citep[e.g.,][]{che11,vor12,gan13} and, especially, an untargeted survey described by \citet{jor13} have significantly increased the number of
sources where class~I maser emission is known to be present in the absence of class~II masers.
The sources without 6.7-GHz masers are likely to correspond to both extremes of the evolutionary stage spectrum (i.e. both the sources too
young and too old to have a 6.7-GHz maser), and, therefore, are particularly important 
for understanding what evolutionary stages the class~I methanol masers can correspond to. 

The sample of sources investigated in this work includes nine targets which do not have a 6.7-GHz maser within
1~arcmin (see Table~\ref{srcoverview}). Three such sources, G305.37$+$0.21, G343.12$-$0.06 and G348.18$+$0.48, show signatures
of an evolved stage of star formation (see section~\ref{notesonsrc}). In the case of G348.18$+$0.48 there is also evidence of triggered
star formation. Therefore, it is hard to attribute this source to a particular evolutionary stage as some of the class~I maser emission may 
correspond to a new generation of stars. Having an OH maser \citep{cas98} and quite large spread class~I maser emission,  
G316.76$-$0.01 most likely also belongs to the same category of evolved sources, although it is not entirely clear. 
Associated with a quite extended H{\sc ii} region, G351.63$-$1.26 is probably the most evolved 
source in our sample which already lacks not only the 6.7-GHz methanol maser, but also the 
OH maser (see section~\ref{notesonsrc}).

Among the remaining sources, at least three, G326.66$+$0.57, G328.21$-$0.59 and G341.19$-$0.23, clearly fit the profile of young 
sources: they all have class~I masers embedded in a dark cloud and no other maser species detected (see section~\ref{notesonsrc}). 
Note, the latter source is also remarkable for its high velocity spread of 36-GHz maser emission and a very compact spatial 
distribution (see section~\ref{notesonsrc}).
The class~I masers in G333.59$-$0.21 are located near the edge of a dark cloud (see section~\ref{notesonsrc}) and probably also
correspond to a very early stage of star formation. However, the morphology of this source resembles that of 
W33-Met \citep[G12.80$-$0.19; see discussion in][]{vor10b} and shows indications for triggered star-formation. Therefore, 
the case for this
source is not as clear cut as the other three examples given above. An additional complication arises because, unlike the
class~II methanol masers, the class~I masers are not exclusively associated with high-mass star formation and may be related
to low- and intermediate-mass YSOs \citep{kal10,kal13,gan13}. In particular, G326.66$+$0.57 has quite weak masers and may be
the site of low- or intermediate-mass star formation. However, both 36- and 44-GHz masers in G341.19$-$0.23 are quite strong
(peaking at about 100~Jy; see Fig.~\ref{map} and Table~\ref{fitresults}), well in excess of the typical flux densities observed in
low- and intermediate-mass star forming regions \citep{kal10,kal13}. Therefore, neither of these issues seem to undermine the general
conclusion that there is a population of the class~I methanol masers pre-dating the onset of the class~II masers at 6.7~GHz.


\subsection{Relative flux densities of 36- and 44-GHz methanol masers}

\begin{figure}
\includegraphics[width=1\linewidth]{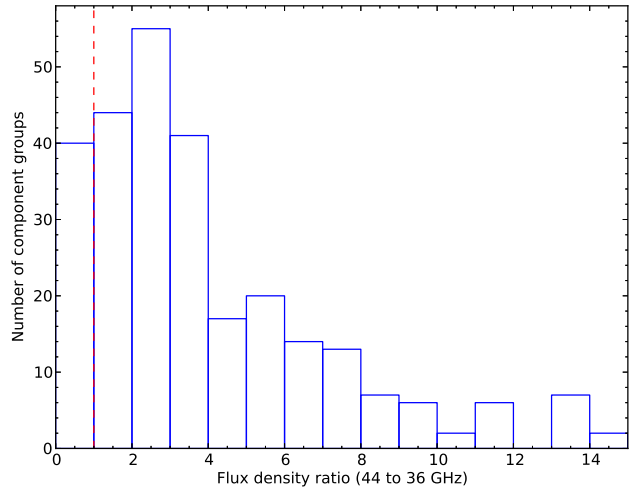}
\caption{The distribution of 44- to 36-GHz flux density ratio ($F_{44}/F_{36}$) across all 292 groups of Gaussian components with 
emission in both transitions present within 3$\sigma$ in position and velocity. The vertical dashed line separates 40 component
groups which are stronger at 36~GHz.}
\label{relfluxhist}
\end{figure}

It is clear from the analysis of individual sources (see, e.g., G333.47$-$0.16 in Fig.~\ref{map} and section~\ref{notesonsrc})
that the two studied transitions can be quite complementary, filling the gaps in morphology when tracing common structures.
In support of this idea, the basic analysis of individual Gaussian components (listed in Table~\ref{fitresults}) concluded that
only 23~per cent of the total number of component groups (collection of components matching in position and velocity within 3$\sigma$) have 
emission simultaneously in both transitions (see the beginning of section~\ref{discuss}). In this section we investigate the properties of
the component groups detected at both 36 and 44~GHz.

In the majority of cases, the emission at 44~GHz was found to be stronger than the matching 36-GHz emission.
Out of 292 component groups, only 40 appeared to be stronger at 36-GHz. However, inspection of individual spectra
suggests that in virtually all cases correct cross-matching between the 36- and 44-GHz emission components may be
questioned due to the complexity of spectral profiles which cannot be resolved with the current data. Noteworthy exceptions are
the strong ($>$200~Jy) 36-GHz emission at location~B in G335.59$-$0.29 and the $-$15~\kss spike at 
location~A in G345.01$+$1.79 (see Fig.~\ref{map}). 
The distribution of the 44- to 36-GHz flux density ratio has a long tail at $F_{44}/F_{36}>4$, which cannot be simply attributed to
incorrect cross-identification of components
in the complex spectral profiles (see Fig.~\ref{relfluxhist}). In several cases contributing to the tail, the spectra contain quite broad 36-GHz 
emission spanning the velocity range of the 44-GHz profile with maser spikes (see, e.g., the spectrum of~B in 
G270.26$+$0.84 shown in Fig.~\ref{map}). Such broad 36-GHz profiles may be related entirely to the quasi-thermal emission or
could contain a significant quasi-thermal contribution. About 62 per cent of the sample (180 out of 292 component groups) have
$F_{44}/F_{36}<4$. This fraction rises to about 94 per cent (274 out of 292 component groups) for $F_{44}/F_{36}<15$ 
(the median of this flux density ratio is 3.21).  It is worth noting that for the reciprocal 36- to 44-GHz flux density ratio,
about 78 per cent of the sample (230 out of 292 component groups) have $F_{36}/F_{44}<0.6$ (the median of this ratio is 0.31). 

We found no reliable relationship between the flux densities of the two transitions. The sources 
without a 6.7-GHz methanol maser within 1~arcmin of any class~I maser emission (see section~\ref{no6src}) do not
stand out in any way from the general population in terms of the individual flux densities or their ratios. We also did not
find any dependence between the flux density of the associated 6.7-GHz maser and the flux densities at either 36 or
44~GHz individually or their ratio.

The maser pumping is a result of a delicate balance between a number of excitation and de-excitation 
transitions, which has been shown to be significantly affected by any elongated geometry
of a maser region \citep[or, more general, the beaming factor describing 
directional variations of the optical depth; see, e.g.,][]{sob94}. The numerical models of \citet{sob05} 
suggest that the maser region elongated along the line of sight is preferred for the 44-GHz masers. 
Therefore, one might expect that the $F_{44}/F_{36}$ ratio could be a good probe of the orientation
of individual maser regions, which are almost certainly anisotropic to some extent, with respect to the observer.
As proposed by \citet{vor10a}, this might be manifested by a dependence in a large statistical ensemble between 
the $F_{44}/F_{36}$ ratio and the distance from the associated 6.7-GHz maser. This is because in an idealised 
case of a single regularly-shaped outflow interacting with the ambient gas of the parent molecular cloud, both the 
orientation of the interface regions and the projected distance from the associated 6.7-GHz maser tracing the YSO
are related to the outflow orientation. However, we did not find any such reliable dependence in our data.
Inspection of individual sources like G333.47-0.16, where the class~I methanol masers trace a regular arc (see Fig.~\ref{map}), reveals
that the 36- and 44-GHz masers are often intermixed along this structure.  This suggests that the orientation of the maser
region is more likely to be related to the turbulence (and the wakes of the shock) at small scales, rather than to the large-scale structure of
the interface.




\subsection {Width distribution}
\label{widthdistr}

\begin{figure}
\includegraphics[width=1\linewidth]{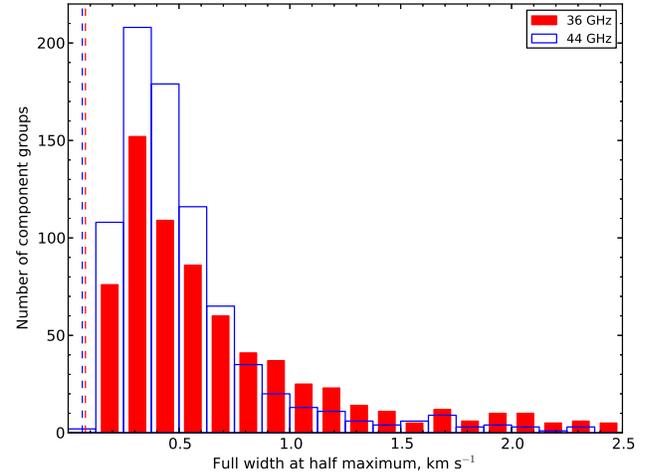}
\caption{The distribution of effective (estimated from the profile integral) full width at half maximum for the groups of Gaussian components at 36 and 44~GHz. 
The vertical dashed lines show spectral resolution at the two frequencies. }
\label{widthhist}
\end{figure}

High spectral resolution data allow us to investigate the distribution of line widths. As before, we analysed groups of Gaussian components that match within 3$\sigma$ in
position and velocity in order to avoid biases caused by the fitting of individual components. For each group, we estimated the effective (scaled to the Gaussian profile) full width at half 
maximum using the profile integral (i.e. the area under the spectral line) and the peak as
\begin{equation}
\label{effwidth}
\mathrm{FWHM}_{\mathrm{eff}} = \frac {\int\limits_{-\infty}^{+\infty}f(v)\;dv}{\strut\max f(v)}\sqrt{\frac{4\ln2}\pi}\mbox{,}
\end{equation}
where $f(v)$ is the spectral profile computed via the sum of individual Gaussian components constituting the group. Defined this way, the effective FWHM matches that
of a single Gaussian component catalogued in Table~\ref{fitresults} for simple groups containing just one Gaussian component. The resulting histogram for 36- and 44-GHz masers
is shown in Fig.~\ref{widthhist}. The individual widths range from 0.15 to 9.3~\kss and from 0.12 to 8.0~\ks, for the 36- and 44-GHz masers, respectively. In all cases, the 
narrowest
components, which are about two spectral channels wide, were found as a result of the decomposition of a complex non-Gaussian profile (see for example $-$44.887~\kss component at C in G341.22$-$0.21), rather
than as stand alone spectral features. Therefore, the extreme narrowness could be a result of the fit systematics. The large ($>$1.5~\ks) widths in Fig.~\ref{widthhist}
mainly correspond to weak wings (see for example location~I in G344.23$-$0.57), although some contribution from quasi-thermal emission cannot be excluded. The histogram
suggests that large FWHM values are more common at 36 than at 44~GHz. This is reflected by the median of 0.53 and 0.43~\kss for the 36- and 44-GHz groups, respectively.
The underlying physics is most likely the same as that which causes the high-velocity components to appear predominantly 
at 36~GHz (see section \ref{hvfeaturesect}), i.e. this transition is inverted under a broader range of conditions. 
For both transitions the distribution peaks at FWHM of 0.3~\kss 
and resembles either a Rayleigh or log-normal distribution. Regardless of the actual shape of the distribution,
the fact that the measured width follows a well defined distribution up to at least 1.5~\kss
suggests that many wide features are indeed masers similar to their narrow counterparts but blended in the spectral domain, and are not quasi-thermal features. 


\subsection{General associations}
\label{genassoc}

High angular resolution observations allow us to probe associations at arcsec scale with other phenomena commonly observed in star forming regions (see also Table~\ref{srcoverview}).
As pointed out in the introduction, the direct observational evidence of associations with outflows at arcsec level have only been demonstrated for a few
sources \citep{kur04,vor06,cyg12}, mainly due to lack of appropriate comparison datasets rather than the maser positions. The maps obtained in the current project
reveal regular structures such as arcs and lines in 17 out of 71 sources ($\sim$24~per cent). Except G305.37$+$0.21 where masers trace a circle around H{\sc ii}
region (see below), the remaining 16 cases are likely to be outflow associations with sufficiently favourable geometry which allows this association to be confirmed with 
future observations of some outflow tracer \cite[e.g. as with the H$_2$ outflow in G343.12$-$0.06;][]{vor06}.

As mentioned in 
section~\ref{evolstagesect}, we found that some class~I methanol maser emission is associated with {\em Spitzer} dark clouds identified by \citet{per09} in 19 sources out of 64 
targets for which comparison data are available ($\sim$30 per cent). In addition to dark clouds, {\em Spitzer} images reveal 4.5-$\mu$m sources
\citep[some of which qualify as EGOs and could trace shocked gas, see][]{cyg08} as well as 8.0-$\mu$m emission features caused by PAH emission, which 
includes 8.0-$\mu$m filaments and PDRs around H{\sc ii} regions.
In 61 targets which fall into the region of the Galactic plane searched for EGOs by \citet{cyg08}, we found close spatial association (within a few arcsec) 
between EGOs and some class~I
maser emission in 33 sources  ($\sim$54 per cent). Interestingly, in 24 sources out of these 61 (or 27 out of the full sample of 64 sources with GLIMPSE data available), 
some class~I masers are 
associated with 4.5-$\mu$m sources not claimed by \citet{cyg08} to be EGOs. If both EGOs and these additional 4.5-$\mu$m sources are considered, the association rate 
increases to 82~per cent (50 out of 61; note seven targets have both EGOs and additional 4.5-$\mu$m sources and are only counted once). It is worth noting 
that \citet{cyg09} reported a converse detection statistic of $\sim$90 per cent (note, the sample size was only 19 sources) in their search for 
the class~I methanol masers towards EGOs. We note that their statistic does not guide us on the question of whether most class~I sites are associated with EGOs, but this
question can be addressed with our data. Although both this work and that of \citet{cyg09} suggests that EGOs
(or 4.5-$\mu$m sources in general) proved to be good targets for class~I maser searches, there are maser sources which do not show this association. 
We also note that our sample of class~I masers does not come from a blind survey, and the fraction of all class~I masers with EGOs might be substantially lower.

Close spatial alignment between some class~I maser emission and 8.0-$\mu$m emission features
was found in 28 sources out of 65 which have {\em Spitzer} data available (43~per cent). In many cases, interpretation of this 8.0-$\mu$m emission is not 
straightforward. In some cases, it could even be caused by diffuse PAH emission in the field which is aligned by chance with the methanol masers (e.g. in G329.18$-$0.31 and G333.13$-$0.56). 
However, in some cases there are regular filaments (see G335.79$+$0.17 in Fig.~\ref{map}) and bubbles (see G341.22$-$0.21 in Fig.~\ref{map}) which may
constitute PDRs of evolved H{\sc ii} regions that we did not detect due to lack of brightness sensitivity in the radio continuum data. This is a new type of
association for the class~I methanol masers which has not been reported before. However, it is not entirely unexpected given that the association 
with expanding H{\sc ii} regions has been previously suggested \citep{vor10b,cyg11,cyg12}. 
Among 71 sources observed in the present survey, we found associations at arcsec scale between class~I maser emission and 
radio continuum emission in 17 sources ($\sim$24 per cent). The ring of masers around an infrared source and an H{\sc ii} region in G305.37$+$0.21 is currently the best available example 
of such an association. Strictly speaking, quite sensitive radio continuum observations with good $uv$-coverage \citep[similar to that of][]{hin12} are required for this analysis. 
Such data are not available at the moment for the majority of our targets. In addition, the continuum data used in our study are very inhomogeneous.
Therefore the association rate given above should be treated as the lower limit.
It is also worth noting that the continuum emission associated with the class~I masers in G337.40$-$0.40 and G343.12$-$0.06 was interpreted as an ionized jet 
and as internal working surface of the jet, respectively \citep[][see also section~\protect\ref{notesonsrc}]{guz12, gar03}, rather 
than free-free emission from an H{\sc ii} region.

 
\section{Conclusions}
\begin{enumerate}

\item We imaged 71 southern class~I methanol maser sources at arcsecond resolution at both 36 and 44~GHz 
quasi-simultaneously (no more than one day apart). Many sources show complex morphology with the maser
emission tracing ordered structures resembling lines or arcs, which may represent outflows. In addition to the outflow
scenario suggested by such geometry or confirmed by 2.12-$\mu$m molecular hydrogen observations, 
we demonstrate associations with shocks caused by expanding H{\sc ii} regions, PDRs traced by 8.0-$\mu$m
emission, and interfaces of cold molecular clouds or shocks represented by the 4.5-$\mu$m emission (EGOs). 

\item Association of the class~I methanol masers with different phenomena in star forming regions means that
these masers are unlikely to trace a unique evolutionary stage and more likely will be present multiple times during the
evolution of a high-mass star forming region. In particular, the sample of sources devoid of a 6.7-GHz
maser is likely to include both evolved sources, which are too old to have a 6.7-GHz maser and 
sources which are too young and potentially trace the earliest stages of star formation.

\item The two maser transitions at 36 and 44~GHz were found to be highly complementary and sometimes crucial for
the correct interpretation of morphology. Only 23 per cent of emission components were detected in both transitions
matching within 3$\sigma$ in both position and velocity. 

\item The high-velocity maser features are a 
phenomenon largely selected against by the existing maser surveys (including this one), which typically lack either
the velocity coverage or the spectral resolution (or both). With a caution about low number statistics, the high-velocity 
features have a tendency to be blue-shifted and stronger at 36~GHz.

\item We suggest a refined rest frequency value of 36169.238$\pm$0.011~MHz for the 4$_{-1}-$3$_0$~E
methanol transition.

\item With the exception of high-velocity features, the velocity distribution with respect to the systemic
velocity (estimated as the middle of the velocity interval of the associated class~II methanol maser 
at 6.7~GHz) was found to be Gaussian with a mean of $-$0.57$\pm$0.06 and $-$0.57$\pm$0.07~\kss 
for the 36- and 44-GHz masers, respectively, and a standard 
deviation of 3.65$\pm$0.05 and 3.32$\pm$0.07~\ks, respectively.

\item We investigated the zone of influence around each YSO location (inferred from a class~II methanol maser at 6.7~GHz).
The probability density function for finding a class~I methanol maser at the given projected linear separation
from the 6.7-GHz methanol maser was found to be well approximated by an exponential drop off with a
scale of 263$\pm$15~milliparsec for both 36 and 44~GHz transition. This observed distribution can be explained
by a uniform spatial density of the class~I masers in the near vicinity of the YSOs (within 0.2$-$0.3~pc) with
a square-law drop off of the spatial density at larger distances. We consider 1~pc to be a good estimate
of the typical size of the zone of influence.

\item The class~I maser emission shows a tendency to have larger spatial and velocity spreads 
when the YSO traced by the class~II methanol maser at 6.7~GHz has an associated hydroxyl maser 
and therefore is considered to be more evolved.

\item The distribution of FWHMs for groups of Gaussian components range 
from 0.12 to 9.3~\ks, and peaks at FWHM of 0.3~\ks, clearly characteristic of masers; the
weak wings extend to large widths which overlap with expectations for thermal lines.
\end{enumerate}

\section*{Acknowledgments}
The authors would like to thank Luke Hindson and Luis Zapata for providing their published
continuum images in machine-readable form.
The Australia Telescope is funded by the
Commonwealth of Australia for operation as a National Facility managed
by CSIRO. This research would not be possible without NASA's
Astrophysics Data System Abstract Service. This research has made use of data products from the
GLIMPSE survey, which is a legacy science program of the {\em Spitzer
Space Telescope}, funded by the National Aeronautics and Space
Administration, and the NASA/IPAC Infrared
Science Archive, which is operated by the Jet Propulsion Laboratory,
California Institute of Technology, under contract with the National
Aeronautics and Space Administration. This research made use of APLpy,
an open-source plotting package for Python hosted at {\it http://aplpy.github.com}.

\bsp

\label{lastpage}

\end{document}